\documentclass[11pt]{article}
\pdfoutput=1

\usepackage{a4wide}
\usepackage{amssymb,amsmath,cite,footnpag}
\usepackage{times,nicefrac}
\usepackage{graphicx}
\allowdisplaybreaks

\numberwithin{equation}{section}

\long\def\symbolfootnote[#1]#2{\begingroup%
\def\thefootnote{\fnsymbol{footnote}}\footnote[#1]{#2}\endgroup}
\usepackage{times}
\usepackage{graphicx}
\usepackage{amsmath}
\usepackage{epsfig}
\usepackage{amssymb,amsfonts}
\def\er{\mathbb{R}}

\def\zi{\mathbb{Z}}
\def\ci{\mathbb{C}}
\def\di{\mathrm{d}}
\def\su{SU(2)}
\def\slr{SL(2,\mathbb{R})}
\def\slc{SL(2,\mathbb{R})/U(1)}

\def\p{\partial}
\def\pb{\bar{\partial}}
\def\Id{{\rm 1\kern-.28em I}}
\def\ds{\displaystyle}
\newcommand{\oao}[2]{{#1\atopwithdelims[]#2}}

\begin{document}

\begin{titlepage}

\rightline{\vbox{\small\hbox{\tt CPHT-092.0909}}}
\vskip 2.5cm

\centerline{\LARGE  Heterotic Resolved Conifolds with Torsion,}
\vskip 0.4cm
\centerline{\LARGE from Supergravity to CFT }

\vskip 1.6cm
\centerline{\bf L.~Carlevaro$^\diamond$ and D.~Isra\"el$^\spadesuit$\symbolfootnote[2]{Email:
carlevaro@cpht.polytechnique.fr,israel@iap.fr}}
\vskip 0.5cm
\centerline{\sl $^\diamond$Centre de Physique Th\'eorique, Ecole Polytechnique,
 91128 Palaiseau, France\footnote{Unit\'e mixte de Recherche
7644, CNRS -- \'Ecole Polytechnique}}

\vskip 0.3cm
\centerline{\sl $^\spadesuit$Institut d'Astrophysique de Paris,
98bis Bd Arago, 75014 Paris, France\footnote{Unit\'e mixte de Recherche
7095, CNRS -- Universit\'e Pierre et Marie Curie}
}

\vskip 1.4cm

\centerline{\bf Abstract} \vskip 0.5cm 

We obtain a family of heterotic supergravity backgrounds describing non-K\"ahler warped conifolds with three-form flux 
and an Abelian gauge bundle, preserving $\mathcal{N}=1$ supersymmetry in four dimensions. At large distance from the singularity 
the usual Ricci-flat conifold is recovered. By performing a $\mathbb{Z}_2$ orbifold of the $T^{1,1}$ base, the conifold singularity 
can be blown-up to a four-cycle, leading to a completely smooth geometry. Remarkably, the throat regions of the solutions, which can be 
isolated from the asymptotic Ricci-flat geometry using a double-scaling limit, possess a worldsheet \textsc{cft} description in 
terms of heterotic cosets whose target space is the warped resolved orbifoldized conifold. Thus this construction provides exact 
solutions of the modified Bianchi identity. By solving algebraically these \textsc{cft}s we compute the exact tree-level heterotic 
string spectrum  and describe worldsheet non-perturbative effects. The holographic dual of these solutions, in particular their confining behavior, and the embedding of these fluxed singularities into heterotic compactifications with torsion are also  discussed.

\noindent

\vfill

\end{titlepage}

\tableofcontents

\section{Introduction}

Heterotic compactifications to four dimensions have acquired over the years a cardinal interest for phenomenological applications, as their geometrical data combined with the 
specification of a holomorphic gauge bundle have played a major role in recovering close relatives to the \textsc{mssm} or intermediate \textsc{gut}s. However, as their type 
\textsc{ii} counterparts,  heterotic Calabi-Yau compactifications are generally plagued with the presence of unwanted scalar degrees of freedom at low-energies. 

A  fruitful strategy to confront this issue has proven to be the  inclusion of fluxes through well-chosen cycles in the compactification manifold. Considerable effort has been successfully 
invested in engineering such constructions in type \textsc{ii} supergravity scenarii (see~\cite{Grana} for a review and references therein). However, if one is eventually to uncover the 
quantum theory underlying these backgrounds, warranting their consistency as string theory vacua, or to evade the large-volume limit where supergravity is valid, 
one has to face the presence of \textsc{rr} fluxes intrinsic to these type \textsc{ii} backgrounds, for which a worldsheet analysis is still lacking. 

In this respect, heterotic geometries with \textsc{nsns} three-form and gauge fluxes  are more likely to allow for such a  description~; the dilaton not being stabilized 
perturbatively the worldsheet theory should be amenable to standard \textsc{cft} techniques. The generic absence of  large-volume limit in heterotic  flux compactifications makes even this 
appealing possibility a {\it necessity}. An attempt in uncovering an underlying worldsheet theory for heterotic flux vacua has been made 
in~\cite{Adams:2006kb,Adams:2009zg} by resorting to  linear sigma-model techniques. This approach however yields 
a fully tractable description only in the \textsc{uv}, while the interacting \textsc{cft}  obtained in the \textsc{ir} is not known explicitly.

A consistent smooth heterotic compactification requires determining a gauge bundle that satisfies a list of consistency conditions. This sheds yet another light on the appearance of 
non-trivial Kalb-Ramond fluxes, now understood as the departure, triggered by the choice of an alternative gauge bundle, from the standard embedding of the spin connection into the 
gauge connection that characterizes Calabi-Yau compactifications. This eventually leads to geometries with torsion. Now, heterotic flux compactifications, although 
known for a long time  (see e.g.~\cite{Strominger:1986uh,Hull:1986kz,Becker:2002sx,Curio:2001ae,Louis:2001uy,LopesCardoso:2003af,Becker:2003yv,Becker:2003sh,Becker:2005nb,Benmachiche:2008ma}) 
are usually far less understood that their type \textsc{iib} counterparts.\footnote{Note that duality can then  applied to specific such heterotic models to map them to type \textsc{ii} 
flux compactifications of interest for moduli stabilization~\cite{Dasgupta:1999ss,Becker:2002sx}.}  

In particular having a non-trivial $\mathcal{H}$-flux threading the geometry results in the metric loosing K\"ahlerity (see~\cite{Goldstein:2002pg} for the analysis of $T^2$ fibrations over $K3$) 
and being conformally balanced instead of Calabi-Yau~\cite{Mich,Ivanov:2000ai,LopesCardoso:2002hd}. This proves as a major drawback for the analysis of such backgrounds, as theorems of K\"ahler 
geometry (such as Yau's theorem) do not hold anymore, making the existence of solutions to the tree-level supergravity equations dubious, let alone their extension to  exact string vacua. 
An additional and general complication for heterotic solutions comes from anomaly cancellation, which requires satisfying the Bianchi identity in the presence of torsion. This usually proves 
notoriously arduous as this differential constraint is highly non-linear. A proof of the existence of a family of smooth solutions to the leading-order Bianchi identity has only 
appeared recently~\cite{Fu:2006vj} (see also~\cite{Goldstein:2002pg} for an earlier discussion of $T^2\times K3$ fibrations, as well as~\cite{Becker:2006et,Fu:2008ga,Becker:2008rc} for developments).

Moduli spaces of heterotic compactifications  have singularities, that arise whenever the gauge bundle degenerates to 'point-like instantons', either at regular points or at singular points of the 
compactification manifold. In the case of $\mathcal{N}=1$ compactifications in six dimensions, the situation is well understood. Point-like instantons at regular points of $K3$ signals the appearance 
of  non-perturbative $Sp(k)$ gauge groups in the case of $Spin(32)/\mathbb{Z}_2$~\cite{Witten:1995gx}, while for $E_8 \times E_8$ one gets  tensionless \textsc{bps} strings~\cite{Ganor:1996mu}, leading 
to interacting \textsc{scft}s. In both cases, 
the near-core 'throat' geometry  of  small instantons is given by the heterotic solitons of Callan, Harvey and Strominger~\cite{Callan:1991dj} (called thereafter \textsc{chs}), that become heterotic 
five-branes in the point-like limit.  
In the case of four-dimensional $\mathcal{N}=1$ \textsc{cy}$_3$ compactifications, let alone torsional vacua, the situation is less understood. For a particular class of \textsc{cy}$_3$ which are $K3$ 
fibrations, one can resort to the knowledge of the six-dimensional models mentioned above --~advocating an 'adiabatic' argument~-- in order to understand the physics in the vicinity of such singularities~\cite{Kachru:1996ci,Kachru:1997rs}.

Recently, a study of heterotic flux backgrounds, supporting an Abelian line bundle, has been initiated~\cite{Carlevaro:2008qf}. 
In a specific double-scaling limit of these torsional vacua, the corresponding worldsheet non-linear sigma model has been shown to admit a solvable 
\textsc{cft} description, belonging to a  particular class of gauged \textsc{wzw} models, whose partition function and low-energy spectrum could be established. 
In the double-scaling limit where this \textsc{cft} description emerges, one obtains non-compact torsional manifolds, that can be viewed  as local models of 
heterotic flux compactifications, in the neighborhood of singularities supporting Kalb-Ramond and magnetic fluxes.  In analogy with the Klebanov--Strassler 
(\textsc{ks}) solution~\cite{Klebanov:2000hb}, which plays a central role in understanding type \textsc{iib} flux backgrounds~\cite{Giddings:2001yu}, these 
local models  give a good handle on degrees of freedom localized in the 'throat' geometries.

The solutions we are considering correspond to the near-core geometry of 'small' gauge instantons  sitting on geometrical singularities, 
and their resolution. Generically, the torsional nature of the geometry can come solely from the {\it local} backreaction of the gauge instanton (as for the \textsc{chs} 
solution that corresponds to a gauge instanton on a $K3$ manifold that is  globally torsionless), or thought of being part of a {\it globally} torsional compactification.\footnote{For certain choices of 
gauge bundle, the Eguchi-Hanson model that we studied in~\cite{Carlevaro:2008qf} could be of both types.} From the point of view 
of the effective four- or six-dimensional theory, these solutions describe (holographically) the physics taking place at  non-perturbative transitions of the 
sort discussed above, or in their neighborhood in moduli space.

In the present work we concentrate on heterotic flux backgrounds preserving $\mathcal{N}=1$ supersymmetry  in four dimensions. More specifically 
we consider codimension four conifold singularities~\cite{Candelas:1989js}, supplemented by a non-standard gauge bundle which induces non-trivial torsion in the $SU(3)$ structure connection. For definiteness we opt for $Spin (32)/\mathbb{Z}_2$ heterotic string theory. The Bianchi identity is satisfied for an appropriate Abelian bundle, which solves the differential constraint in the large charge limit, where the curvature correction to the identity becomes sub-dominant.  Subsequently, numerical solutions to the $\mathcal{N}=1$ supersymmetry equations~\cite{Strominger:1986uh} can be found, which feature non-K\"ahler spaces corresponding to warped torsional conifold geometries with a non-trivial dilaton. At large distance from the singularity, their geometry reproduces the usual Ricci-flat conifold, while in the bulk we observe a squashing of the $T^{1,1}$ base, as the radius of its $S^1$ fiber is varying. 

The topology of this class of torsional spaces allows to resolve the conifold singularity by a blown-up $\mathbb{C}P^1 \times \mathbb{C}P^1$ 
four-cycle, provided we consider a $\mathbb{Z}_2$ orbifold of the original conifold space, that 
avoids the potential bolt singularity. In contrast, in the absence of the orbifold only small resolution by a 
blown-up two-cycle or deformation to a three-cycle remain as possible resolutions of the singularity. The specific de-singularisation we are considering here is particularly amenable to heterotic  or 
type~I constructions, as it leads to a normalizable harmonic two-form which can support an extra magnetic gauge flux (type \textsc{iib} conifolds with blown-up four-cycles and D3-branes were discussed in~\cite{PandoZayas:2001iw,Lu:2002rk,Benvenuti:2005qb}). The numerical supergravity solutions found in this case are perfectly smooth everywhere, and the string coupling can be chosen everywhere small, 
while in the blow-down limit the geometrical singularity is also a strong coupling singularity. 

In the regime where the blow-up parameter $a$ is significantly smaller (in string units) than the norm of the vectors of magnetic charges, one 
can define a sort of 'near-horizon' geometry of this family of solutions, where the warp factor acquires a power-like behavior. This region can  be decoupled from the 
asymptotic Ricci-flat region by defining a double scaling limit~\cite{Carlevaro:2008qf} which sends the asymptotic string coupling $g_s$ to zero, 
while keeping the ratio $g_s/ a^2$ fixed in string units. 

In this limit we are able to find an analytical solution (that naturally gives an accurate 
approximation of the asymptotically Ricci-flat solution in the near-horizon region of 
the latter), where the dilaton becomes asymptotically linear, while the effective 
string coupling, defined at the bolt, can be set to any value by the double-scaling parameter.

Remarkably, the double-scaling limit of this family of torsional heterotic backgrounds admits a
solvable worldsheet \textsc{cft} description, which we construct explicitly in terms of an asymmetric 
gauged \textsc{wzw} model,\footnote{Notice that gauged \textsc{wzw} models for a class 
of $T^{p,q}$ spaces have been constructed in~\cite{PandoZayas:2000he}. However these cosets, that are not heterotic in nature and do not support gauge bundles, 
cannot be used to obtain supersymmetric string backgrounds.}  which is  parametrised by the two vectors $\vec{p}$ and $\vec{q}$ (dubbed hereafter 'shift vectors') 
giving the embedding of the two magnetic fields in the Cartan subalgebra of $\mathfrak{so}(32)$. We establish this correspondence by showing that, integrating out 
classically the worldsheet gauge fields, one obtains a non-linear sigma-model whose  background fields reproduce the warped resolved orbifoldized conifold with flux. 
This result generalizes the \textsc{cft} description for heterotic gauge bundles over Eguchi-Hanson (\textsc{eh}) space or \textsc{eh}$\times T^2$  we achieved in a previous work~\cite{Carlevaro:2008qf}.

The existence of a  worldsheet \textsc{cft} for this class of smooth conifold solutions first implies that these backgrounds are exact heterotic string vacua to all orders in $\alpha'$, once included the worldsheet quantum corrections to the defining gauged \textsc{wzw} models. 
This can be carried out by using the method developed in~\cite{Tseytlin:1992ri,Bars:1993zf,Johnson:2004zq} and usually amounts to a finite correction to the metric. Furthermore, this also entails that the Bianchi identity is exactly satisfied even when the magnetic charges are not large, at least in the near-horizon regime.

Then, by resorting to the algebraic description of coset \textsc{cft}s, we establish the full tree-level string spectrum 
for these heterotic flux vacua, with special care taken in treating both discrete and continuous representations corresponding 
respectively to states whose wave-functions are localized near the singularity, and to states whose wave-functions are delta-function normalizable.

Dealing with arbitrary shift vectors $\vec{p}$ and $\vec{q}$ in full generality turns out to be technically cumbersome, as the arithmetical 
properties of their components play a role in the construction.  We therefore choose to work out the complete solution of the theory 
for a simple class  of shift vectors that satisfy all the constraints. We compute the one-loop partition function in this case 
(which vanishes thanks to space-time supersymmetry), and study in detail the spectrum of localized massless states. 

In addition, the \textsc{cft} construction given here provides information about worldsheet instanton corrections. These worldsheet 
non-perturbative effects are captured by  Liouville-like interactions correcting the sigma-model action, that are expected to correspond to 
worldsheet instantons wrapping one of the $\mathbb{C}P^1$s of the four-cycle.  We subsequently analyze under which conditions the Liouville potentials 
dictated by the consistency of the \textsc{cft} under scrutiny are compatible with the whole construction (in particular with the orbifold 
and~\textsc{gso} projections). This allows to understand known constraints in heterotic supergravity vacua (such as the constraint 
on the first Chern class of the gauge bundle) from a worldsheet perspective. 

Finally, considering that in the double-scaling limit we mentioned above these heterotic torsional vacua feature an asymptotically  linear dilaton, we argue that they should 
admit a holographic description~\cite{Aharony:1998ub}. The dual theory should be a novel kind of little string theory, specified by 
the shift vector $\vec{p}$ in the \textsc{uv}, flowing at low energies to  a four-dimensional $\mathcal{N}=1$ field theory. This theory 
sits on a particular branch in its moduli space, corresponding to the choice of second shift vector $\vec{q}$, and parametrized by 
the blow-up mode.  We use the worldsheet \textsc{cft} description of the gravitational dual in order to study the chiral operators of this 
four-dimensional theory, thereby obtaining the R-charges and representations under the global symmetries for a particular class of them. From the 
properties of the heterotic supergravity solution, we argue that the $Spin(32)/\mathbb{Z}_2$ blown-up backgrounds seem to be confining, 
while for the $E_8\times E_8$ theory the blow-down limit gives an interacting superconformal field theory.

This work is organized as follows. Section~2 contains a short review of supersymmetric heterotic flux compactifications. In section~3 we obtain the heterotic supergravity backgrounds of interest, featuring torsional smooth conifold solutions. 
We provide the numerical solutions for the full asymptotically Ricci-flat vacua together with 
the analytical solution in the double-scaling limit. In addition we study the torsion classes of these solutions and their (non-)K\"ahlerity. 
In section~4 we discuss the corresponding worldsheet \textsc{cft} by identifying the relevant heterotic gauged \textsc{wzw} model. 
In section~5 we explicitly construct the complete one-loop partition function and analyze worldsheet non-perturbative effects. 
Finally in section~6 we summarize our results and discuss two important aspects: the holographic duality 
and the embedding of these non-compact torsional backgrounds in heterotic compactifications. In addition, some details about the gauged \textsc{wzw} models 
at hand and general properties of superconformal characters are given in two appendices. 

\section{$\mathcal{N}$=1 Heterotic vacua with Torsion}
In this section we review some known facts about heterotic supergravity and compactifications to four dimensions preserving $\mathcal{N}=1$ supersymmetry. This will 
in particular fix the various conventions that we use in the rest of this work. 

\subsection{Heterotic supergravity}

The bosonic part of the ten-dimensional heterotic supergravity action reads (in string frame):
\begin{equation}\label{het-lag-sugra}
S=\frac{1}{\alpha'^4}\int \di^{10}x \sqrt{-G}e^{-2\Phi}\,\Big[
R+4|\partial\Phi|^2-\frac{1}{2}|\mathcal{H}|^2+\frac{\alpha'}{4}\big(\mbox{Tr}_{\text{V}}|\mathcal{F}|^2+\text{tr}|\mathcal{R}_+|^2\big)
\Big]\,.
\end{equation}
with the norm of a $p$-form field strength $\mathcal{G}_{[p]}$ defined as $|\mathcal{G}|^2=\nicefrac{1}{p!}\,\mathcal{G}_{M_1..M_p}\mathcal{G}^{M_1..M_p}$. The trace of the Yang-Mills kinetic term is taken in the 
vector representation of $SO(32)$ or $E_8\times E_8$.\footnote{We have chosen to work with anti-hermitian gauge fields, hence the positive sign in front of the gauge kinetic term.}

To be in keep with the modified Bianchi identity below~(\ref{bianchi}), we have included in~(\ref{het-lag-sugra}) the leading string correction to the supergravity Lagrangian. It involves the
generalized curvature two-form $\mathcal{R}(\Omega_{+})^{A}_{\phantom{A}B}$ 
built out of a Lorentz spin connexion $\Omega_+$ that incorporates torsion, generated by the presence of a non-trivial \textsc{nsns} three-form flux:\footnote{Its contribution to~(\ref{het-lag-sugra}) is normalized as
$\text{tr}|\mathcal{R}_+|^2= \tfrac12\, \mathcal{R}(\Omega_+)_{MN\,AB}\mathcal{R}(\Omega_+)^{MN\,AB}
$, the letters $M,N$ and $A,B$ denoting the ten-dimensional coordinate and frame indices, respectively.}
\begin{equation}\label{genO}
\Omega^{\phantom{\pm}A}_{\pm\,\phantom{A}B} =  \omega^{A}_{\phantom{A}B} \pm \tfrac12 \mathcal{H}^{A}_{\phantom{A}B}\,.
\end{equation}
 
In addition to minimizing the action~(\ref{het-lag-sugra}), a heterotic vacuum has to fulfil the generalized Bianchi identity:
\begin{equation}\label{bianchi}
\di \mathcal{H}_{[3]} = 8\alpha'\pi^2 \Big[\text{ch}_2\big(V \big)-p_1\big(\mathcal{R}(\Omega_{+})\big)
 \Big]\,,
\end{equation}
here written in terms of the first Pontryagin class of the tangent bundle and the second Chern character of the gauge bundle $V$. The second topological 
term on the right hand side is the leading string correction to the Bianchi identity required by anomaly cancellation~\cite{Green:1984sg}, and mirrors the one-loop correction on the \textsc{lhs} of~(\ref{het-lag-sugra}).\footnote{
Actually, one can add any torsion piece to the spin connexion $\Omega_+$ without spoiling anomaly cancellation~\cite{Hull:1985dx}.}

By considering gauge and Lorentz Chern-Simons couplings, one can now construct an \textsc{nsns} three-form which exactly solves the modified Bianchi identity (\ref{bianchi}):
\begin{equation}\label{3form}
\mathcal{H}_{[3]}= \di \mathcal{B}_{[2]}+ \alpha' \big( \omega_{[3]}^{L}(\Omega_+)-\omega_{[3]}^{\textsc{ym}}(\mathcal{A})\big) \, ,
\end{equation}
thus naturally including tree-level and one-loop corrections, given by:
\begin{equation}\label{CSform}
\omega_{[3]}^{\textsc{ym}}(\mathcal{A}) = \text{Tr}_{\textsc{v}}\left[ \mathcal{A}\wedge \di\mathcal{A} + \tfrac23 \,\mathcal{A}\wedge\mathcal{A}\wedge\mathcal{A}\right]
\,, \qquad
\omega_{[3]}^{L}(\Omega_{+})=\text{tr}\left[ \Omega_+\wedge \di\Omega_+ + \tfrac23 \,\Omega_+\wedge\Omega_+ \wedge\Omega_+\right]\,.
\end{equation}

\subsection{$\mathcal{N}$=1 supersymmetry and  SU(3) structure}
In the absence of fermionic background, a given heterotic vacuum can preserve a portion of supersymmetry if there exists at least one Majorana-Weyl spinor $\eta$ of $Spin(1,9)$ satisfying
\begin{equation}\label{covar}
\nabla^{-}_{M}\eta\equiv \big( \partial_M  + \tfrac14\, \Omega_{-\phantom{AB}M}^{\phantom{-}AB} \,\Gamma_{AB}\big)\,\eta
=0\,.
\end{equation}
i.e. covariantly constant with respect to the connection with torsion $\Omega_-$ (note that the Bianchi identity is expressed using $\Omega_+$). 
This constraint induces the vanishing of the supersymmetry variation of the graviton, so that in the presence of a non-trivial dilaton and gauge field 
strength extra conditions have to be met, as we will see below.
 
In the presence of flux, the conditions on this globally invariant spinor are related
to the possibility for the manifold in question to possess a reduced structure group, or
$G$-structure, which becomes the $G$ holonomy of $\nabla^-$ when the fluxes vanish (see~\cite{Salomon,Joyce,Gauntlett:2003cy} for details and review).
The requirements for a manifold $\mathcal{M}_d$ to be endowed with a $G$-structure is tied to its
frame bundle admitting a sub-bundle with fiber group $G$. This in turn implies the existence of a set of globally defined $G$-invariant tensors, 
or alternatively, spinors on $\mathcal{M}_d$.
As will be exposed more at length in section~\ref{torsion-cl}, the $G$-structure is specified by the intrinsic torsion of the manifold, 
which measures the failure of the $G$-structure to become a $G$ holonomy of $\nabla^-$. By decomposing the intrinsic torsion into 
irreducible $G$-modules, or torsion classes, we can thus consider classifying and determining the properties of different flux compactifications admitting the same $G$-structure.

\subsubsection*{Manifolds with SU(3) structure}
In the present paper, we will restrict to six-dimensional Riemannian spaces $\mathcal{M}_6$, whose reduced structure group is a subgroup of $SO(6)$, and focus on compactifications preserving 
minimal ($\mathcal{N}=1$) supersymmetry in four dimensions, which calls for an $SU(3)$ structure group.\footnote{As a 
general rule, reducing the dimension of the structure group increases the number of preserved supercharges.}

The structure is completely determined by a real two-form $J$ and a complex three-form $\Omega$,\footnote{The $SU(3)$ structure 
is originally specified the chiral complex spinor $\eta$ solution of~(\ref{covar}),  $J$ and $\Omega$ being then defined as 
$J_{mn}=-i\eta^{\dagger}\Gamma_{mn}\eta$ and $\Omega_{mnp}=\eta^{\top}\Gamma_{mnp}\eta$ respectively. In the following however we will not resort to this formulation.} which are globally defined and
satisfy the relations:
\begin{equation}\label{topcond}
\Omega\wedge \bar\Omega = -\frac{4i}{3}\, J\wedge J\wedge J\,,
\qquad \qquad 
J\wedge \Omega =0\,.
\end{equation}
The last condition is related to the absence of $SU(3)$-invariant vectors or, equivalently, five-forms. 

The 3-form $\Omega$ suffices to determine an almost complex structure $\mathcal{J}_m^{\ n}$, satisfying $\mathcal{J}^2=-\mathbb{I}$, 
such that $\Omega$ is  $(3,0)$ and $J$ is $(1,1)$. The metric  on $\mathcal{M}_6$ is then given by $g_{mn}=\mathcal{J}_m^{\phantom{m}l}J_{ln}$, and the orientation of $\mathcal{M}_6$ is implicit in the choice of volume-form $\text{Vol}(\mathcal{M}_6)=(J\wedge J\wedge J)/6$.

For a background including \textsc{nsns} three-form flux $\mathcal{H}$, the structure $J$ and $\Omega$ is generically not closed anymore, so that $\mathcal{M}_6$ now departs from the usual Ricci-flat 
\textsc{cy}$_3$ background and $SU(3)$ holonomy is lost.

\subsubsection*{Supersymmetry conditions}
We consider a heterotic background in six dimensions specified by a metric $g$, a dilaton $\Phi$, a three-form $\mathcal{H}$ and a gauge field strength $\mathcal{F}$.

Leaving aside the gauge bundle for the moment, it can be shown that preserving $\mathcal{N}=1$ supersymmetry in six dimensions is strictly equivalent 
to solving the differential system for the $SU(3)$ structure:\footnote{
The original and alternative and formulation \cite{Strominger:1986uh} to the supersymmetry conditions (\ref{susy-cond}-\ref{eqH}) replaces the constraint on the top form by
$|\Omega|=e^{-2\Phi}$,
which, inserted in eq.(\ref{susy-cond}), implies that the metric is conformally balanced \cite{Mich,Ivanov:2000ai,Becker:2006et}. The calibration equation for the flux (\ref{eqH}) can
also be rephrased as $\mathcal{H}=i(\pb-\p)J$. 
This latter version of eq.(\ref{eqH}) is however restricted to the $SU(3)$-structure case, and does not lift to a general G-structure and dimension 
$d$, unlike (\ref{susy-cond}a) by replacing $J$ by the appropriate calibration $(d-4)$-form $\Xi$ (see for instance \cite{Gauntlett:2001ur}).}
\begin{subequations}\label{susy-cond}
\begin{align}
\di (e^{-2\Phi}\,\Omega)&=0\,,\\
\di (e^{-2\Phi}\,J\wedge J)&=0\,,
\end{align}
\end{subequations}
with the \textsc{nsns} flux related to the structure as follows \cite{Gauntlett:2001ur}:
\begin{equation}\label{eqH}
e^{2\Phi}\,\di(e^{-2\Phi}\,J) = \star_6 \mathcal{H}\,.
\end{equation}
Let us pause awhile before tackling the supersymmetry constraint on the gauge fields and dwell on the signification of this latter expression. It has been observed that the 
condition~(\ref{eqH}) reproduces a generalized K\"ahler calibration equation for $\mathcal{H}$~\cite{Gutowski:1999iu,Gutowski:1999tu}, since it is defined by the $SU(3)$-invariant $J$. If we adopt a brane interpretation of a background with \textsc{nsns} flux, this equation acquires significance as a minimizing condition for the energy functional of five-branes wrapping  K\"ahler two-cycles in $\mathcal{M}_6$. As noted in~\cite{Gauntlett:2003cy}, this analysis in term of calibration is still valid even when considering the full back-reaction of the brane configuration on the geometry.\footnote{The argument is that we can always add in this case an extra probe five-brane without breaking supersymmetry, provided it wraps a two-cycle calibrated by the same invariant form as the one calibrating the now back-reacted solution, hence the name \it{generalized} calibration.}

\subsection{Constraints on the gauge bundle}
We will now turn  to the conditions the gauge field strength has to meet in order to preserve
$\mathcal{N}=1$ supersymmetry and to ensure the absence of  global worldsheet anomalies.

Unbroken supersymmetry requires the vanishing of the gaugino variation:
\begin{equation}\label{gaugino}
\delta \chi =  \frac14\,\mathcal{F}_{MN} \,\Gamma^{MN} \epsilon = 0\,.
\end{equation}
We see that since the covariantly constant spinor $\eta$ is a singlet of the connection $\nabla^-$,
taking $\mathcal{F}$ in the adjoint of the structure group $SU(3)$ will not break any extra supersymmetry, thus automatically satisfying~(\ref{gaugino}).  
This is tantamount to requiring $\mathcal{F}$ to be an instanton of $SU(3)$:
\begin{equation}\label{Finstanton}
\mathcal{F}_{mn}= -\frac{1}{4}\,\big(J\wedge J\big)_{mn}^{\phantom{mn}kl}\,\mathcal{F}_{kl}
\qquad \Longleftrightarrow\qquad 
\star_6 \mathcal{F}=-\mathcal{F}\wedge J\,.
\end{equation}
As pointed out in \cite{Strominger:1986uh}, this condition is equivalent to require the gauge bundle $V$ to satisfy the zero-slope limit of the Hermitian Yang-Mills equation:
\begin{subequations}\label{hym} 
\begin{align}
\mathcal{F}^{(2,0)}= \mathcal{F}^{(0,2)}&=0\,, \\ \mathcal{F}^{a\bar b}J_{a\bar b}&=0\,.
\end{align}
\end{subequations}
The first equation entails that the gauge bundle has to be a holomorphic gauge bundle while the second is the tree-level Donaldson-Uhlenbeck-Yau (\textsc{duy}) condition which is satisfied for $\mu$-stable bundles.

In addition, a line bundle is subject to a condition ensuring the absence of global anomalies in the heterotic worldsheet sigma-model~\cite{Witten:1985mj,Freed:1986zx}. This condition 
(also known as K-theory constraint in type~I) amounts to a Dirac quantization condition for the $Spin(32)$ spinorial representation  of positive chirality, that 
appears in the massive spectrum of the heterotic string. It forces the first Chern class of the gauge bundle $V$ over $\mathcal{M}_6$ to be in the second even integral cohomology group. 
In this work we consider only Abelian gauge backgrounds, hence the bundle needs to satisfy the condition:
\begin{equation}\label{Kth}
c_1(V) \in H^2(\mathcal{M},2\zi) \, \Longrightarrow \,\sum_{i=1}^{16} \int_{\Sigma_I} \frac{\mathcal{F}^i}{2\pi}\equiv 0 \mbox{ mod }2\,,\quad I=1,..,h_{1,1} \, .
\end{equation}

\section{Resolved Heterotic Conifolds with Abelian Gauge Bundles}

The supergravity solutions we are interested in are given as 
a non-warped product of four-dimensional Minkowski space with a six-dimensional 
non-compact manifold supporting \textsc{nsns} flux and an Abelian gauge bundle. 
They preserve minimal supersymmetry ($\mathcal{N}=1$) in four dimensions and can be viewed 
as local models of flux compactifications. For definiteness we choose $Spin(32)/\mathbb{Z}_2$ heterotic strings.

More specifically we take as metric ansatz a warped conifold geometry~\cite{Candelas:1989js}. The singularity is resolved by a K\"ahler deformation corresponding to blowing up a  
$\ci P^1\times \ci P^1$ four-cycle on the conifold base. This is topologically possible only for a $\mathbb{Z}_2$ orbifold of the conifold, see below.\footnote{Without an orbifold the conifold singularity can be smoothed 
out only by a two-cycle (resolution) or a three-cycle (deformation).}  The procedure is 
similar to that used in~\cite{PandoZayas:2001iw,Benvenuti:2005qb} to construct a smooth Ricci-flat orbifoldized conifold  
by a desingularization \`a la Eguchi-Hanson. In our case however we have in addition non-trivial flux back-reacting on the geometry and deforming it away from Ricci-flatness by generating torsion in the background.

The geometry  is conformal to a six-dimensional smoothed cone over a $T^{1,1}$ space.\footnote{We recall that $T^{1,1}$ is the  coset 
space $(SU(2)\times SU(2))/U(1)$ with the $U(1)$ action embedded symmetrically in the two $SU(2)$ factors.} It has therefore an 
$SU(2)\times SU(2)\times U(1)$ group of continuous isometries. Considering $T^{1,1}$ as an  $S^1$ fibration over a $\ci P^1\times \ci P^1$  base, the metric component in front of the fiber will be dependent on the radial coordinate of the cone, hence squashing $T^{1,1}$ away from the Einstein metric.

The metric and \textsc{nsns} three-form ans\"atze of the heterotic supergravity solution are chosen of the following form:
\begin{subequations} \label{sol-ansatz}
\begin{align}
\di s^2 & = \eta_{\mu\nu}\di x^{\mu}\di x^{\nu} +\tfrac{3}{2}\,H(r) \,\biggl[
\frac{\di r^2}{f^2(r)} + 
 \frac{r^2}{6}\big(\di\theta_1^2+\sin^2\theta_1\,\di\phi_1^2 + \di\theta_2^2+\sin^2\theta_2\,\di\phi_2^2\big) \notag\\
 & \hspace{2cm} +\,\frac{r^2}{9} f(r)^2 \big(\di \psi + \cos \theta_1 \,\di \phi_1 + \cos \theta_2 \,\di \phi_2 \big)^2  \biggr]\,,\\
\mathcal{H}_{[3]} &  = \frac{\alpha'k}{6}\,g_1(r)^2\,\big( \Omega_1+\Omega_2\big)\wedge  \tilde{\omega}_{[1]} \,,
\end{align}
\end{subequations}
with the volume forms of the two $S^2$s and the connection one-form $\tilde{\omega}_{[1]}$ defined by
\begin{subequations} 
\begin{align}
\Omega_i&=\sin \theta_i \,\di \theta_i\wedge \di \phi_i\,,\quad \text{for }i=1,2\,,\qquad
\tilde{\omega}_{[1]} = \di \psi + \cos \theta_1 \,\di \phi_1 + \cos \theta_2 \,\di \phi_2\,.
\end{align}
\end{subequations}
In addition, non-zero \textsc{nsns} flux induces a nontrivial dilaton $\Phi(r)$, while satisfying the Bianchi identity requires an Abelian gauge bundle, which will be discussed below.

The resolved conifold geometry in~(\ref{sol-ansatz}a), denoted thereafter by $\tilde{\mathcal{C}}_6$, is topologically equivalent to the 
total space of the line bundle $\mathcal{O}(-K)\rightarrow \ci P^1 \times \ci P^1$. The resolution of the singularity is governed by the 
function $f(r)$ responsible for the squashing of  $T^{1,1}$.  Indeed the zero locus  of this function defines the blowup mode $a$ of the conifold, 
related to the product of the volumes of the two $\ci P^1$'s.

Asymptotically in $r$, the numerical solutions that will be found below are such that both $f$ and $H$ tend to constant values, 
according to $\text{lim}_{r\rightarrow \infty}f=1$ and $\text{lim}_{r\rightarrow \infty} H=H_{\infty}$, hence the known Ricci-flat conifold metric is restored at infinity (however without 
the standard embedding of the spin connexion in the gauge connexion (see below).

To determine the background explicitly, we impose the supersymmetry conditions~(\ref{susy-cond}) and the Bianchi identity~(\ref{bianchi}) on the the ansatz~(\ref{sol-ansatz}), 
which implies~\cite{deWit:1986xg,Gauntlett:2002sc} solving the equations of motion for the Lagrangian~(\ref{het-lag-sugra}). In addition, one has to implement the condition~(\ref{Kth}), 
thereby constraining the magnetic charges specifying the Abelian gauge bundle.

\subsection{The supersymmetry equations}

To make use of the supersymmetry equations~(\ref{susy-cond}) and the calibration condition for the flux~(\ref{eqH}), we choose the following complexification of the vielbein:
\begin{equation}\label{complex-vielbein}
E^1=e^2+ie^3\,, \qquad E^2=e^4+ie^5\,, \qquad E^3=e^1+ie^6\,.
\end{equation}
written in terms of the left-invariant one-forms on $T^{1,1}$:
\begin{equation}\label{vielbn}
\begin{array}{ll}
e^1=\sqrt{\frac{3H}{2}}\frac1f\,\di r\, & 
\quad e^6= \frac{r\sqrt{H}f}{\sqrt{6}}\,\tilde\omega\\[5pt] 
e^2=  \frac{r \sqrt{H}}{2}\,\big(\sin\frac{\psi}{2}\,\di \theta_1- \cos\frac{\psi}{2}\sin\theta_1\,\di\phi_1\big)\,, &\quad
e^3=  -\frac{r\sqrt{H}}{2}\,\big(\cos\frac{\psi}{2}\,\di \theta_1+ \sin\frac{\psi}{2}\sin\theta_1\,\di\phi_1\big)\,, \\[5pt] 
e^4=  \frac{r\sqrt{H}}{2}\,\big(\sin\frac{\psi}{2}\,\di \theta_2- \cos\frac{\psi}{2}\sin\theta_2\,\di\phi_2\big) \,,& \quad
e^5= -\frac{r\sqrt{H}}{2}\,\big(\cos\frac{\psi}{2}\,\di \theta_2+ \sin\frac{\psi}{2}\sin\theta_2\,\di\phi_2\big)\,.
\end{array}
\end{equation}
The corresponding $SU(3)$ structure then reads:
\begin{subequations}
\label{complexbasis}
\begin{align}
\Omega_{[3,0]} &= E^1\wedge E^2 \wedge E^3 \equiv e^{124}-e^{135}-e^{256}-e^{346}+i
\big(e^{125}+e^{134}+e^{246}-e^{356}\big)\,, \\
J_{[1,1]} &=\frac{i}{2} \sum_{a=1}^3 E^a\wedge \bar E^a \equiv e^{16}+e^{23}+e^{45}\,.
\end{align}
\end{subequations}
Imposing the supersymmetry conditions~(\ref{susy-cond}) leads the following system of first order differential equations:
\begin{subequations} \label{susy-system}
\begin{align}
&f^2 H' = f^2 H\,\Phi'  = -\frac{2\alpha'k \,g_1^2}{r^3} \,,\\
&r^3H f f'+3r^2 H\,(f^2-1) +\alpha'k\, g_1^2=0\,.
\end{align}
\end{subequations}

\subsection{The Abelian gauge bundle}

To solve the Bianchi identity~(\ref{bianchi}), at least in the large charge limit, one can consider an Abelian gauge bundle, 
supported both on the four-cycle $\ci P^1\times \ci P^1$ and on the $S^1$ fiber of the squashed $T^{1,1}/\zi_2$:
\begin{equation}\label{gauge-ansatz}
\mathcal{A}_{[1]}=\tfrac{1}{4}\Big(\left(\cos\theta_1\,\di\phi_1 - \cos \theta_2\,\di \phi_2 \right)\vec{p} + g_2(r)\,\tilde\omega\,\vec{q}\Big) \cdot \vec{H}\,.
\end{equation}
where $\vec{H}$ spans the 16-dimensional Cartan subalgebra of $\mathfrak{so}(32)$ and the $H^i$, $i=1,..,16$ are chosen anti-Hermitean, with Killing 
form $K(H^i,H^j)=-2\delta_{ij}$. The solution is characterized by two {\it shift vectors}\footnote{This terminology is borrowed from the 
orbifold limit of some line bundles over singularities, see $e.g.$~\cite{Nibbelink:2007rd}.} $\vec{p}$ and $\vec{q}$ that 
specify the Abelian gauge bundle and are required to satisfy $\vec{p}\cdot\vec{q}=0$. The function $g_2(r)$
will be determined by the \textsc{duy} equations.

The choice~(\ref{gauge-ansatz}) is the most general ansatz of line bundle over the manifold~(\ref{sol-ansatz}a)
satisfying the holomorphicity condition~(\ref{hym}a). Then, to fulfil the remaining supersymmetry condition, we rewrite:
\begin{equation}\label{FS}
\begin{array}{rcl}
\mathcal{F}_{[2]} & = &
{\ds -\tfrac{1}{4} \Big[\big(\Omega_1-\Omega_2\big)\,\vec{p}
 +\big( g_2(r)(\Omega_1+\Omega_2)
  -g_2'(r)\,\di r\wedge \tilde{\omega}\big)\,\vec{q} \,\Big]\cdot \vec{H} } \\[7pt]
   & =& -{\ds \frac{i}{r^2H}\Big[
        \big(E^1\wedge \bar E^1 - E^2\wedge \bar E^2 \big)\,\vec{p} + 
         \big( g_2 \,(E^1\wedge \bar E^1 + E^2\wedge \bar E^2) +  \tfrac12 rg_2'\, E^3\wedge \bar
         E^3\big)\vec q
          \,\Big] \cdot \vec{H}}
\end{array}
\end{equation}
so that imposing (\ref{hym}b) fixes:
\begin{equation}\label{g2-prof}
g_2(r)=\left(\frac a r\right)^4 \,.
\end{equation}
In defining this function we have introduced a scale $a$ which is so far a free real parameter of the solution.  
It will become clear later on that $a$ is the blow-up mode related to the unwarped volume of the four-cycle. 

The function (\ref{g2-prof}) can also be determined in an alternative fashion by observing that the standard singular Ricci-flat conifold possesses two harmonic two-forms, which are also shared by the 
resolved geometry $\tilde{\mathcal{C}}_6$ (see~\cite{Lu:2002rk} for a similar discussion about the Ricci-flat orbifoldized conifold), where they can be written locally as:
\begin{equation}\label{2form}
\varpi_1 = \frac{1}{4\pi}\,\di\big(\cos\theta_1\,\di\phi_1 - \cos \theta_2\,\di \phi_2\big)\,,\qquad
\varpi_2 = \frac{a^4}{4\pi}\,\di\left( \frac{\tilde\omega}{r^4}\right)\,,
\end{equation}
and form a base of two-forms that completely span the gauge field strength:
\begin{equation}\label{F-tf}
\mathcal{F}= \pi\big( \varpi_1\,\vec{p}+ \varpi_2\,\vec{q}\,\big)\cdot \vec{H}
\end{equation}
Note in particular that $\varpi_2$ is normalizable on the warped resolved conifold, while $\varpi_1$ is not, since we have
\begin{equation}
(4\pi)^2\varpi_m \wedge \star_6 \varpi_m =h_m(r)\,\di r\wedge\Omega_1\wedge \Omega_2\wedge \di\psi
\end{equation}
characterized by the functions
\begin{equation}
h_1(r)= r H(r)\,,\qquad h_2(r)= \frac{3 a^8 H(r)}{r^7}\,
\end{equation}
and the conformal factor $H$ is monotonously decreasing with no pole at $r=a$ and asymptotically constant. Thus, contrary to the four-dimensional heterotic solution with a line bundle over  warped
Eguchi-Hanson space~\cite{Carlevaro:2008qf}, the fact that the $\varpi_1$ component of the gauge field is non-normalizable implies that $\mathcal{F}$ has non vanishing charge at infinity, due 
to $\int_{\infty}\varpi_1\neq 0$. 

\subsubsection*{Constraints on the first Chern class of the bundle}

The magnetic fields arising from the gauge background~(\ref{FS}) lead to Dirac-type quantization conditions associated with the compact two-cycles of the geometry. 
We first observe that the second homology $H_2(\tilde{\mathcal{C}}_6,\er)$ of the resolved conifold is spanned by two representative two-cycles
related to the two blown-up $\ci P^1$s pinned at the bolt of $\tilde{\mathcal{C}}_6$:
\begin{equation}
\Sigma_i=\{r=a,\theta_i=\text{const},\phi_i=\text{const},\psi=0\}\,,\quad i=1,2\,.
\end{equation}
One then constructs a dual basis of two-forms, by taking the appropriate combinations of the harmonic two-forms~(\ref{2form}):
\begin{equation}\label{cohom}
L_1 = \tfrac12\big(\varpi_2 -\varpi_1\big)\,,\qquad 
L_2 = \tfrac12\big(\varpi_1 +\varpi_2\big)\,,
\end{equation}
which span the second cohomology $H^2(\tilde{\mathcal{C}}_6,\er)=\er\oplus\er$.\footnote{The K\"ahler form $J$ being non-integrable is
absent from the second cohomology of $\tilde{\mathcal{C}}_6$.} If we now develop the gauge field-strength~(\ref{F-tf}) on the cohomology base~(\ref{cohom}), one gets that 
\begin{equation}
\int_{\Sigma_{1}} \frac{\mathcal{F}}{2\pi} = \frac{1}{2} (\vec{q}-\vec{p})\cdot \vec{H} \quad , \qquad 
\int_{\Sigma_{2}} \frac{\mathcal{F}}{2\pi} = \frac{1}{2} (\vec{q}+\vec{p})\cdot \vec{H} \, .
\end{equation}
Imposing a Dirac quantization condition for the adjoint (two-index) representation leads to the possibilities
\begin{subequations}\label{constr-shiftvect}\begin{align}
&q_\ell\pm p_\ell \equiv 0 \mod 2 \qquad \forall \ell=1,\ldots,16  \qquad  \text{or} &\nonumber  \\ & q_\ell \pm p_\ell \equiv 1 \mod 2 \qquad \forall \ell=1,\ldots,16 \, , 
\end{align}
\end{subequations}
$i.e.$ the vectors $(\vec{p}\pm \vec{q})/2$ have  either all entries integer or all entries half-integer.  
The former corresponds to bundles 'with vector structure' and the latter to  bundles 'without vector structure'~\cite{Berkooz:1996iz}. 
The distinction between these types of bundles is given by the generalized Stiefel-Whitney class $\tilde{w}_2 (V)$, measuring the obstruction to associate the 
bundle $V$ with an $SO(32)$ bundle.  

The vectors $\vec{p}$ and $\vec{q}$ being orthogonal, we choose them to be of the form  
$\vec{p}=(p_\ell,0^{n})$ with $\ell=0,\ldots,16-n$ and 
$\vec{q}=(0^{16-n},q_\ell)$ with $\ell=16-n+1,\ldots,16$. This gives the separate conditions
\begin{equation}
\left\{
\begin{array}{lr}
q_\ell \equiv 0 \mod 2 \ , \ \  p_\ell \equiv 0 \mod 2 \, , & \quad\text{for} \quad \tilde{w}_2 (V) = 0  \, ,\\[4pt]
q_\ell \equiv 1 \mod 2 \ , \ \ p_\ell \equiv 1 \mod 2 \, , & \quad \text{for} \quad\tilde{w}_2 (V) \neq   0\, , 
\end{array} 
\right. \qquad \forall \ell\, .
\label{constr-pandq}
\end{equation}

In addition, as the heterotic string spectrum contains massive states transforming in the spinorial representation of $Spin(32)$ of, say, positive chirality, the shift vectors $\vec p$ and $\vec q$ specifying
the gauge field bundle~(\ref{FS}) have to satisfy the extra constraint~(\ref{Kth}). It yields two conditions:
\begin{equation}\label{Kth2}
\sum_{\ell=1}^{16}\, (p_\ell \pm  q_\ell) \equiv 0 \text{ mod }4\, , 
\end{equation}
which are in fact equivalent for bundles with vector structure. In section~\ref{wnpe}, these specific constraints will be re-derived from  non-perturbative corrections to the worldsheet theory.

\subsection{The Bianchi identity at leading order}
\label{bianchisect}
To determine the radial profile of the three-form $\mathcal{H}$, $i.e.$ the function $g_2(r)$ in the ansatz~(\ref{sol-ansatz}), we need to solve the 
Bianchi identity (\ref{bianchi}); this is generally a difficult task. In the large charges limit $\vec{p}^{\, 2} \gg 1$ (corresponding in the blow-down limit to 
considering the back-reaction of a large number of wrapped heterotic five-branes, see latter), the tree-level contribution to the \textsc{rhs} of the Bianchi identity is 
dominant and the higher derivative (curvature) term can be neglected. Using the gauge field strength ansatz~(\ref{FS}), equation~(\ref{bianchi}) becomes:
\begin{equation}\label{Bianchi1}
\frac{1}{\alpha'}\, \di \mathcal{H}_{[3]}  = 
{\ds \frac{1}{4}\Big(\big[\vec{q}^{\, 2} g_2^2 -\vec{p}^{\, 2}\big]\,\Omega_1\wedge\Omega_2 - 
\vec{q}^{\, 2}\, g_2\, g_2'\,\di r\wedge\big(\Omega_1+\Omega_2\big)\wedge\tilde\omega\Big) +  \mathcal{O}\left(1\right)}\, . 
\end{equation}
Then, using the solution of the \textsc{duy} equations~(\ref{g2-prof}), we obtain:
\begin{equation}\label{g3}
g_1^2(r)=\tfrac34\big[1-g_2^2(r)\big]= 
\tfrac34\Big[1-\left(\frac{a}{r}\right)^8\Big]
\end{equation}
and the norm of the shift vectors are constrainted to satisfy:
\begin{equation}\label{pqk}
\vec{p}^{\, 2} =  \,\vec{q}^{\, 2} = k\,,
\end{equation}
such that the tree-level $F^2$ term on the \textsc{rhs} of the Bianchi identity~(\ref{Bianchi1}) is 
indeed the leading contribution. The relevance of one-loop corrections to $\mathcal{H}$ coming from generalized Lorentz Chern-Simons couplings~(\ref{CSform}) will be discussed below.

Finally, one can define a quantized five-brane charge as asymptotically the geometry is given by a cone over 
$T^{1,1}/\zi_2 \sim \er P_3\times S^2$:
\begin{equation}\label{N5flux}
 Q_5 = \frac{1}{2\pi^2 \alpha'} \int_{\er P_3 , \, \infty}\!\!\!\!\!\! \mathcal{H}_{[3]} \, = \frac{k}{2} \, .
\end{equation}

\subsubsection*{The orbifold of the conifold}
Having determined the functions $g_1(r)$ and $g_2(r)$ governing the $r$ dependence of the torsion three-form and of the  gauge bundle 
respectively, one can already make some important observation. Since function $g_1 (r)$ (\ref{g3}) vanishes for $r=a$, assuming that the 
conformal factor $H(r)$ and its derivative do not vanish there (this will be confirmed by the subsequent numerical analysis),  
eq.~(\ref{susy-system}a) implies that the squashing function $f^2(r)$ also vanishes for $r=a$. Therefore the manifold exhibits a 
$\mathbb{C}P^1\times \mathbb{C}P^1$ bolt, with possibly a conical singularity. 

Then evaluating the second supersymmetry condition (\ref{susy-system}b) at the bolt (where both $f^2$ and $g_1$ vanishes) 
we find that $(f^2)' |_{r\to a_+} = \nicefrac{6}{a}$. With this precise first order expansion of $f^2$ near the bolt, 
the conical singularity  can be removed by restricting the periodicity of the $S^1$ fiber in $T^{1,1}$, as $\psi\sim\psi + 2\pi$ instead of 
the original $\psi\in[0,4\pi[$.  In other words we need to consider a $\mathbb{Z}_2$ orbifold of the conifold, as studied e.g.  in~\cite{Bershadsky:1995sp} in 
the Ricci-flat torsionless case. 
Following the same argument as in~\cite{PandoZayas2}, the deformation parameter $a$ can be related to the volume of the blown-up four-cycle $\ci P^1\times \ci P^1$, 
and thus represents a {\it local} K\"ahler deformation.

One may wonder whether this analysis can be spoiled by the higher-order $\alpha'$ corrections (as we solved only the Bianchi identity at leading order). 
However we will prove in the following that the $\mathbb{Z}_2$ orbifold is also necessary in the full-fledged heterotic worldsheet theory.

\subsection{Numerical solution}

Having analytical expressions for the functions $g_1$ and $g_2$, we can consider solving the first order 
system~(\ref{susy-system}) for the remaining functions $f$ and $H$ that arises from the supersymmetry conditions. 
If we ask the conformal factor $H$ to be asymptotically constant, as expected from a brane-type solution in 
supergravity, the system~(\ref{susy-system}) can only be solved numerically. 

\begin{figure}[!t]
\centering
\includegraphics[width=75mm]{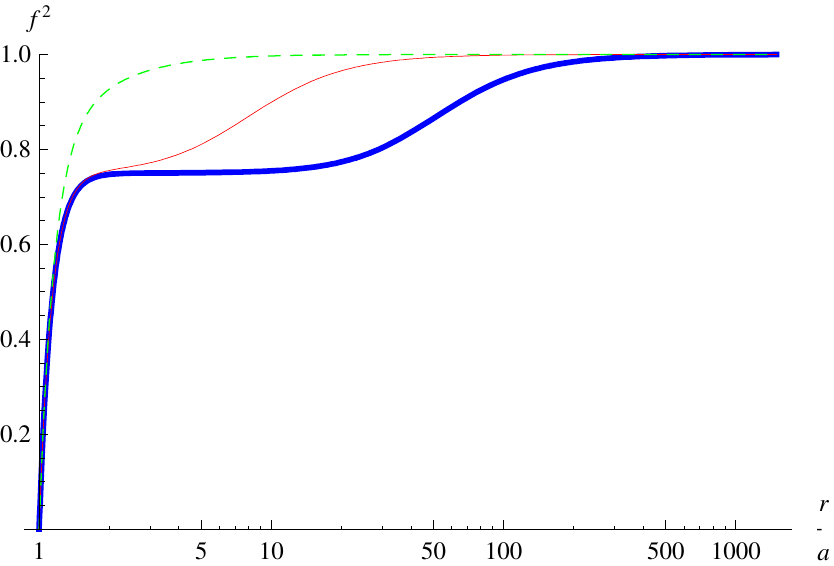}\hskip5mm
\includegraphics[width=75mm]{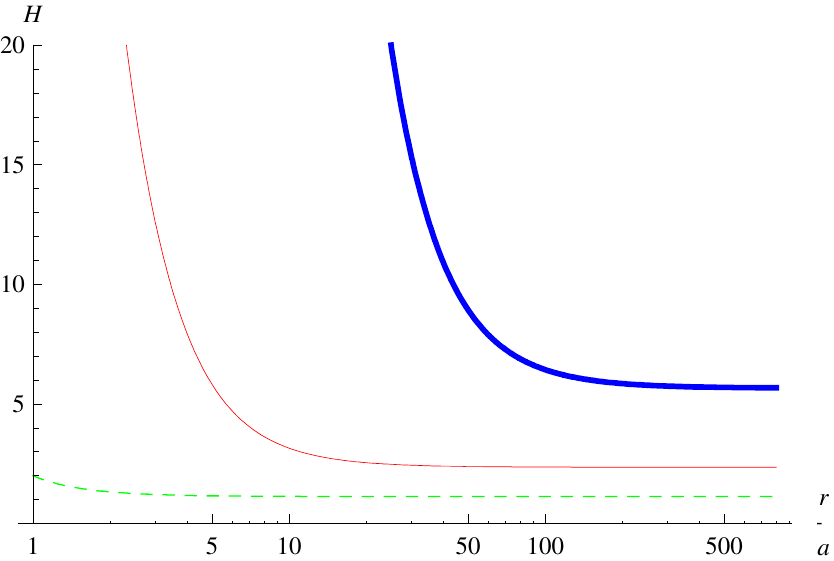}
\caption{Numerical solution for $f^2(r)$ and $H(r)$,  with the choice of parameters $k=10000$ and $a^2/\alpha'k=\{0.0001,0.01,1\}$, respectively thick, thin and dashed lines.}
\label{tab1}
\end{figure}

In figure~\ref{tab1}, we represent a family of such solutions with conformal
factor having the asymptotics:
\begin{equation}\label{Hprofile}
H(r) \stackrel{r\to a^+}{\sim} 1+\frac{\alpha'k}{r^2}\,,\qquad \lim_{r\rightarrow \infty }H(r)=H_{\infty}\, ,
\end{equation}
and a function $f^2$ possessing a bolt singularity at $r=a$ (where the blow-up parameter $a$ has been set previously in defining the gauge bundle). 
The dilaton is then  determined by the conformal factor, up to a constant, by integrating eq.(\ref{susy-system}a):
\begin{equation}\label{ephi}
e^{2\Phi(r)}=e^{2\Phi_0}H(r)^2\,.
\end{equation}

We observe in particular that since $\lim_{r\to \infty} f^2=1$, the solution interpolates between the squashed resolved conifold at finite $r$ and the usual cone 
over the Einstein space $T^{1,1}/\zi_2$ at infinity, thus restoring a Ricci-flat 
background asymptotically. In figure~\ref{tab1} we also note that in the regime where $a^2$ is small compared to $\alpha'k$, the function
$f^2$ develops a saddle point that disappears when their ratio tends to one.

As expected from this type of torsional backgrounds, in the blow-down limit the gauge bundle associated with $\vec{q}$ becomes  a kind of point-like instanton, 
leading to a five-brane-like solution. The appearance of five-branes manifests itself by a singularity in the conformal factor $H$ in 
the $r\to0$ limit, hence of the dilaton. In this limit the solution behaves as the backreaction of heterotic five-branes wrapping some supersymmetric vanishing
two-cycle, together with a gauge bundle turned on. As we will see later on this singularity is not smoothed out by the $\mathcal{R}^2$ curvature correction to the 
Bianchi identity. 


\subsection{Analytical solution in the double-scaling limit}
\label{sec:analytic}

The regime $a^2/\alpha'k\ll 1$ in parameter space allows for a limit where the system~(\ref{susy-system}) admits an analytical solution, which corresponds to a sort of 'near-bolt' or 
throat geometry of the family of torsional backgrounds seen above.\footnote{In the blow-down limit where the bundle degenerates to a wrapped five-brane-like solution, this regime should be called a 'near-brane' geometry.} 
This solution is valid in the coordinate range:
\begin{equation}
a^2\leqslant r^2 \ll \alpha' k \,.
\end{equation}
Note that this is not a 'near-singularity' regime as the location $a$ of the bolt is chosen hierarchically smaller than the scale $\sqrt{\alpha' k}$ at which
one enters the throat region.

This geometry can be extended to a full solution of heterotic supergravity by means of a {\it double scaling limit}, defined as 
\begin{equation}
\label{DSL}
g_s \to 0 \,, \qquad\qquad \mu=\frac{g_s \alpha'}{a^2} \quad \text{fixed}\,,
\end{equation}
and given in terms of the asymptotic string coupling $g_s=e^{\Phi_0}H_{\infty}$ set by the $r\to\infty$ limit of expression~(\ref{ephi}).
This isolates the dynamics  near the four-cycle of the resolved singularity, without going to the blow-down limit, i.e. keeping 
the transverse space to be conformal to the non-singular resolved conifold.\footnote{For this limit to make sense, one needs to check 
that the asymptotic value of the conformal factor $H_\infty$ stays of order one in this regime. We checked with the numerical 
solution that this is indeed the case.} 
 
One obtains an interacting theory whose effective string coupling constant is set by the double-scaling parameter $\mu$. 
The metric is determined by solving (\ref{susy-system}) in this limit, yielding
the analytic expressions:
\begin{equation}\label{Hfg}
H(r)=\frac{\alpha'k}{r^2}\,,\qquad f^2(r)=g_1^2(r)=\tfrac{3}{4}\Big[1-\left(\frac{a}{r}\right)^8\Big]\, .
\end{equation}

\begin{figure}[!t]
\centering
\includegraphics[width=75mm]{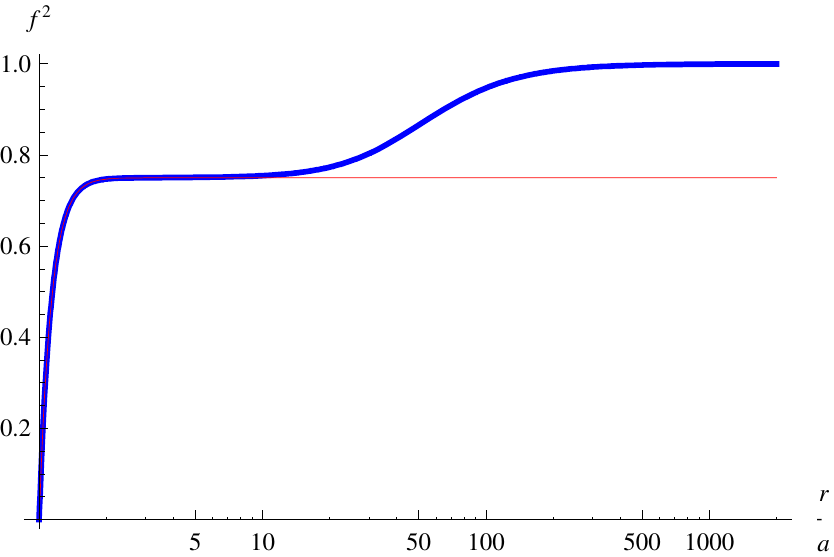}\hskip5mm
\includegraphics[width=75mm]{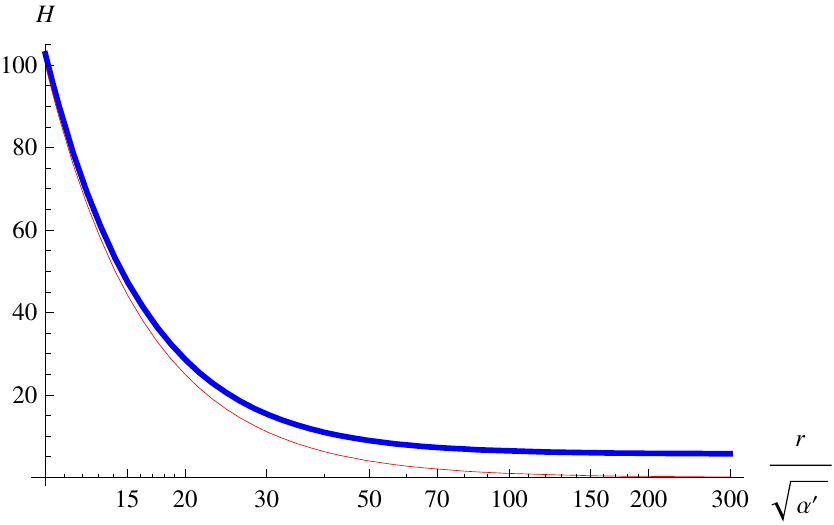}
\caption{Comparison of the profiles of $f(r)$ and $H(r)$  for the asymptotically flat supergravity solution (thick line) and its double scaling limit (thin line), for $k=10000$ and $a^2/\alpha'k=0.0001$.}
\label{tab2}
\end{figure}

To be more precise in defining the double-scaling limit one requests to stay at fixed distance from the bolt. We use then the rescaled 
dimensionless radial coordinate $R=r/a$, in terms of which one obtains the double scaling limit of the background (\ref{sol-ansatz},\ref{gauge-ansatz},\ref{ephi}):
\begin{subequations}\label{sol-nhl}
\begin{align}
\di s^2 & = \eta_{\mu\nu}\di x^{\mu}\di x^{\nu} + \frac{2\alpha' k}{R^2} \,\biggl[
\tfrac{\di R^2}{1-\tfrac{1}{R^8}} + \frac{R^2}{8}\,
\Big( \di\theta_2^2+\sin^2\theta_1\,\di\phi_1^2 + \di\theta_2^2+\sin^2 \theta_2\,\di \phi_2^2 \Big) \nonumber \\
&  \hspace{4.5cm}+\,\frac{R^2}{16} (1-\tfrac{1}{R^8}) \big(\di \psi + \cos \theta_1 \,\di \phi_1 + \cos  \theta_2 \,\di  \phi_2 \big)^2  \biggr]\, , \label{met-sol-nhl}
\\
\mathcal{H}_{[3]} & =  \frac{\alpha' k}{8}\, \Big(1-\tfrac{1}{R^8}\Big)\,\big(\Omega_1+\Omega_2\big)\wedge\tilde\omega \, ,\\
e^{\Phi(r)} & = \frac{\mu}{H_{\infty}} \left( \frac{k}{R^2}\right)\, , \\
\mathcal{A}_{[1]} & =\tfrac{1}{4}\Big[\big(\cos\theta_1\,\di\phi_1 - \cos \theta_2\,\di \phi_2 \big)\, \vec{p} 
+ \tfrac{1}{R^4}\,\tilde\omega\,\vec{q}\, \Big]\cdot \vec{H} \, , 
\end{align}
\end{subequations}

The warped geometry is a six-dimensional torsional analogue of  Eguchi-Hanson space, as anticipated before in subsection~\ref{bianchisect}. 
We observe that (as for the double-scaling limit of the warped Eguchi-Hanson space studied in~\cite{Carlevaro:2008qf}) the blow-up parameter $a$ disappears from the metric, being absorbed 
in the double-scaling parameter $\mu$, hence in the dilaton zero-mode that fixes the effective string coupling. 

As can be read off from the asymptotic form of the metric (\ref{sol-nhl}), the metric of its $T^{1,1}$ base is  non-Einstein even at infinity, so that the space is not 
asymptotically Ricci-flat, contrary to the full supergravity solution corresponding figure~\ref{tab1}. 
But as expected, in the regime where $a^2\ll \alpha'Q_5$ both the supergravity and the the near-horizon background agree perfectly in the vicinity of the bolt, as shown in figure~\ref{tab2}.

Finally we notice that taking the near-brane limit of blown-down geometry (which amounts to replace $f^2$ by one in the metric~(\ref{met-sol-nhl}), and turning off the gauge bundle associated with $\vec{q}$) the six-dimensional metric factorizes into a linear dilaton direction and a non-Einstein $T^{1,1}/\mathbb{Z}_2$ space. 

\subsection{One-loop contribution to the Bianchi identity}

The supergravity solution (\ref{sol-ansatz}) is valid in the large charges regime $k\gg 1$, where higher derivative (one-loop) corrections to the 
Bianchi identity~(\ref{bianchi}) are negligible. Given the general behaviour of the function $f^2$ and $H$ as plotted in figure~\ref{tab1}, we must still 
verify that the curvature contribution $\text{tr}\,\mathcal{R}_+\wedge \mathcal{R}_+$ remains finite for large 
$k$ and arbitrary value of $a$, for any $r\geqslant a$, with coefficients of order one, so that the truncation performed on the Bianchi identity is 
consistent and the solution obtained is reliable. 

We can give an 'on-shell' expression of the one-loop contribution in~(\ref{bianchi}) by using the supersymmetry equations (\ref{susy-system}) to 
re-express all first and second derivatives of $f$ and $H$ in terms of the functions $g_1$, $f$ and $H$ themselves. We obtain:
\begin{equation}\label{TrRR}
\begin{array}{l}
{\ds \mbox{tr}\, \mathcal{R}(\Omega_+)\wedge \mathcal{R}(\Omega_+) \,=}\\[6pt]
 {\ds \qquad  - 4\Big(
1-\frac{4 f^2}{3} \big(2-f^2\big)
-\frac{2g_1^2(1-f^2)}{3} \Big[\frac{\alpha'k}{r^2 H}\Big]
+\frac{2 g_1^4}{3f^2}  \Big[\frac{\alpha'k}{r^2 H}\Big]^2 +\frac{2 g_1^6}{9f^2}  \Big[\frac{\alpha'k}{r^2 H}\Big]^3
\Big)
\,\Omega_1\wedge\Omega_2}
   \\[10pt]
  {\ds \qquad \quad
  -8 \Big( 4(1-f^2)^2 
 +(1-f^2)(1-4g_1^2)\Big[\frac{\alpha'k}{r^2 H}\Big] 
 +\frac{g_1^2\big(-6 f^2 + g_1^2 (3+2f^4+6f^2)}{3 f^4} \Big[\frac{\alpha'k}{r^2 H}\Big]^2}
 \\[8pt]
 {\ds \qquad  \qquad
  +\frac{g_1^4\big(-3f^2+2g_1^2 (1+2f^2)\big)}{3 f^4} \Big[\frac{\alpha'k}{r^2 H}\Big]^3 
     +\frac{2 g_1^8}{9 f^4} \Big[\frac{\alpha'k}{r^2 H}\Big]^4 \Big)
  \frac{\di r}{r} \wedge\big(\Omega_1+\Omega_2\big)\wedge \tilde\omega \,.}
\end{array}
\end{equation}
We observe from the numerical analysis of the previous subsection that $f\in[0,1]$ while $H$ is monotonously
decreasing from $H_{\text{max}}=H(a)$ finite to $H_{\infty}>0$. So expression (\ref{TrRR}) remains finite at $r\rightarrow \infty$, 
since all overt $r$ contributions come in powers of $\alpha'k/(r^2 H)$, which vanishes at infinity. 

Now, since $f$ and $g_1$ both vanish at $r=a$, there might also arise a potential divergences in (\ref{TrRR}) in the vicinity of the bolt. However:
\begin{itemize}
	\item At $r=a$, all the potentially divergent terms appear as ratios:
	$g_1^{2n}\,f^{-2m}$, with $n\geq m$, and are thus zero or at most finite, 
	since $g_1$ and $f$ are equal at the bolt.
	\item The other contributions all remain finite at the bolt, since they are all expressed as
	powers of $\alpha'k/(r^2 H)$, which is maximal at $r=a$, with:
	$$
	\text{Max}\Big[\frac{\alpha'k}{r^2 H}\Big] = \left(1+\frac{a^2}{\alpha'k}\right)^{-1}\leq1\,.
	$$
\end{itemize}

Taking the double-scaling limit, the expression~(\ref{TrRR}) simplifies to:
\begin{equation}\label{TrRR2}
\mbox{tr}\, \mathcal{R}(\Omega_+)\wedge \mathcal{R}(\Omega_+)   = -\big(4-8g^2+5g^4\big)\,\Omega_1\wedge\Omega_2
    - 2\big(16-34g^2+23g^4\big)\,\frac{\di r}{r} \wedge\big(\Omega_1+\Omega_2\big)\wedge \tilde\omega\,,
\end{equation}
where $g_1$ has been rescaled to $g(r)=\sqrt{1-(a/r)^8}$ for simplicity. 
We see that this expression does not depend on $k$, because of the particular profile of $H$ in this limit (\ref{Hfg}), and is clearly finite of $\mathcal{O}(1)$ for $r\in[a,\infty[$.

\subsubsection*{Bianchi identity at the bolt}
By using the explicit form for $\text{tr}\,\mathcal{R}_+\wedge \mathcal{R}_+$ determined above, we can evaluate the full  Bianchi identity~(\ref{bianchi}) at the bolt. 
At $r=a$, the \textsc{nsns} flux $\mathcal{H}$ vanishes, and the tree-level and one-loop contributions are both on the same footing.
The Bianchi identity can be satisfied at the form level for (\ref{TrRR}):
\begin{equation}
0= \text{Tr}\,\mathcal{F}\wedge \mathcal{F}-\text{tr}\,\mathcal{R}(\Omega_+)\wedge \mathcal{R}(\Omega_+)=\big(
\vec{p}^{\, 2} -  \vec{q}^{\, 2} + 4 \big)\,\Omega_1\wedge \Omega_2\, 
\end{equation}
provided:
\begin{equation}
\vec{p}^{\, 2} = \vec{q}^{\, 2} - 4\,.
\end{equation}
As we will see in section~\ref{sec:param} when deriving the worldsheet theory for the background~(\ref{sol-nhl}), this result will be precisely reproduced in the \textsc{cft} by 
the worldsheet normally cancellation condition. It suggests that the $\alpha'$ corrections to the supergravity solution vanish at the bolt, as the worldsheet result is exact. 

\subsubsection*{Tadpole condition at infinity}
In order to view the solution~(\ref{sol-ansatz}) as part of a compactification manifold, it is useful to consider the tadpole condition associated to it, 
as it has non-vanishing charges at infinity.

One requests at least to cancel the leading term in the asymptotic expansion of  the modified Bianchi identity at infinity, where the metric becomes Ricci-flat, and 
the five-brane charge can thus in principle be set to zero (not however that the gauge bundle $V$ is different from the standard embedding). In this limit, only the first gauge 
bundle specified by the shift vector $\vec{p}$ contributes, so that~(\ref{bianchi}) yields
the constraint:
\begin{equation}\label{tad}
6Q_5= 3\vec{p}^{\, 2} - 4\,.
\end{equation}
Since $\vec{p}\in \zi^{16}$, we can never set the five-brane charge to zero and fulfil this condition. 
Furthermore, switching on the five-brane charge could only balance the instanton number of the gauge bundle, but never the curvature  contribution, for
elementary numerological reasons. Again, eq.~(\ref{tad}) can only be satisfied in the large charge
regime, where the one-loop contribution is subleading.

In the  warped Eguchi-Hanson solution tackled in~\cite{Carlevaro:2008qf}, the background was locally torsional but for some appropriate choice 
of Abelian line bundle the five-brane charge could consistently be set to zero;
here no such thing occurs.\footnote{The qualitative difference between the two types of solutions is that Eguchi-Hanson space is asymptotically 
locally flat, while the orbifold of the conifold is only asymptotically locally Ricci-flat.} This amounts to say that in the present case torsion is
always present to counterbalance tree-level effects, while
the only way to incorporate higher order contributions is to compute
explicitly the one-loop correction to the background~(\ref{sol-ansatz}) from the Bianchi identity, 
as in~\cite{Fu:2008ga}. In the double-scaling limit~(\ref{sol-nhl}), this could in principle be carried out by
the worldsheet techniques developed in~\cite{Tseytlin:1992ri,Bars:1993zf,Johnson:2004zq}, using the gauged \textsc{wzw} model description we discuss  in the next section.

\subsection{Torsion classes and effective superpotential}\label{torsion-cl}

In this section we will delve deeper into the $SU(3)$ structure of the background as a way of characterizing the geometry and the flux background we are dealing with. We will briefly go through some elements of the classification of $SU(3)$-structure that we will need in the following (for a more detailled and general presentation, {\it cf.} \cite{Salomon,Gauntlett:2003cy,Grana}). On general grounds, as soon as it departs from Ricci-flatness, a given space acquires intrinsic torsion, which classifies the 
$G$-structure it is endowed with. According to its index structure, the intrinsic torsion $T^{i}_{\phantom{i}jk}$ takes value in $\Lambda^1 \otimes \mathfrak{g}^{\perp}$, where $\Lambda^1$ is the space of one-forms, and $\mathfrak{g}\oplus\mathfrak{g}^{\perp}=\mathfrak{spin}(d)$, with $d$ the dimension of the manifold, and it therefore decomposes into irreducible $G$-modules $\mathcal{W}_i$.

\subsubsection*{Torsion classes of SU(3)-structure manifolds}
The six-dimensional manifold of interest has $SU(3)$-structure, and can therefore be classified in terms of the following decomposition of $T$ into of irreducible representations of $SU(3)$:
\begin{equation}
\begin{array}{rcl}
T\in \Lambda^1 \otimes \mathfrak{su}(3)^{\perp}
   & =& \mathcal{W}_1\oplus \mathcal{W}_2\oplus \mathcal{W}_3\oplus \mathcal{W}_4\oplus \mathcal{W}_5\\[4pt]
(\boldsymbol{3}+\bar{\boldsymbol{3}})\times (\boldsymbol{1}+\boldsymbol{3}+\bar{\boldsymbol{3}}) & =&
 (\boldsymbol{1}+\boldsymbol{1})+(\boldsymbol{8}+\boldsymbol{8})+ (\boldsymbol{6}+\bar{\boldsymbol{6}})
 +(\boldsymbol{3} + \bar{\boldsymbol{3}}) +(\boldsymbol{3} + \bar{\boldsymbol{3}})\,.
\end{array}
\end{equation}
This induces a specific decomposition of the exterior derivatives of the $SU(3)$ structure $J$ and $\Omega$ onto the components of the intrinsic torsion $W_i\in \mathcal{W}_i$: 
\begin{subequations}\label{intrinsic}
\begin{align}
\di J &= -\tfrac32 \,\text{Im}(W_1^{(1)}\,\bar\Omega)+ W_4^{(3+\bar{3})}\wedge J + W_3^{(6+\bar{6})}\,,\\[4pt]
\di \Omega & = W_1^{(1)}\,J\wedge J + W_2^{(8)}\wedge J + W_5^{(\bar{3})}\wedge \Omega\,,
\end{align}
\end{subequations}
which measures the departure from the Calabi-Yau condition $\di J=0$ and $\di\Omega=0$ ensuring Ricci-flatness. 

We have in particular $W_1$ a complex $0$-form, $W_2$ a complex $(1,1)$-form and $W_3$ a real primitive $[(1,2)+(2,1)]$-form. $W_4$ is a real vector and $W_5^{(\bar{3})}$ is the anti-holomorphic
part of the real one-form $W_5^{(3+\bar{3})}$, whose holomorphic piece is projected out in expression (\ref{intrinsic}b).
In  addition $W_2$ and $W_3$ are {\it primitives}, i.e. they obey $J\lrcorner W_i=0$, with the generalized inner product of a $p$-form $\alpha_{[p]}$ and $q$-form $\beta_{[q]}$ for $p\leq q$ given by $\alpha\lrcorner\beta=\tfrac{1}{p!}\alpha_{m_1..m_p}\beta^{m_1..m_p}_{\phantom{m_1..m_p}m_{p+1}..m_{q}}$.  

The torsion classes can be determined by exploiting the primitivity of $W_2$ and $W_3$ and the defining relations (\ref{topcond}) of the $SU(3)$ structure. Thus, we can recover $W_1$ from both
equations (\ref{intrinsic}). In our conventions, we  have then
\begin{equation}
W_1^{(1)} = \tfrac{1}{12}\,J^2\lrcorner \di \Omega \equiv \tfrac{1}{36}\,J^3\lrcorner(\di J\wedge \Omega)\,.
\end{equation}
Likewise, one can compute $W_4$ and $W_5$, by using in addition the relations $J\lrcorner\Omega=J\lrcorner\bar\Omega=0$:
\begin{equation}\label{W4}
W_4^{(3+\bar 3)} = \frac{1}{2}\, J\lrcorner \di J\,,\qquad
\bar W_5^{(3+\bar 3)} =-\tfrac{1}{8} \big(\bar\Omega \lrcorner \di\Omega + \bar\Omega \lrcorner \di\Omega\big)\,.
\end{equation}
This in particular establishes $W_4$ as what is known as the {\it Lee form} of $J$, while, by rewriting $\bar W_5$ as 
$\bar W_5=-\tfrac12\text{Re}\Omega\lrcorner \di \text{Re}\Omega=-\tfrac12\text{Im}\Omega\lrcorner \di \text{Im}\Omega$, we
observe that $W_5$ is the Lee form of $\text{Re}\Omega$ or $\text{Im}\Omega$, indiscriminately \cite{Gauntlett:2003cy}. This alternative formulation in terms of the Lee form 
is characteristic of the classification of almost Hermitian manifolds.

The torsion class $W_3^{(6+\bar{6})}=W_3^{(6)}+W_3^{(\bar{6})}$ is a bit more involved to compute, but may be determined in components by contracting with the totally 
antisymmetric holomorphic and anti-holomorphic tensors of $SU(3)$, which projects to the $\boldsymbol{6}$ or $\bar{\boldsymbol{6}}$ of $SU(3)$:
\begin{equation}\label{W3}
(\star_3 W_3^{(6)})_{\bar a\bar b}=\,(W_3)^{\bar c \bar d}_{\phantom{cd}[\bar a}\, \bar\Omega_{\bar b]\bar c \bar d}\,,\qquad
(\star_{\bar 3}W_3^{(\bar 6)})_{ab}=(W_3)^{cd}_{\phantom{cd}[a}\, \Omega_{b] cd}\,,
\end{equation}
with the metric $\eta_{ab}=2\delta_{a}^{\phantom{a}\bar b}$ and the "Hodge star products" in three dimensions given by $\star_{3}E^{\bar a \bar b} = \epsilon^{\bar a \bar b}_{\phantom{ab}c}\,E^{c}$, and $\star_{\bar{3}}(\bullet)$ applying to the complex conjugate of the former expression.

The \textsc{nsns} flux also decomposes into $SU(3)$ representations:
\begin{equation}\label{flux-class}
\mathcal{H}= -\tfrac32 \,\text{Im}(H^{(1)}\,\bar\Omega)+ H^{(3+\bar3)}\wedge J + H^{(6+\bar6)}\,.
\end{equation}
As a general principle, since torsion is generated by flux, supersymmetry requires that the torsion classes (\ref{intrinsic}) be supported by flux classes in
the same representation of $SU(3)$. Thus, we observe in particular that there is no component of $\mathcal{H}$ in the $(\boldsymbol{8}+\boldsymbol{8})$, which implies that $W_2=0$, for our type of backgrounds.

\subsubsection*{The torsion classes of the warped resolved conifold}
After this general introduction we hereafter give the torsion classes for the warped six-dimensional background~(\ref{sol-ansatz}) studied in this work. They can be extracted from the following differential conditions, which have been established using the supersymmetry equations~(\ref{susy-system}) and the relation~(\ref{g3}):
\begin{subequations}\label{torsion-warp}
\begin{align}
\di\Omega & = 2 \mu(r)\,\big( e^{1256}+e^{1346}+i(e^{1356}-e^{1246})\big)\,,\\
\di J & = -\mu(r)\, \big(e^{123}+e^{145}\big)\,,\\
\mathcal{H} & = \mu(r)\, (e^{236}+e^{456})\,,
\end{align}
\end{subequations}
with the function:
\begin{equation}\label{mu}
\mu(r)= \sqrt{\frac{2}{3}}\,\frac{2\alpha'kg_1^2}{r^3 H^{3/2}f} = -\sqrt{\frac{2}{3}}\frac{f}{\sqrt{H}}\,\Phi'\,.
\end{equation}
Since relations~(\ref{torsion-warp}) imply satisfying the first supersymmetry condition~(\ref{susy-system}a), this induces automatically
$W_1=W_2=0$ (this can be checked explicitly in~(\ref{torsion-warp})),
which in turn entails that the manifold~(\ref{sol-ansatz}a) is complex, since the complex structure is now integrable\footnote{For a six dimensional manifold to be complex, 
the differential $\di\Omega$ can only comprise a $(3,1)$ piece, which leads to $W_1=W_2=0$. This condition can be shown to be equivalent to the vanishing of the Nijenhuis tensor, ensuring the integrability of the complex structure.}.

Then, using relations~(\ref{W4}) and~(\ref{W3}), one determines the remaining torsion classes:
\begin{equation}
W_1=W_2=W_3 =0\,, \label{wt1}
\end{equation}
and 
\begin{equation}
W_4^{(3+\bar 3)} = \tfrac12\,W_5^{(3+\bar 3)} = -\mu(r)\,\text{Re}E^3\,. \label{wt2}
\end{equation}
They are supported by the flux:
\begin{equation}\label{H33}
H^{(3+\bar 3)}= -\mu(r)\,\text{Im}E^3\,.
\end{equation}

Two remarks are in order. First, combining~(\ref{intrinsic}a) and~(\ref{susy-cond}b) leads to the generic relation $W_4=\di\Phi$, which is indeed satisfied by the Lee form~(\ref{wt2}) by taking into account expression~(\ref{mu}). Secondly, the relation $W_5=2 W_4$ in~(\ref{wt2}) is a particular case of the formula $W_5=(-1)^{n+1} 2^{n-2} W_4$~\cite{LopesCardoso:2002hd,Gauntlett:2003cy} which holds for a manifold with $SU(n)$ structure.

\subsubsection*{Effective superpotential}
The effective superpotential of four-dimensional $\mathcal{N}=1$ supergravity for this particular solution, viewing the throat solution we consider as part of  some heterotic flux compactification.  
It can be derived  from a  generalization of the Gukov-Vafa-Witten superpotential~\cite{Gukov:1999ya}, which includes the full contribution from torsion and
$\mathcal{H}$-flux~\cite{Grana:2005ny}, or alternatively using generalized calibration methods~\cite{Koerber:2007xk}. The general expression reads:
\begin{equation}\label{superpot}
\mathcal{W}=\tfrac14\int_{\mathcal{M}_6}\Omega\wedge (H+i\di J)\, .
\end{equation}

We evaluate this expression on the solution~(\ref{sol-ansatz}) by using the results obtained in~(\ref{wt1}-\ref{H33}). This leads to the 'on-shell' complexified K\"ahler structure
\begin{equation}
\mathcal{H}+i\,\di J = iW_5^{(\bar 3)}\wedge J = -i\mu(r)\,\bar E_3\wedge J\,,
\end{equation}
which together with the first relation in~(\ref{topcond}) entails
\begin{equation}\label{superpot2}
\mathcal{W}=0
\end{equation}
identically.\footnote{As explained in~\cite{Vafa:2000wi} and systematized later in~\cite{Martucci:2006ij}, one can determine
the superpotential~(\ref{superpot}) without knowing explicitly the full background, 
by introducing a resolution parameter determined by a proper calibration of the 'off-shell' superpotential, and subsequently minimizing the 
latter with respect to this parameter~(see \cite{Maldacena:2009mw} for a related discussion).} 

In Vafa's setup of ref.~\cite{Vafa:2000wi}, corresponding 
to D5-branes wrapping the two-cycle of the resolved conifold, this  leads to an $\mathcal{N}=1$ Veneziano-Yankielowicz superpotential 
(where the resolution parameter is identified with the glueball superfield of the four dimensional super Yang-Mills theory), showing that the background is holographically dual to a confining theory, with a gaugino  condensate. In our case having a vanishing superpotential  means that the blow-up parameter $a$ corresponds to a modulus of the holographically dual $\mathcal{N}=1$ four-dimensional theory. More aspects 
of the holographic duality are discussed in subsection~\ref{holo}.

\subsubsection*{A K\"ahler potential for the non-Ricci-flat conifold}

In the following, we will show that the manifold corresponding to the metric~(\ref{sol-ansatz}a) is conformally K\"ahler. This can be readily established  by means of the differential conditions~(\ref{intrinsic}), as the characteristics of a given space are related to the vanishing of certain torsion classes or specific constraint relating them (see~\cite{Grana} for a general overlook).

For this purpose, we now have to determine the torsion classes for the  resolved conifold space conformal to the geometry~(\ref{sol-ansatz}a):
\begin{equation}\label{metric-unwarp}
\di s^2_{\tilde{\mathcal{C}}_6} = \frac{\di r^2}{f^2} + \frac{r^2}{6}\big(\di\Omega_1^2 +\di\Omega_2^2 \big)+
\frac{r^2 f^2}{9}\tilde\omega^2\,.
\end{equation}
Again, these can be read from the differential conditions:
\begin{subequations}\label{torsion-unwarp}
\begin{align}
\di\tilde\Omega & = \tfrac32 \,\tilde\mu(r)\,\big( \tilde{e}^{1256}+\tilde{e}^{1346}+i(\tilde{e}^{1356}-\tilde{e}^{1246})\big)\,\\
\di \tilde J & = 0 \,,
\end{align}
\end{subequations}
with now
\begin{equation}
\tilde\mu(r)= \frac{\alpha'k\left(1-\left(\frac{a}{r}\right)^8\right)}{r^3 f(r)}\,.
\end{equation}
and the new set of vielbeins given by:
\begin{equation}
\tilde{e}^i = \sqrt{\frac{2}{3H}}\,e^i\,,\qquad \forall i=1,..,6\,.
\end{equation}
Repeating the analysis carried out earlier, the torsion classes are easily established:
\begin{eqnarray}
&W_1=W_2=W_3=W_4 =0\,,& \label{tr1}\\
&W_5^{(3+\bar 3)} = -2\tilde\mu(r)\, \text{Re}E^3\,,& \label{tr2}
\end{eqnarray}
The first relation~(\ref{tr1}) tells us that the manifold is complex, since $W_1=W_2=0$, and symplectic, 
since the K\"ahler form $\tilde J$ is closed. Fulfilling both these conditions gives precisely  a K\"ahler manifold, and the Levi Civita connection is in this case endowed with $U(3)$ holonomy.

\paragraph{The K\"ahler potential}

The K\"ahler potential for the conifold metric~(\ref{metric-unwarp}) is most easily computed by 
starting from the generic definition of the (singular) conifold as a quadratic on $\ci^4$,
 whose base is determined by the intersection of this quadratic with a three-sphere of
radius $\varrho$. These two conditions are summarized in~\cite{Candelas:1989js}: 
\begin{equation}\label{conif1}
\mathcal{C}_6\stackrel{\mbox{def}}{=}\sum_{A=1}^4 (w_A)^2=0\,\qquad 
\sum_{A=1}^4 |w_A|^2=\varrho^2\,.
\end{equation}
One can rephrase these two conditions in terms of a $2\times 2$ matrix $W$ parametrizing the $T^{1,1}$
base of the conifold, viewed as the coset $(SU(2)\times SU(2))/U(1)$, as $W=\tfrac{1}{\sqrt{2}}\sum_{A}w^A\sigma_A$. In  this language, the defining equations~(\ref{conif1}) take the form:
\begin{equation}\label{conif2}
\text{det}\,W=0\,,\qquad \text{tr}\,W^{\dagger}W = \varrho^2\,.
\end{equation}

For the K\"ahler potential $\mathcal{K}$ to generate the metric~(\ref{metric-unwarp}), it has to be invariant under the action of the rotation group $SO(4)\simeq SU(2)\times SU(2)$ of~(\ref{conif1}) and can thus only depend on $\varrho^2$. In terms of $\mathcal{K}$ and $W$, the metric on the conifold reads:
\begin{equation}\label{metric-su2}
\di s^2 = \Dot{\mathcal{K}}\,\text{tr}\,\di W^{\dagger}\di W + \Ddot{\mathcal{K}}\,|\text{tr}\,W^{\dagger}\di W|^2\,,
\end{equation}
where the derivative is $\dot{(\bullet)}\equiv \frac{\partial}{\partial\rho^2}(\bullet)$.
By defining the function $\gamma(\varrho)=\varrho^2\,\Dot{\mathcal{K}}$, the metric~(\ref{metric-su2}) can be recast into the form:
\begin{equation}
\di s^2 = \Dot{\gamma}\,\di\varrho^2 + \frac{\gamma}{4}\big(\di\Omega_1^2 +\di\Omega_2^2 \big)+
\frac{\varrho^2 \Dot{\gamma}}{4}\,\tilde\omega^2\,.
\end{equation}
Identifying this expression with the metric~(\ref{metric-unwarp}) yields two independent first order differential equations, one of them giving the expression of the radius of the $S^3$ in~(\ref{conif1}) in terms of the radial coordinate in~(\ref{metric-unwarp}): 
\begin{equation}\label{ro}
\varrho = \varrho_0\,e^{{\ds \tfrac32\int\frac{\di r}{r f^2}}}\,,\qquad \gamma(r)=\tfrac23\,r^2\,.
\end{equation}
From these relations, one derives the K\"ahler potential as a function of $r$:
\begin{equation}\label{kahpot2}
\mathcal{K}(r)=\mathcal{K}_0 + \int\frac{\di(r^2)}{f^2}\,.
\end{equation}

In particular, we can work out $\mathcal{K}$ explicitly in the near horizon limit~(\ref{DSL}):
\begin{equation}\label{kahlerpot}
\mathcal{K}(r)=\mathcal{K}_0 + \frac{4a^2}{3}\, \left[\left(\frac{r}{a}\right)^2 -\frac12\text{arctan}\left(\frac{r}{a}\right)^2+ \frac14\log\left(\frac{r^2-a^2}{r^2+a^2}\right)\right]\,.
\end{equation}

\begin{figure}[!ht]
\centering 
\includegraphics[width=70mm]{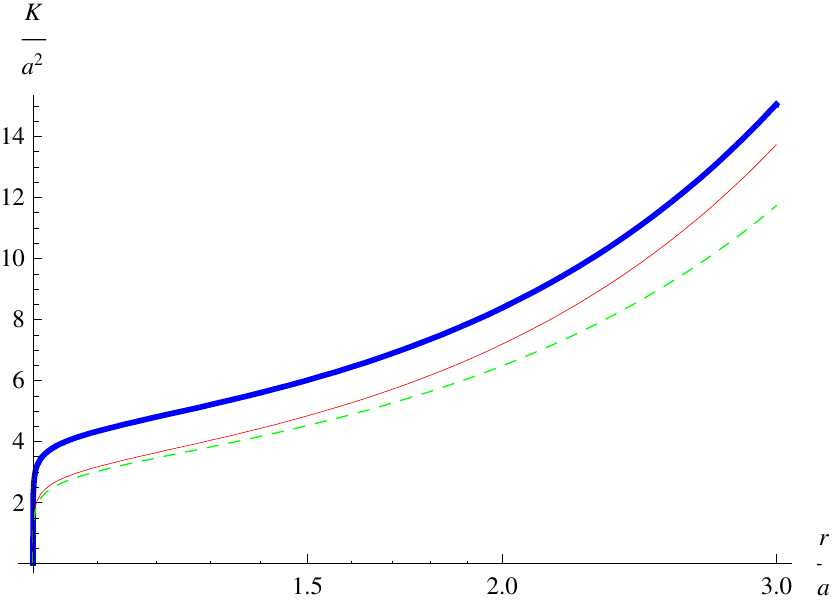}
\caption{The K\"ahler potential for the asymptotically flat supergravity solution with $k=10000$ and $a^2/\alpha'k=\{0.0001,0.01,1\}$.}
\label{kahlerpotfig}
\end{figure}

Choosing $\varrho_0=1$, we have $\varrho=\sqrt[4]{r^8-a^8}$, which varies over $[0,\infty[$, as expected. With an
exact K\"ahler potential at our disposal, we can make an independent check that
the near horizon geometry~(\ref{sol-nhl}) is never conformally Ricci flat. Indeed,
by establishing the Ricci tensor  $R_{i\bar \jmath}=\partial_i\partial_{\bar \jmath}\ln\sqrt{|g|}$ for the K\"ahler manifold~(\ref{metric-su2}), we observe that the condition for Ricci flatness
imposes the relation $\partial_{\varrho^2}[(\varrho^2 \dot{\mathcal{K}})^3]= 2\varrho^2 $~\cite{PandoZayas:2001iw}, which is
never satisfied by the potential~(\ref{kahlerpot}).

In figure~\ref{kahlerpotfig} we plot the K\"ahler potential~(\ref{kahpot2}) for the asymptotically Ricci-flat
supergravity backgrounds given in figure~\ref{tab1}. We represent $\mathcal{K}$ only for small values of $r$, since for large
$r$ it universally behaves like $r^2$. One also verifies that, for small $r$, the analytic expression~(\ref{kahlerpot}) determined in the double-scaling limit fits perfectly 
the numerical result.

\section{Gauged WZW model for the Warped Resolved Orbifoldized Conifold}
\label{gaugedsec}

The heterotic supergravity background obtained in the first section has been shown to admit a double scaling limit, isolating the throat  region where an analytical solution can be found. The manifold is conformal to a cone over a non-Einstein $T^{1,1}/\mathbb{Z}_2$ base with a blown-up four-cycle, and features an asymptotically linear dilaton. The solution is parametrized by two 'shift vectors' $\vec{p}$ and $\vec{q}$ which determine the Abelian gauge bundle, and are orthogonal to each other. They are related to the \textsc{nsns} flux number $k$ as $k= \vec{p}^{\, 2} = \vec{q}^{\, 2}$. 
These conditions, as well as the whole solution~(\ref{sol-nhl}), are valid in the large charge limit $\vec{p}^{\, 2} \gg 1$.

The presence of an asymptotic linear dilaton is a hint that an exact worldsheet \textsc{cft} description may exist. We will show in this section that it is indeed the case; for any 
consistent choice  of line bundle, a gauged \textsc{wzw} model, whose background fields are the same as the supergravity solution~(\ref{sol-nhl}), exists. Before dealing 
with the details let us stress  important points of the worldsheet construction:
\begin{enumerate}
\item In the blow-down limit $a\to 0$, the dependence of the metric on the radial coordinate simplifies, 
factorizing the space into the (non-Einstein) $T^{1,1}$ base times the linear dilaton direction $r$.
\item The $T^{1,1}$ space is obtained as an asymmetrically gauged $SU(2)_k\times SU(2)_k$ \textsc{wzw}-model involving the 
right-moving current algebra of the  heterotic string. 
\item In order to find the blown-up solution the linear dilaton needs to be replaced by an auxiliary $SL(2,\mathbb{R})_{k/2}$ \textsc{wzw}-model. It is gauged together with the  
$SU(2)\times SU(2)$ factor, also in an asymmetric way.
\item The 'shift vectors' $\vec{p}$ and $\vec{q}$ define the embedding of the both gaugings in the $Spin (32)/\mathbb{Z}_2$ lattice
\item These two worldsheet gaugings are anomaly-free if $k = \vec{p}^{\, 2}=\vec{q}^{\, 2}-4$ and $\vec{p} \cdot \vec{q}=0$. These relations are exact in $\alpha'$. 
\end{enumerate}
A detailed study of a related model, based on a warped Eguchi-Hanson space, is given in ref.~\cite{Carlevaro:2008qf}. We refer the reader to this work for more details on the techniques
used hereafter.

\subsection{Parameters of the gauging}\label{sec:param}
We consider an $\mathcal{N}=(1,0)$ \textsc{wzw} model for the group $SU(2)\times SU(2)\times \slr$, 
whose element we denote by $(g_1,g_2,h)$. The associated levels of the $\mathcal{N}=1$ affine 
simple algebras are respectively chosen to be\footnote{
It should be possible to generalize the construction starting with $SU(2)$ WZW models at non-equal levels. 
Note also that the $\slr$ level does not need to be an integer.} $k$, $k$ and $k'$. 
The left-moving central charge reads
\begin{equation}
c = 9 -\frac{12}{k}+\frac{6}{k'}\, , 
\end{equation}
therefore the choice $k'=k/2$ ensures that the central charge has the requested value $c=9$ for any $k$, 
allowing to take a small curvature supergravity limit $k\to \infty$.

The first gauging, yielding a $T^{1,1}$ coset space with 
a non-Einstein metric, acts on $SU(2)\times SU(2)$ as 
\begin{equation}
\Big(g_1 (z,\bar z),g_2(z,\bar z)\Big) \longrightarrow \Big(e^{i\sigma_3\alpha (z,\bar z) }g_1 (z,\bar z), 
e^{-i\sigma_3\alpha (z,\bar z) } g_2(z,\bar z)\Big)\,.
\label{gaugingI}
\end{equation}
This gauging is highly asymmetric, acting only by left multiplication. It has to 
preserve $\mathcal{N}=(1,0)$ superconformal symmetry on the worldsheet, hence the worldsheet gauge fields 
are minimally coupled to the left-moving worldsheet fermions of the super-\textsc{wzw} model. 

In addition, the classical anomaly from this gauging 
can be cancelled by minimally coupling some of the 32  right-moving worldsheet fermions of the heterotic worldsheet theory. 
We introduce a sixteen-dimensional  vector $\vec{p}$ that gives the embedding of the gauging in the $\mathfrak{so}(32)$ Cartan sub-algebra.  
The anomaly cancellation condition gives the constraint\footnote{This condition involve the supersymmetric levels, as the gauging only acts 
on the left-moving supersymmetric side in the $SU(2)_k \times SU(2)_k$ \textsc{wzw} model.}
\begin{equation}
k+k = 2 \vec{p}^{\, 2}  \implies k = \vec{p}^{\, 2} \, .
\label{anomalyconstr0}
\end{equation}
On the left-hand side, the two factors correspond to the gauging in both $SU(2)_k$ models. We denote the components 
of the worldsheet gauge field as $(A,\bar A)$.

The second gauging, leading to the resolved conifold, also acts on the $\slr_{k'}$ factor, along the 
elliptic Cartan sub-algebra (which is time-like). Its action is given as follows:
\begin{multline}
\Big(g_1 (z,\bar z),g_2(z,\bar z),h(z,\bar z )\Big)\\ \longrightarrow\ \Big(e^{i\sigma_3\beta_1 (z,\bar z) }g_1 (z,\bar z), 
e^{i\sigma_3\beta_1 (z,\bar z) } g_2(z,\bar z),e^{2i\sigma_3\beta_1 (z,\bar z) } h(z,\bar z) e^{2i\sigma_3\, \beta_2 (z,\bar z) }\Big)\,,
\label{gaugingII}
\end{multline}
and requires a pair of worldsheet gauge fields $\mathbf{B}=(B_1,B_2)$. 
The left gauging, corresponding to the gauge field 
$B_1$, is anomaly-free (without the need of right-moving fermions) for 
\begin{equation}
2k = 4k' \, ,
\label{matchinglevels}
\end{equation}
which is satisfied by the choice $k'=k/2$ that was assumed above.\footnote{Note that the generator of the 
$U(1)$ isometry in the $\slr$ group was chosen to be $2i\sigma_3$, which explains the factor of four in the right-hand side of equation~(\ref{matchinglevels}).} 
The other gauging, corresponding to the gauge field $B_2$, 
acts only on $\slr$, by right multiplication. This time the coupling to the worldsheet 
gauge field need not be supersymmetric, as we are dealing with a $\mathcal{N}=(1,0)$ (heterotic) worldsheet.  

The anomaly is again cancelled by minimally coupling worldsheet fermions from the gauge sector. Denoting the corresponding shift vector $\vec{q}$ one gets the condition
\begin{equation}
4\left(\frac{k}{2}+2\right) = 2\vec{q}^{\, 2} \implies k = \vec{q}^{\, 2}-4\, ,
\label{anomalyconstrI}
\end{equation}  
which involves the bosonic level of $\slr$, as explained above; the constant  term on the \textsc{rhs} 
corresponds to the renormalization of the background fields by $\alpha'$ corrections, exact to all orders. In order to avoid the appearance of mixed anomalies in the full gauged \textsc{wzw} model, 
one chooses the vectors defining the two gaugings to be orthogonal to each other
\begin{equation}
\vec{p} \cdot \vec{q} = 0 \, .
\label{anomalyconstrII}
\end{equation}

\subsection{Worldsheet action for the gauged WZW model}
The total action for the gauged \textsc{wzw} model defined above is given as follows:
\begin{equation}\label{Stot}
S_{\textsc{wzw}}(A,{\bf B}) = S_{\slr_{k/2+2}} + S_{\su_{k-2},\, 1} +  S_{\su_{k-2},\, 2} + S_\text{gauge}(A,{\bf B}) + S_\text{Fer} (A,{\bf B})\, ,
\end{equation}
where the first three factors correspond to bosonic \textsc{wzw} actions, the fourth one to the bosonic terms involving the gauge fields and the 
last one to the action of the minimally coupled fermions. As it proves quite involved, technically speaking, to tackle the general case for generic 
values of the shift vectors $\vec{p}$ and $\vec{q}$, we restrict, for simplicity, to the 'minimal' solution of the 
constraints~(\ref{anomalyconstrI},\ref{anomalyconstrII}) given by 
\begin{equation}
\vec{p}=(2\ell,0^{15}) \ , \quad 
\vec{q}=(0,2\ell,2,0^{13})\qquad \text{with} \quad \ell > 2\, ,
\label{minimalsol}
\end{equation}
implying in particular $k=4\ell^2$. This choice ensures that $k$ is even, which will later on show to be necessary when considering the orbifold.   
The coset theory constructed with these shift vectors involves overall six Majorana-Weyl right-moving fermions from the sixteen participating in the fermionic representation of the $Spin(32)/\mathbb{Z}_2$ lattice.

We parametrize the group-valued worldsheet scalars $(g_1,g_2,h)\in \su\times\su\times\slr$ in terms of Euler angles as follows:
\begin{subequations}
\begin{align}
g_\ell &= e^{\frac{i}{2}\sigma_3 \psi_\ell} e^{\frac{1}{2}\sigma_1 \theta_\ell} e^{\frac{i}{2}\sigma_3 \phi_\ell}\ , \quad \ell=1,2\\
h&= e^{\frac{i}{2}\sigma_3 \phi_L} e^{\frac{1}{2}\sigma_1 \rho} e^{\frac{i}{2}\sigma_3 \phi_R}\,,\quad
\end{align}
\end{subequations}
where $\sigma_i$, $i=1,..,3$, are the usual Pauli matrices. 

The action for the worldsheet gauge fields, including the couplings to the bosonic affine currents of the \textsc{wzw} models, is given by:\footnote{The left-moving purely bosonic $SU(2)\times SU(2)$ currents of the Cartan considered here are normalized as $j^3_1=i\sqrt{k-2}\,(\p\psi_1+\cos\theta_1\,\p\phi_1)$ and $j^3_2=i\sqrt{k-2}\,(\p\psi_2+\cos\theta_2\,\p\phi_2)$, while the $\slr$ left- and right-moving ones read $k^3=i\sqrt{\tfrac{k}{2}+2}\,(\p\phi_L+\cosh\rho\,\p\phi_R)$ and $\bar k^3=i\sqrt{\tfrac{k}{2}+2}\,(\pb\phi_R+\cosh\rho\,\pb\phi_L)$.}
\begin{multline}\label{S-gauge}
S_\text{gauge}(A,{\bf B}) = \frac{1}{8\pi} \int \di^2 z \,
\Big[2i\big(j^3_1 -j^3_2 \big)\bar A \,+\, 2(k-2) A\bar A
 \,+\,2 B_1 i\bar k^3 +2 i\big(j^3_1 +j^3_2 + 2 k^3 \big)\bar B_2
\\ +\, 2(k-2) B_2\bar B_2
 \,-\,\left(\tfrac k 2 + 2\right)\big( B_1\bar B_1 + 4 B_2\bar B_2 + 4 \cosh\rho\, B_1\bar B_2\big)
  \Big] \,.
\end{multline}
The action for the worldsheet fermions comprises the left-moving Majorana-Weyl fermions coming
from the $\su\times\su\times\slr$ $\mathcal{N}=(1,0)$ 
super-\text{wzw} action,\footnote{We did not include the fermionic superpartners of the gauged currents, as they are gauged away.}  
respectively ($\zeta^1, \zeta^2)$, ($\zeta^3,\zeta^4)$ and ($\zeta^5,\zeta^6)$, supplemented by six 
right-moving Majorana-Weyl fermions coming from the $Spin(32)_1/\mathbb{Z}_2$ sector, that we denote $\bar\xi^a$, $a=1,..,6$:
\begin{multline}\label{S-Fer}
S_{\text{Fer}} (A,{\bf B}) =
\frac{1}{4\pi} \int \di^2 z \,\Big[
\sum_{i=1}^6 \zeta^i \pb \zeta^i
 - 2  \big(\zeta^1  \zeta^2  - \zeta^3 \zeta^4 \big) \bar A \\
 - 2  \big(\zeta^1 \zeta^2  + \zeta^3 \zeta^4 + 2\zeta^5 \zeta^6 \big) \bar{B}_2
 +\sum_{a=1}^6 \bar{\xi}^a \p \bar{\xi}^a - 2 \ell A \, \bar{\xi}^1\bar{\xi}^2
 - 2 \bar B_1 \big(\bar{\xi}^3\bar{\xi}^4 + \ell\bar{\xi}^5\bar{\xi}^6\big)
 \Big]\,.
\end{multline}
Note in particular that both actions (\ref{S-gauge}) and (\ref{S-Fer}) are in keep with
the normalization of the gauge fields required by the peculiar form of the second (asymmetric)
gauging (\ref{gaugingII}).

\subsection{Background fields at lowest order in $\alpha'$}
Finding the background fields corresponding to a heterotic coset theory is in general more tricky than for the usual bosonic 
or type \textsc{ii} cosets, because of the worldsheet anomalies generated by the various pieces of the asymmetrically gauged 
\textsc{wzw} model. In our analysis, we will closely follow the methods used in~\cite{Johnson:1994jw,Johnson:2004zq}. 
A convenient way of computing the metric, Kalb-Ramond and gauge field background from a heterotic 
gauged \textsc{wzw} model consists in bosonizing the fermions 
before integrating out the gauge field. 

One will eventually need to refermionize the appropriate
scalars to recover a heterotic sigma-model in the standard 
form, {\it i.e.} (see~\cite{Sen:1985eb,Hull:1985jv}):
\begin{multline}
S= \frac{1}{4\pi} \int\!\! \di^2 z\, \Big[ \tfrac{2}{\alpha'} (g_{ij} +\mathcal{B}_{ij}) \p X^i \pb X^j 
+ g_{ij} \zeta^i \bar{\nabla} (\Omega_+) \zeta^j +  \bar{\xi}^A \nabla(\mathcal{A})_{AB} \bar{\xi}^B
+\tfrac{1}{4}   \mathcal{F}^{AB}_{ij} \bar{\xi}_A \bar{\xi}_B \zeta^i \zeta^j \Big]
\label{generic_sigma_action}
\end{multline}
where the worldsheet derivative $\bar{\nabla}  (\Omega_+)$ is defined with respect to the spin connexion 
$\Omega_+$  with torsion  and the derivative $\nabla (\mathcal{A})$ 
with respect to the space-time gauge connexion $\mathcal{A}$.

The details of this bosonization-refermionization procedure for the coset under scrutiny are given in appendix~\ref{AppBoson}. At leading order in $\alpha'$ (or more precisely 
at leading order in a $1/k$ expansion)  we thus obtain, after integrating out classically the gauge fields, the bosonic part of the total action as follows:
\begin{multline}\label{Sbos-final}
S_{\textsc{b}}= \frac{k}{8\pi} \int \di^2 z\,\left[ \frac{1}{2}\p\rho \pb \rho+
\p\theta_1 \pb \theta_1 + \p\theta_2 \pb \theta_2 + \sin^2\theta_1 \,\p\phi_1 \pb \phi_1+
\sin^2\theta_2\,\p\phi_2 \pb \phi_2\right. \\ 
+\tfrac{1}{2}\tanh^2\rho\,(\p\psi+\cos\theta_1\,\p\phi_1+\cos \theta_2\,\p \phi_2 )(\pb\psi+\cos\theta_1\,\pb\phi_1+\cos \theta_2\,\pb \phi_2 ) \\
\left. +\frac{1}{2}\big(\cos\theta_1\,\p\phi_1+\cos \theta_2\,\p\phi_2\big)\pb\psi- \frac{1}{2}
 \p\psi \big(\cos\theta_1\,\pb\phi_1+\cos \theta_2\,\pb\phi_2\big) \right]\,,
\end{multline}
while the fermionic part of the action is given by
\begin{multline}\label{Sfer-final}
S_{\textsc{f}}= \frac{k}{4\pi} \int \di^2 z\,\Big[ \sum_{i=1}^{6}\zeta^i\pb\zeta^i
+ (\bar\zeta^1,\bar\zeta^2)\big[ \Id_2\,\p +(\cos\theta_1\, \p\phi_1-\cos\theta_2\, \p\phi_2)\,i\sigma^2\big]\left(\begin{matrix} \bar\xi^1\\ \bar\xi^2 \end{matrix}\right) \\
+ \bar\Xi^\top \left[ \Id_4\,\p +\frac{\ell}{\cosh\rho}\big(\p \psi+\cos\theta_1\, \p\phi_1+\cos \theta_2\, \p\phi_2 \big)\,i\sigma^2\otimes \left( \begin{matrix}   1 & 0 \\  0 & \ell \end{matrix}\right)\right] \bar\Xi \\
- \tfrac{1}{\ell}\,\bar\xi^1\bar\xi^2\,\big(\zeta^1\zeta^2-\zeta^3\zeta^4\big) +\frac{1}{\ell^2\cosh\rho}\,
\big(\bar\xi^3\bar\xi^4+\ell\bar\xi^5\bar\xi^6\big) \big(\zeta^1\zeta^2+\zeta^3\zeta^4+2\zeta^5\zeta^6\big) \Big]\, ,
\end{multline}
with $\bar\Xi^\top=(\bar\xi^3,\bar\xi^4,\bar\xi^5,\bar\xi^6)$. 
In addition, a non-trivial dilaton is produced by the integration of the worldsheet gauge fields
\begin{equation}
\Phi = \Phi_0 -\tfrac12\ln\cosh\rho\,.
\end{equation}

The background fields obtained above exactly correspond to the double-scaling limit of the supergravity solution~(\ref{sol-nhl}) for a particular choice of vectors $\vec{p}$ and $\vec{q}$, after the change of coordinate
\begin{equation}
\cosh \rho = (r/a)^4 = R^4\,.
\end{equation}
As noticed in section~\ref{sec:analytic}, the blow-up parameter, which is not part of the definition of the coset \textsc{cft}, is  absorbed in the dilaton zero-mode. It is straightforward --~but cumbersome~-- to extend the computation to 
a more generic choice of bundle.  This would lead to the background fields reproducing the generic supergravity solution~(\ref{sol-nhl}).

In this section we left aside the discussion of the necessary presence of a $\mathbb{Z}_2$ orbifold acting on the $T^{1,1}$ base of the conifold. Its important consequences will 
be tackled below. 


\section{Worldsheet Conformal Field Theory Analysis}
In this section we provide the algebraic construction of the worldsheet \textsc{cft} corresponding to the $\mathcal{N}=(1,0)$ gauged \textsc{wzw} model defined in section~\ref{gaugedsec}. We have shown previously that the non-linear sigma model with 
the warped deformed orbifoldized conifold as  target space is given by the asymmetric coset: 
\begin{equation}
\frac{\slr_{k/2} \times\, \left(\text{\small \raisebox{-1mm}{$U(1)_\textsc{l}$}\! $\backslash$ \!\!\raisebox{1mm}{$SU(2)_k \times SU(2)_k$}}\right)}{U(1)_\textsc{l} \times U(1)_\textsc{r}} \,,
\label{cosetdef}
\end{equation}
which combines a left gauging of $SU(2) \times SU(2)$ with a pair of chiral gaugings which also involve the $\slr$ \textsc{wzw} model.  In addition, the full worldsheet \textsc{cft} comprises 
a flat $\mathbb{R}^{3,1}$ piece, the right-moving heterotic affine algebra and an  $\mathcal{N}=(1,0)$ superghost system. We will see later on that the coset~(\ref{cosetdef}) has an enhanced worldsheet $\mathcal{N}=(2,0)$ superconformal symmetry, which allows to achieve $\mathcal{N}=1$ target-space supersymmetry. 

In the following, we will segment our algebraic analysis of the worldsheet \textsc{cft} for clarity's sake, and deal separately with the  singular conifold case, before moving on to treat the resolved geometry. This was somehow prompted by fact that the singular construction appears as a non-trivial building block of the 'resolved' \textsc{cft}, as we shall see below.

\subsection{A \textsc{cft} for the $T^{1,1}$ coset space}
\label{sec:acft}

For this purpose, we begin by restricting our discussion to the \textsc{cft} underlying the non-Einstein $T^{1,1}$ base of the conifold, which is captured by the  coset space $[SU(2)\times SU(2)]/U(1)$. 
In addition, this space supports a  gauge bundle specified by the vector of magnetic charges $\vec{p}$. Then, the full quantum theory describing the throat region of heterotic strings on 
the torsional {\it singular} conifold, can be constructed by tensoring this  \textsc{cft} with $\mathbb{R}^{3,1}$, the heterotic current algebra and a linear dilaton $\mathbb{R}_\mathcal{Q}$ with background charge\footnote{In the near-brane regime of~(\ref{sol-ansatz}), the conformal factor 
$H \sim Q_5/r^2$ cancels out the $r^2$ factor in front of the $T^{1,1}$ metric, hence the latter factorizes in the blow-down limit.} 
\begin{equation}
\mathcal{Q}=\sqrt{\frac{4}{k}}\,.
\end{equation}  

Focusing now on the $T^{1,1}$  space, we recall the action (\ref{gaugingI}) of the first gauging on the group element $(g_1,g_2)\in SU(2) \times SU(2)$, supplemented with an action on the left-moving fermions dictated by $\mathcal{N}=1$ worldsheet supersymmetry. As seen in section~\ref{gaugedsec}, the anomaly following from this gauging is compensated by a minimal coupling to the worldsheet fermions of the gauge sector of the heterotic string, 
specified by the shift vector $\vec{p}$. 

By algebraically solving the coset \textsc{cft} associated with this gauged \textsc{wzw} model, we are led to the following constraints on the zero-modes of the affine currents $J^{3}_{1.2}$ of the $SU(2) \times SU(2)$ 
Cartan subalgebra:\footnote{These are the total currents of 
the $\mathcal{N}=1$ affine algebra, including contributions of the worldsheet fermion bilinears.}
\begin{equation}
\label{zeromodcond}
(J^3_1)_0-(J^{3}_2)_0 = \vec{p}\cdot \vec{Q}_\textsc{f}\, ,
\end{equation}
where $\vec{Q}_\textsc{f}$ denotes the $\mathfrak{so}(32)$ weight of a given state. The affine currents of the  $\widehat{\mathfrak{so}(32)}$ algebra can be alternatively written in the fermionic or bosonic representation as
\begin{equation}
\bar{\jmath}^{i} (\bar z) =   \bar{\xi}^{2i-1} \bar{\xi}^{2i}   (\bar z) = \sqrt{\frac{2}{\alpha'}}\, \bar \partial X^i (\bar z)
\quad , \qquad i=1,\ldots 16 \,,
\label{cartanbosfer}
\end{equation}
and the components of $\vec{Q}_\textsc{f}$ can be identified with the corresponding fermion number (mod 2). 

In order to explicitly solve the zero-mode constraint~(\ref{zeromodcond}) at the level of the one-loop partition function, 
it is first convenient to split the left-moving supersymmetric $SU(2)$ characters in terms of the characters of an
$SU(2)/U(1)$ super-coset:\footnote{These super-cosets correspond to $\mathcal{N}=2$ minimal models. Some details about their characters $ C^j_m \oao{a}{b}$ are given in appendix~\ref{appchar}.} 
\begin{equation}
\chi^j \vartheta \oao{a}{b} = \sum_{m \in \mathbb{Z}_{2k}} C^j_m \oao{a}{b} \Theta_{m,k}\, .
\end{equation}
Next, to isolate the linear combination of Cartan generators  appearing in~(\ref{zeromodcond}), 
one can combine the two theta-functions at level $k$ corresponding to the Cartan generators 
of the two $\widehat{\mathfrak{su}(2)}_k$ algebras by 
using the product formula:
\begin{equation}
\label{thetasum}
\Theta_{m_1,k}\Theta_{m_2,k} = \sum_{s \in \mathbb{Z}_2} 
\Theta_{m_1-m_2+2ks,2k} \Theta_{m_1+m_2+2ks,2k}\, .
\end{equation}
Thus, the gauging yielding the $T^{1,1}$ base will effectively 'remove' the $U(1)$ corresponding to the first theta-function. For simplicity, we again limit ourselves to the same minimal choice
of shift vectors as in (\ref{minimalsol}), namely $\vec{p} = (2\ell,0^{15})$, $\ell \in \mathbb{Z}$, which implies by (\ref{anomalyconstr0})\footnote{We will see later 
that the evenness of $k$ is a necessary condition to the resolution of the conifold by a blown-up four-cycle.} 
\begin{equation}
k=4\ell^2 \, .
\end{equation}
Then  the gauging will involve only  a single right-moving Weyl fermion.  
Its contribution to the partition function is given by a standard fermionic theta-function:
\begin{equation} 
\vartheta \oao{u}{v} (\tau ) = \sum_{N \in \mathbb{Z}} q^{\frac{1}{2}(N+\frac{u}{2})^2} e^{i\pi v(N+\frac{u}{2})}\, ,
\label{fermitheta}
\end{equation}
where $\oao{u}{v}$ denote the spin structure on the torus. The solutions of the zero-mode constraint~(\ref{zeromodcond}) can 
be obtained from the expressions~(\ref{thetasum}) and~(\ref{fermitheta}). It gives (see~\cite{Berglund:1995dv,Israel:2004vv} for  simpler cosets of the same type):
\begin{equation}
m_1-m_2= 2\ell(2M+u) \quad , \qquad M \in \mathbb{Z}_{2\ell}\, .
\label{zeroconstr}
\end{equation}
We are then left, for given $SU(2)$ spins $j_1$ and $j_2$, with contributions to the coset partition function of the form 
\begin{equation}
\sum_{m_1 \in \mathbb{Z}_{8\ell^2}} C^{j_1}_{m_1} \oao{a}{b}\, \bar{\chi}^{j_1}\, 
\sum_{M \in \mathbb{Z}_{2\ell}} e^{i\pi v (M+\tfrac{u}{2})}C^{j_2}_{m_1-2\ell(2M+u)} \oao{a}{b} \,\bar{\chi}^{j_2}  \, \sum_{s \in \mathbb{Z}_2} 
\Theta_{2m_1-2\ell(2M+u)+8\ell^2 s,8\ell^2}\,.
\end{equation}
One can in addition simplify this expression using the identity
\begin{equation}
\sum_{s \in \mathbb{Z}_2}
\Theta_{2m_1-2\ell(2M+u)+8\ell^2s,8\ell^2} = \Theta_{m_1-\ell(2M+u),2\ell^2}\, .
\end{equation}
Note that the coset partition function by itself cannot be modular invariant, since fermions from the gauge sector of the heterotic string were used in the coset construction.

\subsection{Heterotic strings on the singular conifold}
\label{sec:hetstr}
The full modular-invariant partition function for the {\it singular} torsional conifold case
can now be established by adding (in the light-cone gauge) the $\mathbb{R}^{2}\times \mathbb{R}_\mathcal{Q}$ contribution, together with the remaining gauge fermions. 
Using the coset defined above, one then obtains the following one-loop amplitude:
\begin{multline}
\label{conepf}
Z (\tau ,\bar \tau) = \frac{1}{(4\pi \tau_2 \alpha')^{5/2}} \frac{1}{\eta^3 \bar \eta^3} \, 
\frac{1}{2}\sum_{a,b}(-)^{a+b} \frac{\vartheta \oao{a}{b}^2}{\eta^2} 
 \sum_{m_1 \in \mathbb{Z}_{2k}} \sum_{M \in \mathbb{Z}_{2\ell}}\, 
\frac{1}{2}\sum_{u,v \in \mathbb{Z}_2} \frac{\Theta_{m_1-2\ell(M+u/2),2\ell^2}}{\eta} \, \times \\[4pt] 
\times \,  \sum_{2j_1,2j_2=0}^{k-2} 
C^{j_1}_{m_1} \oao{a}{b}
C^{j_2}_{m_1-2\ell(2M+u)} \oao{a}{b} e^{i\pi v (M+\tfrac{u}{2})} \bar{\chi}^{j_1} \bar{\chi}^{j_2} 
\, 
\frac{\bar{\vartheta} \oao{u}{v}^{15}}{\bar{\eta}^{15}}\,.
\end{multline}
The terms on the second line correspond to the contribution of the  $\mathbb{R}^2\times \mathbb{R}_\mathcal{Q} \times U(1)$ piece with the 
associated left-moving worldsheet fermions. Their spin structure  is given by $\oao{a}{b}$, with $a=0$ (resp. $a=1$) corresponding to the \textsc{ns} (resp. \textsc{r}) sector. 
Again, the spin structure of the right-moving heterotic fermions for the 
$Spin(32)/\mathbb{Z}_2$ lattice is denoted by $\oao{u}{v}$ (see the last term in this partition function). One may as well consider the $E_8\times E_8$ heterotic string theory, by changing the spin structure accordingly.

We notice that the full right-moving $SU(2)\times SU(2)$ affine symmetry, corresponding to the isometries of the $S^2 \times S^2$ part of the geometry, is preserved, while the surviving left-moving $U(1)$ current represents translations along the $S^1$ fiber. In the partition 
function~(\ref{conepf}), the $U(1)$ charges are given by the argument of the theta-function at level $2\ell^2$.  Later on, we will realize this $U(1)$ in terms of the canonically normalized free chiral boson $X_\textsc{l} (z)$.  

\subsubsection*{Space-time supersymmetry}
The left-moving part of the \textsc{cft} constructed above, omitting the flat space piece, can be described as an orbifold of the superconformal theories:
\begin{equation}
\left[ \mathbb{R}_{1/\ell} \times U(1)_{2\ell^2}\right] \times \frac{SU(2)_k}{U(1)} \times \frac{SU(2)_k}{U(1)} \,.
\end{equation}
The term between the brackets corresponds to a linear dilaton $\rho$ with background charge $\mathcal{Q}=\tfrac{1}{\ell}$, together with a $U(1)$ at level $2\ell^2$ (associated with the bosonic field $X_\textsc{l}$) and a Weyl fermion. This system has $\mathcal{N}=(2,0)$ supersymmetry, as it can be viewed as the holomorphic part of  $\mathcal{N}=2$ Liouville theory at zero coupling. The last two factors are $SU(2)/U(1)$ super-cosets which are $\mathcal{N}=2$ minimal models. One then concludes that the left-moving part of the \textsc{cft} has an  $\mathcal{N}=2$ 
superconformal symmetry. The associated R-current reads~:
\begin{equation}
J_R (z) = i \psi^\rho \psi^\textsc{x} +\sqrt{\frac{2}{\alpha'}} \frac{i\partial X_\textsc{l}}{\ell} +  i \zeta^1 \zeta^2-\frac{J^3_1}{2\ell^2} +
 i \zeta^3 \zeta^4-\frac{J^3_2}{2\ell^2} \, .
\end{equation}
One observes from the partition function~(\ref{conepf}) that the $U(1)$ charge under the holomorphic current 
$i\sqrt{2/\alpha'} \partial X_\textsc{l}/\ell$, given by the argument 
of the theta-function at level $2\ell^2$, is always such that the total R-charge is an integer of definite parity. Therefore, with the usual 
fermionic \textsc{gso} projection, this theory preserves $\mathcal{N}=1$ supersymmetry in four dimensions {\it \`a la} Gepner~\cite{Gepner:1987qi}.

\subsection{Orbifold of the conifold}
\label{orb-sec}
The worldsheet \textsc{cft} discussed in sections \ref{sec:acft} and \ref{sec:hetstr}, as it stands, defines a  singular heterotic string background, at least at large $\rho$ where the string coupling constant is small. In addition, it is licit to take an orbifold of the $T^{1,1}$ base in a way that preserves $\mathcal{N}=1$  supersymmetry. If one resolves the singularity with a four-cycle,  a $\mathbb{Z}_2$ orbifold is actually needed. From the supergravity point of view, this removes the conical singularity at the bolt, while from the \textsc{cft} perspective, the presence of the orbifold is related to worldsheet non-perturbative effects, as will be discussed below.

Among the possible supersymmetric orbifolds of the conifold, we consider here a half-period shift along the $S^1$ fiber of $T^{1,1}$ base~:
\begin{equation}
\mathcal{T}\, : \ \psi \sim \psi +2 \pi\,,
\end{equation}
which amounts to a shift orbifold in the lattice of the chiral $U(1)$ at level $||\vec{p}^{\,2}||/2$. As the coordinate $\psi$ on the fiber is identified with corresponding  coordinates on the Hopf fibers of the two three-spheres, $i.e.$ $\psi/2=\psi_1=\psi_2$, the modular-invariant action of the orbifold can be conveniently derived by orbifoldizing on the left one of the two $SU(2)$ \textsc{wzw} models along the Hopf fiber (which gives the $\mathcal{N}=(1,0)$ worldsheet \textsc{cft} for a Lens space), {\it before} performing the gauging~(\ref{gaugingI}). This orbifold is consistent provided  $k$ is even, which is clearly satisfied for the choice $\vec{p}=(2\ell,0^{15})$ we have made so far. Then, the coset \textsc{cft} constructed from this orbifold theory will automatically yield a modular-invariant orbifold of the $T^{1,1}$ \textsc{cft}.

The partition function for the singular orbifoldized conifold is derived as follows. We should first make in the partition function~(\ref{conepf}) the following substitution
\begin{equation}
C^{j_2}_{m_1-2\ell(2M+u)} \oao{a}{b} \ 
\to \ \frac{1}{2}\sum_{\gamma,\delta \in \mathbb{Z}_2} e^{i\pi \delta(m_1 + 2\ell^2 \gamma)}\,
C^{j_2}_{m_1+4\ell^2 \gamma-2\ell(2M+u)} \oao{a}{b} \,,
\end{equation}
which takes into account the geometrical action of the orbifold. As expected, the 
orbifold projection, given by the sum over $\delta$, constrains the momentum along the fiber to be even,  both 
in the untwisted sector ($\gamma=0$) and 
in the twisted sector ($\gamma=1$). Using the reflexion symmetry~(\ref{reflsym}), this expression is equivalent to  
\begin{equation}
\frac{1}{2}\sum_{\gamma,\delta \in \mathbb{Z}_2}  e^{i\pi \delta(2j_2 + (2\ell^2-1) \gamma)} \,
C^{j_2+ \gamma(k/2-2j_2-1)}_{m_1-2\ell(2M+u)} \oao{a}{b} \,(-)^{\delta a +\gamma b +\gamma \delta}\, .
\label{orbchar}
\end{equation}
The phase factor $(-)^{\delta a +\gamma b +\gamma \delta}$ gives the action of a $(-)^{F_{\textsc{l}}}$  
orbifold, $F_{\textsc{l}}$ denoting the left-moving space-time fermion number. Therefore the orbifold by itself is not supersymmetric, 
as space-time supercharges are constructed out of $SU(2)/U(1)$ primaries with $j_1=j_2=0$ in the \textsc{r} sector ($a=1$). 
In order to obtain a supersymmetric orbifold one then needs to supplement this identification with 
a $(-)^{F_{\textsc{l}}}$ action in order to offset this projection. Then, we will instead quotient by $\mathcal{T} (-)^{F_{\textsc{l}}}$, which preserves space-time supersymmetry. 

The last point to consider is the possible action of the orbifold on the $Spin (32)/\mathbb{Z}_2$ lattice. In this case, there is a specific constraint to be satisfied that will guide us in the selection of the right involution among all the possible ones. From the form of the orbifold projection in expression~(\ref{orbchar}) one notices that in the twisted sector ($\gamma=1$) the $SU(2)$ spin $j_2$ needs to be half-integer. As we will discuss below, if we consider the worldsheet \textsc{cft} for the resolved conifold, 
this leads to an inconsistency due to worldsheet non-perturbative effects. Note that this 
problem is only  due to the particular choice of shift vectors $\vec{p}$ of the form~(\ref{minimalsol}) satisfying $\vec{p}^{\, 2}\equiv 0 \mod 4$, rather than  $\vec{p}^2\equiv 2 \mod 4$ which is more natural in supergravity.\footnote{This choice was made for convenience, as it involves the minimal number of right-moving fermions. One 
can check that all coset models with $\vec{p}\equiv 2 \mod 4$ involve a larger number of right-moving worldsheet fermions. In such cases, one cannot obtain a partition function explicitly written in terms of standard fermionic characters (although the \textsc{cft} is of course well-defined).}

However, as one would guess, the situation is not hopeless. In this example, as in other models with $\vec{p}^{\, 2}\equiv 0 \mod 4$, one way to obtain the correct projection in the twisted sector is to supplement the $\mathbb{Z}_2$ geometrical action with a $(-)^{\bar S}$ projection in the  $Spin(32)/\mathbb{Z}_2$ lattice, defined such that spinorial representations of $Spin(32)$ are odd.\footnote{It has a similar effect as the $(-)^{F_{\textsc{l}}}$ projection on the left-movers.} This has the effect of adding an extra monodromy 
for the gauge bundle, around the orbifold singularity. Overall one mods out the conifold \textsc{cft} by the  $\mathbb{Z}_2$ symmetry 
\begin{equation}
\mathcal{R} = \mathcal{T} (-)^{F_\textsc{l}+\bar S}\, .
\label{orbdef}
\end{equation}
Combining the space-time orbifold as described in eq.~(\ref{orbchar}) with the $(-)^{\bar S}$ action, one obtains a \textsc{cft} for orbifoldized conifold, 
which is  such that  states in the left \textsc{ns} sector have integer $SU(2)\times SU(2)$ spin in the orbifold twisted sector. The full partition 
function of this theory reads:
\begin{multline}
\label{orbconepf}
Z  (\tau, \bar \tau) =\frac{1}{(4\pi \tau_2 \alpha')^{5/2}} \frac{1}{\eta^3 \bar \eta^3} \, 
\frac{1}{2}\sum_{a,b}(-)^{a+b} \frac{\vartheta \oao{a}{b}^2}{\eta^2} 
\sum_{2j_1,2j_2=0}^{k-2} \sum_{m_1 \in \mathbb{Z}_{2k}}C^{j_1}_{m_1} \oao{a}{b}\, \frac{1}{2}\sum_{u,v \in \mathbb{Z}_2}  
\frac{\bar{\vartheta} \oao{u}{v}^{15}}{\bar{\eta}^{15}}
\, \times \\[4pt]
\times \, 
\frac{1}{2}\sum_{\gamma,\delta \in \mathbb{Z}_2}   (-)^{\delta(2j_2 + 2\ell^2 \gamma+u) +v\gamma} 
\sum_{M \in \mathbb{Z}_{2\ell}}C^{j_2+ \gamma(k/2-2j_2-1)}_{m_1-2\ell(2M+u)} \oao{a}{b} e^{i\pi v (M+\tfrac{u}{2})}
\frac{\Theta_{m_1-2\ell(M+u/2),2\ell^2}}{\eta}\, \bar{\chi}^{j_1} \bar{\chi}^{j_2}  \,.
\end{multline}
 
To conclude, we insist that if one chooses a gauge bundle with $\vec{p}^{\, 2}\equiv 2 \mod 4$, no orbifold action on the gauge bundle is needed in order 
to obtain a consistent worldsheet \textsc{cft} for the resolved orbifoldized conifold.

\subsection{Worldsheet CFT for the Resolved Orbifoldized Conifold}
In this section, we move on to construct the worldsheet \textsc{cft} underlying the resolved orbifoldized conifold with torsion~(\ref{sol-nhl}), which possesses a non-vanishing four-cycle at the tip of the cone. As a reminder, this theory 
is defined by both gaugings~(\ref{gaugingI},\ref{gaugingII}), where the second one now also involves an $\slr$ $\mathcal{N}=(1,0)$  \textsc{wzw} model at level $k/2$ and comprises an action on the $Spin(32)/\mathbb{Z}_2$ lattice parametrized by the vector $\vec{q}$.  

Denoting by $K^3$ the left-moving total affine current corresponding to the elliptic Cartan of $\mathfrak{sl}(2,\mathbb{R})$ and by $\bar{k}^3$ the right-moving purely bosonic one, the gauging leads to two constraints on their zero modes~:
\begin{equation}
K^3_0 = \frac{\sqrt{\alpha ' k'}}{2} p_\textsc{x} \, , \qquad\qquad 2\bar{k}^3_0 = -\vec{q} \cdot \vec{Q}_\textsc{f}\,,
\label{sl2constr}
\end{equation}
where $p_\textsc{x}$ is the momentum of the chiral boson $X_\textsc{l}$.  
As for the first gauging, these constraints can be solved by decomposing the $\slr$ characters in terms of the (parafermionic) characters of the coset $\slc$ and of the time-like $U(1)$ which is gauged. 

We consider from now on the model obtained for the choice of shift vectors $\vec{p}$ and $\vec{q}$ given by eq.~(\ref{minimalsol}), minimally solving the anomaly cancellation conditions~(\ref{anomalyconstrI},\ref{anomalyconstrII}). This choice  implies also that the $\slr$ part of the gauged \textsc{wzw} model will be the same as for an $\mathcal{N}=(1,1)$ model (as the third entry of $\vec{q}$ 
corresponds to the worldsheet-supersymmetric coupling of fermions to the gauged \textsc{wzw} model). The supersymmetric 
level of $\slr$ in this example is $k'=2\ell^2$. Conveniently one can then use the characters of the super-coset both 
for the left- and right-movers.\footnote{These characters, identical to the ones of $\mathcal{N}=2$ Liouville theory, are described in appendix~\ref{appchar}.} 
Then, the third entry of the shift vector $\vec{q}$~(\ref{minimalsol}) corresponds to the minimal coupling of the  gauge field to an extra right-moving Weyl fermion of charge $\ell$.

Solving for the constraints~(\ref{sl2constr}), one obtains the partition function for $Spin(32)/\mathbb{Z}_2$ heterotic strings on the resolved orbifoldized conifold with torsion. The first contribution comes 
from continuous representations, of $\slr$ spin $J=\tfrac{1}{2} + iP$, whose wave-function is delta-function normalizable. It reads
\begin{multline}
\label{resolvedcontpf}
Z_c (\tau ,\bar \tau) =\, \frac{1}{(4\pi \tau_2 \alpha')^{2}} \frac{1}{\eta^2 \bar \eta^2} 
\frac{1}{2}\sum_{a,b}(-)^{a+b} \frac{\vartheta \oao{a}{b}}{\eta} 
\sum_{2j_1,2j_2=0}^{4\ell^2-2} \, \sum_{m_1 \in \mathbb{Z}_{8\ell^2}}\, C^{j_1}_{m_1} \oao{a}{b} \, \\
\times
\frac{1}{2}\!\!\sum_{u,v \in \mathbb{Z}_2}  
\frac{1}{2}\!\sum_{\gamma,\delta \in \mathbb{Z}_2}   (-)^{\delta(2j_2 + 2\ell^2 \gamma+u) +v\gamma}
\!\!\!\sum_{M,N \in \mathbb{Z}_{2\ell}}\!\! \!(-)^{ v (M+N+u)} C^{j_2+ \gamma(k/2-2j_2-1)}_{m_1-2\ell(2M+u)} \oao{a}{b}
\bar{\chi}^{j_1} \bar{\chi}^{j_2} \frac{\bar{\vartheta} \oao{u}{v}^{13}}{\bar{\eta}^{13}}\\[4pt]
\times \frac{4}{\sqrt{\alpha' k}} \int_0^\infty\! \di P \, 
Ch_c \oao{a}{b}\left(\tfrac{1}{2}+iP, \tfrac{m_1}{2}-\ell(M+\tfrac{u}{2});\, \tau\right) \, 
Ch_c \oao{u}{v} \left(\tfrac{1}{2}+iP,\ell(N+\tfrac{u}{2});\, \bar{\tau}\right)\,.
\end{multline}
By using the explicit expression for the characters $Ch_c \oao{a}{b} (\tfrac{1}{2}+iP,n)$ of the continuous representations of $\slr$ (see eq.~(\ref{extcontchar})), one can show that 
this contribution to partition function is actually 
identical to the partition function~(\ref{orbconepf}) for the orbifoldized singular conifold. 
This is not suprising, as the one-loop amplitude (\ref{resolvedcontpf}) captures the modes that are not localized close to the singularity and
hence are not sensitive to its resolution.\footnote{The effect of the resolution can be however observed  in the sub-dominant term of the density of continuous representations, that does not scale 
with the infinite volume of the target space and is related to the reflexion amplitude by the Liouville potential discussed below, see\cite{Maldacena:2000kv,Israel:2004ir}.}

More interestingly, we have {\it discrete representations} appearing in the spectrum, labelled by their  $\slr$ spin $J>0$. They correspond to states whose wave-function is localized near the resolved singularity, $i.e.$ for $r\sim a$. Their contribution to the partition function is as follows
\begin{multline}
\label{resolveddiscpf}
Z_d (\tau ,\bar \tau) =\, \frac{1}{(4\pi \tau_2 \alpha')^{2}} \frac{1}{\eta^2 \bar \eta^2} 
\frac{1}{2}\sum_{a,b}(-)^{a+b} \frac{\vartheta \oao{a}{b}}{\eta} 
\sum_{2j_1,2j_2=0}^{4\ell^2-2} \, \sum_{m_1 \in \mathbb{Z}_{8\ell^2}}\, C^{j_1}_{m_1} \oao{a}{b} \, \\ \times
\frac{1}{2}\!\sum_{u,v \in \mathbb{Z}_2}  
\frac{1}{2}\!\sum_{\gamma,\delta \in \mathbb{Z}_2}   (-)^{\delta(2j_2 + 2\ell^2 \gamma+u) +v\gamma}
\!\!\!\sum_{M,N \in \mathbb{Z}_{2\ell}}\!\! (-)^{ v (M+N+u)} C^{j_2+ \gamma(k/2-2j_2-1)}_{m_1-2\ell(2M+u)} \oao{a}{b}
\bar{\chi}^{j_1} \bar{\chi}^{j_2} \frac{\bar{\vartheta} \oao{u}{v}^{13}}{\bar{\eta}^{13}}\\ \times\sum_{2J=2}^{2\ell^2+2}
Ch_d \oao{a}{b}\left(J, \tfrac{m_1}{2}-\ell(M+\tfrac{u}{2})-J-\tfrac{a}{2};\, \tau\right) \, 
Ch_d \oao{u}{v} \left(J, \ell(N+\tfrac{u}{2})-J-\tfrac{u}{2};\, \bar{\tau}\right) \\ \times \ 
\delta^{[2]}_{m_1-\ell(2M+u)-a , 2J}\ \delta^{[2]}_{\ell(2N+u)-u, 2J}\,,
\end{multline}
where the mod-two Kronecker symbols ensure that relation~(\ref{Rchargecoset}) holds. These discrete states break part of the gauge symmetry which was left unbroken by the first gauging. 

As can be checked from the partition function~(\ref{resolveddiscpf}), the resolution of the singularity preserves $\mathcal{N}=1$ space-time supersymmetry. Indeed, the left-moving part of the one-loop amplitude consists in a tensor product of $\mathcal{N}=2$ superconformal theories 
(the $\slc$ and two copies of $SU(2)/U(1)$ super-cosets) whose worldsheet R-charges add up to integer values of definite parity. 

Getting the explicit partition function for generic shift vectors $\vec{p}$ and $\vec{q}$ is not conceptually more difficult, but technically more involved. 
One needs to introduce the string functions associated with the coset \textsc{cft} $[Spin(32)/\mathbb{Z}_2]/[U(1)\times U(1)]$, where the embedding of the two gauged affine $U(1)$ factors are specified by 
$\vec{p}$ and $\vec{q}$.  In the fermionic representation, this amounts to repeatedly use
product formulas for theta-functions. The actual form of the results will clearly depend on the arithmetical properties of the shift vectors' entries.

\subsection{Worldsheet non-perturbative effects}\label{wnpe}
The existence of a worldsheet \textsc{cft} description for the heterotic resolved conifold background gives us in addition a handle on worldsheet instantons effects. As for the warped Eguchi-Hanson background analyzed in~\cite{Carlevaro:2008qf}, at least part of these effects are captured by worldsheet non-perturbative corrections to the $\slc$ super-coset part of the \textsc{cft}. In the present context, these corrections should correspond to string worldsheets wrapping the $\mathbb{C}P^1$'s of the blown-up four-cycle.

It is actually known~\cite{fzz,Kazakov:2000pm,Hori:2001ax} that the $\slc$ coset receives non-perturbative 
corrections in the form of a sine-Liouville potential (or an $\mathcal{N}=2$ Liouville 
potential in the supersymmetric case). Thus, to ensure that the worldsheet \textsc{cft} is 
non-perturbatively consistent, one needs to check whether the operator corresponding to this 
potential, in its appropriate form, is part of the physical spectrum of the theory. Whenever this is not the case, the resolution of the conifold singularity with a four-cycle is not possible. 

The marginal deformation corresponding to this Liouville potential can be written in an asymptotic free-field description, valid in the large $\rho$ region far from the bolt. There, $\rho$ can be viewed as 
a linear dilaton theory, as for the singular conifold theory.  
Let us begin with the specific choice of gauge bundle corresponding to the model~(\ref{resolvedcontpf}).
The appropriate Liouville-type interaction reads in this case (using the bosonic representation of the Cartan generators in~(\ref{cartanbosfer})):\footnote{We set here $\alpha'=2$ for convenience. The bosonic fields $X$,$Y^i$ and $\rho$, as well 
as the fermionic superpartners, are all canonically normalized.}
\begin{equation}
\label{Liouvint}
\delta S = \mu_\textsc{l} \int \di^2 z \, (\psi^\rho +i\psi^\textsc{x})(\bar{\xi}^5+i\bar{\xi}^6) e^{-\ell(\rho + iX_\textsc{l}+iY^2_\textsc{r})} 
+ c.c.\,.
\end{equation} 
Note that the contribution of the $SU(2)/U(1)$ coset is trivial. One now requires the operator  appearing in the deformation (\ref{Liouvint}) to be part of the 
physical spectrum, at super-ghost number zero.  If so, it can be used to de-singularize the 
background. 

We proceed to determine the quantum numbers of this operator to be able to identify its contribution in the partition function~(\ref{orbconepf}). Let us begin by looking at the holomorphic 
part. We denote by $p_\textsc{x} = -\ell$ the  momentum  of the compact boson $X_\textsc{l}$. Looking at the partition function for the singular conifold~(\ref{orbconepf}), a state with such momentum for $X_\textsc{l}$ obeys the condition
\begin{equation}
m_1-\ell(2M+u) \equiv -2\ell^2 \mod 4\ell^2\,.
\end{equation}
For this operator to be in the right-moving \textsc{ns} sector we require $u=0$. Secondly we want the contributions of both 
$SU(2)/U(1)$ super-cosets to be isomorphic to the identity. The solution to these constraints is given by\footnote{Note 
that the two $SU(2)/U(1)$ cosets seem naively to play inequivalent roles; this simply comes
from the fact that we are solving the coset constraint~(\ref{zeroconstr}) in a way that is not explicitly invariant under permutation of the 
two cosets.}
\begin{equation}
m_1=0 \quad , \qquad M = \ell
\end{equation}
In order to obtain the identity operator, one selects the 
representations $j_1=0$ and  $j_2=0$ respectively. The reflexion symmetry~(\ref{reflsym}) maps the contribution of the second $SU(2)/U(1)$ super-coset
--~which belongs to the twisted sector of the $\mathbb{Z}_2$ orbifold~(\ref{orbdef})~-- to the identity.
This property also ensures  that the Liouville potential in~(\ref{Liouvint}) is even under the left-moving \textsc{gso} projection.\footnote{Indeed, as a $(-)^b$ factor appears in the right-hand side of the identity~(\ref{reflsym}), the left \textsc{gso} projection is reversed.}  

On the right-moving side, one first needs to choose the momentum of $Y_\textsc{r}^2$ to be 
$\bar{p}_\textsc{y} = -\ell$. This implies that the state under consideration has $N=-\ell$ in the partition function~(\ref{orbconepf}). Secondly, having $j_1=j_2=0$ ensures that the right $SU(2)_k\times SU(2)_k$ contribution is trivial. This would not have be possible without the $\mathbb{Z}_2$ orbifold. This shows that, as in~\cite{Carlevaro:2008qf}, 
the presence of the orbifold is dictated by the non-perturbative consistency of the worldsheet \textsc{cft}. This illustrates in a remarkable way how the condition in supergravity guaranteeing
the absence of a conical singularity at the bolt manifests itself in a fully stringy description.

A last possible obstruction to the presence of the Liouville potential (\ref{Liouvint}) in the spectrum 
comes from the right-moving \textsc{gso} projection, defined in the fermionic representation of the 
$Spin (32)/\mathbb{Z}_2$ lattice,  given in~(\ref{orbconepf}) by the sum over $v$. Now, the 
right worldsheet fermion number of the Liouville potential~(\ref{Liouvint}) is given by
\begin{equation}
\bar{F}= \ell +1 \mod 2\,,
\end{equation}
and, in addition, the right-moving \textsc{gso} projection receives a contribution related to the momentum $p_X$,  which can be traced back to the coset producing the $T^{1,1}$ base of the conifold (see the phase $(-)^{vM}$  in the partition function~(\ref{orbconepf}) of our model). 

As we are in the twisted sector of the $\mathbb{Z}_2$ orbifold, the 
heterotic \textsc{gso} projection is reversed (because of the $(-)^{v\gamma}$ factor). Overall, the right \textsc{gso} parity of the Liouville 
operator~(\ref{Liouvint}) 
is then $2\ell \mod 2$. Therefore the Liouville potential~(\ref{Liouvint}) is part of the physical spectrum for any $\ell$. 

In the \textsc{cft} for the resolved conifold, the operator corresponding to the Liouville potential belongs to the discrete representation of $\slr$ spin 
$J=\ell^2$. One can check from the partition function of the discrete states~(\ref{resolveddiscpf}) that it is indeed physical. This operator 
is also chiral w.r.t. both the left and right $\mathcal{N}=2$ superconformal algebras of $\slc\times SU(2)/U(1) \times SU(2)/U(1)$.

\subsubsection*{Non-perturbative corrections for generic bundles}
This analysis can be extended to a generic Abelian gauge bundle over the resolved conifold, 
{\it i.e.} for an arbitrary shift vector $\vec{q}$ leading to
a consistent gauged \textsc{wzw} model. One can write the necessary Liouville potential in a free-field description as
\begin{equation}
\label{Liouvintgen}
\delta S = \mu_\textsc{l} \int \di^2 z (\psi^\rho +i\psi^\textsc{x})e^{-\frac{\sqrt{\vec{q}^{\,2}-4}}{2} (\rho + i X_\textsc{l})} 
\, e^{\frac{i}{2}\vec{q} \cdot \vec{Y}_\textsc{r}}
+ c.c. \,.
\end{equation}
Again we require this operator to be part of the physical spectrum of the heterotic coset \textsc{cft} (\ref{cosetdef}), 
taking into account the \textsc{gso} and orbifold projections. 

We have to discuss two cases separately:\\ 

\noindent $\bullet$ {\it Bundles with} $c_1(V) \in H^2 (\mathcal{M}_6,2\mathbb{Z})$\\[4pt]
Let us first start by looking at bundles with $\vec{p}^{\, 2}\equiv 2 \mod 4$, for which the orbifold allows the Liouville operator to 
be in the spectrum  without any action in the $Spin(32)/\mathbb{Z}_2$ lattice (see the discussion in subsection~\ref{orb-sec}). 
On top of the parity under the orbifold projection, on also needs to check that the right \textsc{gso} projection is satisfied.  
The right worldsheet fermion number of this operator is given by 
\begin{equation}
\bar{F}= \frac{1}{2}\sum_{i=1}^{16} q_i\, .
\end{equation}
As for the particular example above,  the right \textsc{gso} projection also receives a contribution from the $X_\textsc{l}$ momentum. 
The generalization of the $(-)^{v\ell}$  phase found there to a generic Abelian bundle can be shown to be:
\begin{equation}
e^{ \frac{i \pi }{2} v \sum\limits_{i=1}^{16} p_i}\, .
\end{equation} 
Therefore,  one concludes that the gauge bundle associated with the resolution of the conifold needs to satisfy the constraint
\begin{equation}
\frac{1}{2}\sum_{i=1}^{16} (q_i - p_i) \equiv 0 \mod 2 \,.
\label{fwcondcft}
\end{equation}
We observe (as for the warped Eguchi-Hanson heterotic \textsc{cft}, see~\cite{Carlevaro:2008qf}) that this condition  is similar to one of the two conditions given by eq.~(\ref{Kth2}). Considering only bundles with vector structure, the constraints~(\ref{fwcondcft}) 
and~(\ref{Kth2}) are just the same. If we choose instead a bundle without vector structure, the entries of $\vec{q}$ are all 
odd integers, see~(\ref{constr-pandq}). Therefore the condition of right \textsc{gso} invariance of the complex conjugate Liouville operator actually reproduces 
the second constraint of eq.~(\ref{Kth2}).

To make a long story short, this means that, in all cases, requiring the existence of a Liouville operator invariant under the right \textsc{gso} projection in the physical spectrum is equivalent to the condition~(\ref{Kth}) on the first Chern class of the gauge bundle, {\it i.e.} that $c_1 (V) \in H^2(\mathcal{M}_6,2\mathbb{Z})$. This remarkable 
relation between  topological properties of the gauge bundle and the \textsc{gso} parity of worldsheet instanton corrections may 
originate from modular invariance, that relates the existence of spinorial representations of the gauge group to the projection with the right-moving worldsheet fermion number.\\

\noindent $\bullet$ 
{\it Bundles with} $c_1(V) \in H^2 (\mathcal{M}_6,2\mathbb{Z}+1)$\\[4pt]
We now consider bundles with $\vec{p}^{\, 2}\equiv 0 \mod 4$, for which an orbifold action  in the $Spin(32)/\mathbb{Z}_2$ lattice is necessary for the Liouville operator to be part of the physical spectrum. The $(-)^{\bar{S}}$ action in the orbifold has the effect of reversing  the \textsc{gso} projection in the twisted sector. Hence we obtain the condition
\begin{equation}
\frac{1}{2}\sum_{i=1}^{16} (q_i - p_i) \equiv 1 \mod 2\,,
\label{fwcond}
\end{equation}
which now entails $c_1 (V) \in H^2(\mathcal{M}_6,2\mathbb{Z}+1)$. This condition on the first Chern class is the opposite (in evenness) to the standard condition on $c_1(V)$ appearing in the previous case~(\ref{fwcondcft}); this fact can be traced back to the extra monodromy 
of the gauge bundle around the resolved orbifold singularity.

\subsection{Massless spectrum}
In this section, we study in detail the massless spectrum of the resolved heterotic conifold with torsion. As in~\cite{Carlevaro:2008qf}, the gauge bosons 
corresponding to the unbroken gauge symmetry are non-normalizable, hence do not have support near the resolved singularity. In contrast, the spectrum 
of normalizable, massless states consists in chiral multiplets of $\mathcal{N}=1$ supersymmetry in four dimensions. 

As all the states in the right Ramond sector are massive, we restrict ourselves to the \textsc{ns} sector ($u=0$). In this case the orbifold projection 
enforces $j_2 \in \mathbb{Z}$.  One first looks for chiral operators w.r.t. the left-moving $\mathcal{N}=2$ superconformal algebra of the 
coset~(\ref{cosetdef}) of worldsheet R-charge $Q_R = \pm 1$.\footnote{Note that states with $Q_R=0$ in the conifold \textsc{cft} cannot give 
massless states, as the identity operator is not normalizable in the $\slc$ \textsc{cft}.}  
Then, one must pair them with a right-moving part of conformal dimension $\bar \Delta = 1$.  
In the special case studied here, which also comprises a right $\mathcal{N}=2$ superconformal algebra for the $\slc$ factor, one can  start with right chiral primaries 
of $\slc$, tensored with conformal primaries of the bosonic $SU(2)_{k-2} \times SU(2)_{k-2}$, which
overall yields a state of dimension $\bar \Delta=1/2$. 
A physical state of dimension one can then be constructed either by:
\begin{itemize}
\item adding a fermionic oscillator $\bar{\xi}^a_{-1/2}$  from the free $SO(26)_1$ gauge sector. This gives a state in the fundamental  representation of $SO(26)$. 
\item taking the right superconformal descendant of the $(1/2,1/2)$ state using the global 
right-moving superconformal algebra of the $\slc$ coset ({\it i.e.} acting with $\bar{G}_{-1/2}$). This leads to a singlet of $SO(26)$. 
\end{itemize}
In both cases, one needs to check, using the discrete part of the partition function~(\ref{resolveddiscpf}), that such physical states actually exist.

The $U(1)$ symmetry corresponding to translations along the $S^1$ fiber of $T^{1,1}$ (of coordinate $\psi$) corresponds to an  R-symmetry in space-time 
(of four-dimensional $\mathcal{N}=1$ supersymmetry).  In the  worldsheet \textsc{cft}  for the singular conifold, the associated affine $U(1)$ symmetry is realized in terms of the chiral boson 
$X_\textsc{l}$.  Therefore the R-charge $R$ in space-time is given by the argument of the theta-function at level $\vec{p}^{\, 2}/2$ (see the partition function~(\ref{orbconepf}).\footnote{In order to correctly
normalize the space-time R-symmetry charges, one needs to ensure that the space-time supercharges have R-charges $\pm 1$. The latter are constructed from vertex operators in the  Ramond sector ($a=1$), with $j_1=j_2=0$, $m_1 = \pm 1$ and $M=0$.} In the resolved geometry it is 
broken to a $\mathbb{Z}_{\vec{q}^2/2-2}$ discrete subgroup by the Liouville potential~(\ref{Liouvintgen}).

\subsubsection*{Untwisted sector}
Let us begin by discussing the untwisted sector. On the left-moving side, one can first consider states of the 
$(a,a,a)$ type, $i.e.$ antichiral w.r.t. the $\mathcal{N}=2$ superconformal 
algebras of the  $\slc$ and the two $SU(2)/U(1)$ super-cosets. For properties of these chiral primaries we refer the reader to appendix~\ref{appchar}. States of this type have  conformal dimension one-half provided the $\slr$ spin obeys
\begin{equation}
J=1+\frac{j_1+j_2}{2}\,.
\end{equation}
The condition relating the R-charges of the three coset theories, as can be read from the partition function~(\ref{resolveddiscpf}), imply that:\footnote{ 
These three equations correspond respectively to the $\slc$ factor, to the first $SU(2)/U(1)$ super-coset, with spin $j_1$ and to the second one, with spin $j_2$.}
\begin{equation}
\left\{ \begin{array}{lcl} m_1 -2\ell M &=&2(J-1)\, =\, j_1 + j_2 \\[4pt]
m_1 & =& 2j_1\\[4pt]
m_1-4\ell M & =& 2j_2 
\end{array}
\right. \implies j_1 - j_2 \,=\, 2\ell M \,.
\end{equation}
Then, one can first tensor states of this kind with right chiral primaries  of $\slc$ (denoted $\bar{c}$). The conformal dimension 
of the  conformal primary obtained by adding the $ SU(2)_{k-2} \times SU(2)_{k-2}$ contribution has the requested dimension 
$\bar \Delta = 1/2$, provided that 
\begin{equation}
 (j_1+1)^2 + (j_2+1)^2 = 2 \ell^2 \, , 
\end{equation}
and the R-charge of $\slc$ is such that $j_1+j_2+2 = 2\ell N$. 

There exists a single solution to all these constraints for $N=1$ and $M=0$, leading to a 
$(a,a,a)_\textsc{u} \otimes \bar{c}$ state with  $J=\ell$ and $j_1=j_2=\ell-1$.  Starting instead with a right anti-chiral primary 
of $\slc$ (denoted $\bar{a}$), we arrives at the two constraints
\begin{equation}
\left\{ \begin{array}{lcl} j_1^2 + j_2^2 &=& 0\\[4pt]
j_1+j_2 &=& 2\ell N
\end{array}
\right. \, , 
\end{equation}
which can simultaneously be solved by setting $J=1$ and $j_1=j_2=0$. 

One can attempt to obtain other massless states in the untwisted sector of the theory by considering left chiral primaries of the $(c,c,a)$ or $(c,a,c)$ type. In those cases, however, one finds that there are no solutions to the corresponding system of constraints, and so no corresponding physical states.

To summarize, the untwisted sector spectrum contains only the following states, that are all even under the left and right \textsc{gso} projections~:
\begin{itemize} 
\item Two chiral multiplets in space-time from $(a,a,a)_\textsc{u} \otimes \bar{c}$ worldsheet chiral primaries
with spins $j_1=j_2=\ell-1$, one in the singlet and the other one in the fundamental of $SO(26)$. These states both have space-time R-charge 
$R=2(\ell-1)$. 
\item Two chiral multiplets from $(a,a,a)_\textsc{u} \otimes \bar{a}$ primaries
with spins $j_1=j_2=0$, one in the singlet and the other one in the fundamental of $SO(26)$. These states both have vanishing space-time R-charge. 
\end{itemize}

\subsubsection*{Twisted sector}
The analysis of the twisted sector is along the same lines, except that the spin of the second $SU(2)/U(1)$ is 
different, and that the right \textsc{gso} projection is reversed. 
One can first consider states of the $(a,a,a)_\textsc{t}$ type. The $\slr$ spin takes the values
\begin{equation}
J=\ell^2+\frac{1}{2}+\frac{j_1-j_2}{2}\,.
\end{equation}
Then, the relation between the left R-charges entails that
\begin{equation}
\left\{ \begin{array}{lcl} m_1 -2\ell M &=&2(J-1)\, =\, 2\ell^2-1+j_1-j_2 \\[4pt]
m_1 & =& 2j_1\\[4pt]
m_1-4\ell M & = & 4\ell^2-2j_2-2 
\end{array}
\right. \implies j_1 + j_2+1 \,=\, 2\ell( M+\ell)\,.
\end{equation}
Now, tensoring the states under consideration with a right chiral primary of $\slc$ does not give any solution. Instead, tensoring with a right anti-chiral primary of the same leads to the two constraints:
\begin{equation}
\left\{ \begin{array}{lcl} j_1^2 + (j_2+1)^2 &=& 2\ell^2\\[4pt]
j_1-j_2-1+2\ell^2 &=& 2\ell N
\end{array}
\right. \, ,
\end{equation}
which are simultaneously solved by $N=\ell$ and $M=1-\ell$. This corresponds to a state with spins 
$j_1= \ell$, $j_2=\ell-1$ and $J= \ell^2+1$. 

A second kind of physical states is obtained by starting from a left $(c,a,c)_\textsc{t}$ chiral primary, with $\slr$ spin obeying
\begin{equation}
J=\ell^2-\frac{j_1+j_2}{2}\,.
\end{equation}
Repeating the previous analysis, the relation between the R-charges dictates
\begin{equation}
\left\{ \begin{array}{lcl} m_1 -2\ell M &=&2J = 2\ell^2-j_1-j_2\\[4pt]
m_1 & = & 2j_1\\[4pt]
m_1-4\ell M & = & 4\ell^2-2j_2
\end{array}
\right. \implies \left\{ \begin{array}{lcl}  j_1&=&0 \\[4pt] j_2&=&2\ell(\ell+M)  \end{array}
\right. \,.
\end{equation}
Then for a right chiral primary $\bar{c}$ of $\slc$, this leads to the conditions:
\begin{equation}
\left\{ \begin{array}{lcl} 4\ell^2(M+\ell)^2 &=&0\\[4pt] \ell (M+N) & =& 0
\end{array}
\right. \, ,
\end{equation}
with a single solution for $M=-\ell$ and $N=\ell$. This implies $j_1=0$, $j_2 = 0$ and $J=\ell^2$. One can check that no other combinations of left and right 
chiral primaries leads to any new massless physical state.

To summarize, we have found that the twisted sector spectrum only contains the following states:\footnote{These states 
are even under the \textsc{gso} 
projection  because the latter is reversed in the twisted sector of the orbifold.} 
\begin{itemize} 
\item Two chiral multiplets in space-time from $(a,a,a)_\textsc{t} \otimes \bar{a}$ worldsheet chiral primaries
with spins $j_1=j_2+1=\ell$ and $J=\ell^2+1$,  in the singlet and fundamental of $SO(26)$. 
\item Two chiral multiplets from $(c,a,c)_\textsc{t} \otimes \bar{c}$ primaries
with spins $j_1=j_2=0$ and $J=\ell^2$, in the singlet and  fundamental of $SO(26)$.
\end{itemize}
All these states have space-time R-charge $R=2\ell^2$. Note that the singlet $(c,a,c)_\textsc{t} \otimes \bar{c}$ state corresponds to the vertex operator that 
appears in the Liouville interaction~(\ref{Liouvint}). 

\begin{table}
\centering 
\begin{tabular}{|l|l|l|l|}
\cline{2-4}
\multicolumn{1}{c|}{ } & Worldsheet chirality & $SU(2) \times SU(2)$ spin & Spacetime R-charge \\
\hline
Untwisted sector & $(a,a,a) \otimes \bar{c}$ & $j_1=j_2 = \ell-1$ & $R=2(\ell-1)$\\
& $(a,a,a) \otimes \bar{a}$ & $j_1=j_2 = 0$ & $R=0$\\
\hline
Twisted sector & $(a,a,a) \otimes \bar{a}$ & $j_1=j_2+1=\ell$ & $R=2\ell^2$\\
& $(c,a,c) \otimes \bar{c}$ & $j_1=j_2=0$ & $R=2\ell^2$\\
\hline
\end{tabular}
\caption{\it Massless spectrum of chiral multiplets in space-time. For each entry of the table one has one singlet and one fundamental 
of $SO(26)$.}
\label{masslesstable}
\end{table}

We have summarized the whole massless spectrum found in our particular example in table~\ref{masslesstable}.

\section{Conclusion and Discussion}

In this work, we have constructed a new class of conifold backgrounds in heterotic string 
theory, which exhibit non-trivial torsion and support an Abelian gauge bundle. The supersymmetry 
equations and the Bianchi identity of heterotic supergravity also imply a non-trivial dilaton and a conformal factor for the conifold metric.

By implementing a $\mathbb{Z}_2$ orbifold on the $T^{1,1}$ base, one can consider resolving the conifold singularity (which is in the present case also a strong coupling singularity) by a four-cycle, leading to a smooth solution. 
This is a natural choice of resolution in the heterotic context, as the resolution 
is then naturally supported by a gauge flux proportional to the normalizable harmonic two-form implied by 
Hodge duality. It is of course perfectly possible 
that, in addition, a deformation of the conifold singularity is also allowed in the presence of torsion and of a line bundle. This would be an interesting follow-up of this work, having in mind heterotic conifold transitions.  

Numerical solutions for the metric have been found in the large charge limit, such that at infinity one recovers the 
Ricci-flat, K\"ahler conifold, while at finite values of the radial coordinate the conifold is squashed and warped, and acquires intrinsic 
torsion, leading to a complex but non-K\"ahler space. 

Remarkably, the region near the resolved conifold singularity, that can be cleanly isolated from the asymptotically Ricci-flat region by means of a double scaling limit, is found to admit a worldsheet \textsc{cft} description in terms of a gauged \textsc{wzw} model. This allows in principle to obtain the background fields to all orders in $\alpha'$, providing by construction an exact 
solution to the  Bianchi identity beyond the large charge limit. We did not explicitly calculate the expressions for the 
exact background fields, which is straightforward but technically involved. 

Instead, we used the algebraic worldsheet \textsc{cft} to compute the full string spectrum of the theory, focusing on a particular class of shift vectors. We found  a set of states localized near the resolved singularity, that give four-dimensional massless $\mathcal{N}=1$ chiral multiplets in space-time. We also emphasized the role of non-perturbative $\alpha'$ effects, 
or worldsheet instantons, that manifest themselves as sine--Liouville-like interactions, for generic bundles. We showed in particular how the conditions necessary for the existence of the corresponding operator in the physical spectrum of the quantum theory are related to the $\zi_2$ orbifold in the geometry, and how
the constraint on the first Chern class of the Abelian bundle can be exactly reproduced 
from worldsheet instanton effects.

There are other interesting aspects of this class of heterotic solutions that we did not develop in the previous sections. We would therefore like to comment here on their holographic interpretation  and their embedding in heterotic flux compactifications.

\subsection{Holography}\label{holo}
In the blow-down limit $a\to 0$ of the solutions~(\ref{sol-ansatz}), the dilaton becomes linear in the whole throat region, 
hence a strong coupling singularity appears for $r\to 0$. As reviewed in the introduction, this breakdown of perturbation theory 
generically expresses itself in the appearance of heterotic five-branes, coming from the zero-size limit of some gauge instanton.

In the present context, where the transverse space geometry is the warped conifold, the heterotic five-branes should be wrapping the vanishing two-cycle on the $T^{1,1}$ base, to eventually give rise to a four-dimensional theory. The $\mathcal{H}$-flux is indeed supported by the three-cycle orthogonal to it, see~(\ref{sol-ansatz}b). In addition, we have a non-trivial magnetic gauge flux (characterized by the shift vector $\vec{p}$) threading the two-cycle, which is necessary to satisfy the Bianchi identity at leading order. Hence we can understand this brane configuration as the heterotic analogue of fractional D3-branes on the conifold (which are actually D5-branes wrapped on the vanishing two-cycle). However here the number of branes, or the flux number, is not enough to characterize the theory, as one should also specify the actual gauge bundle intervening in the construction.  

Adding a $\mathbb{Z}_2$ orbifold to the $T^{1,1}$ base of the conifold, one can consider resolving the singularity by blowing up a $\ci P^1\times \ci P^1$, which, in the heterotic theory, requires turning on a second Abelian gauge bundle (with shift vector $\vec{q}$). This  does not change the asymptotics of the solution, hence the dilaton is still asymptotically linear; however the solution is now smooth everywhere. As for the flat heterotic five-brane solution of \textsc{chs}~\cite{Callan:1991dj}, this amounts, from the
supergravity perspective, to give a finite size to the gauge 
instanton.\footnote{Unlike for non-Abelian instantons, in the present case there is 
no independent modulus giving the size of the instanton.}

 From the perspective of the compactified four-dimensional  heterotic string, 
one leaves the singularity in moduli space by moving along a perturbative  branch of the 
compactification moduli space, changing the vacuum expectation value of the geometrical moduli field associated with the 
resolution of the conifold singularity.

Both in the blow-down and in the double-scaling limit, the dilaton is  asymptotically linear, hence a holographic interpretation is expected~\cite{Aharony:1998ub}. 
The dual theory should be a four-dimensional $\mathcal{N}=1$ 'little string theory'~\cite{Seiberg:1997zk}, living on the worldvolume of the wrapped five-branes. 
Unlike usual cases of type \textsc{iia}/\textsc{iib} holography, one does not have a good understanding of the  dual theory at hand, from a weakly coupled brane construction. 
Therefore, one should guess its properties from the heterotic supergravity background. First, its global symmetries  can be read from the isometries of the solution. 

As for ordinary heterotic five-branes~\cite{Gremm:1999hm}, the  gauge symmetry of the heterotic supergravity becomes a global symmetry. In the present case, 
$SO(32)$ is actually broken to a subgroup. The breaking pattern is specified by the shift vector $\vec{p}$ which is in some sense 
defined at an intermediate  \textsc{uv} scale of the theory, as the corresponding gauge flux in supergravity is not supported by a normalizable two-form. 

Second, the isometries of the conifold  itself become global symmetries of the gauge theory, as 
in \textsc{ks} theory~\cite{Klebanov:2000hb}. The $SU(2) \times SU(2)$ isometries of $T^{1,1}$ are kept unbroken at the string level, since
they correspond to the right-moving affine $\mathfrak{su}(2)$ algebras at level $\vec{p}^{\, 2}-2$.\footnote{However, 
the spins of the allowed $SU(2) \times SU(2)$ representations are bounded from above, as $j_1,\, j_2 \leqslant \vec{p}^{\, 2}/2-1$.}  
As in \textsc{ks} theory, the latter should be a flavour symmetry. 

More interestingly, the $U(1)$ isometry along the fiber of $T^{1,1}$ is expected to give an R-symmetry in the dual theory. When the singularity is resolved (in the orbifold theory) by a blown-up four-cycle, this  symmetry is broken by the Liouville potential~(\ref{Liouvintgen}) to a discrete $\mathbb{Z}_{\vec{q}^{\,2}/2-2}$ subgroup.  From the point of view of the dual four-dimensional theory, it means that one considers at the singular point a theory with an unbroken
$U(1)_\textsc{r}$ symmetry. The supergravity background is then deformed by adding a normalizable gauge bundle, corresponding to $\vec{q}$,
 without breaking supersymmetry. By usual AdS/CFT arguments, this corresponds in the dual theory to giving a vacuum expectation value to some 
chiral operator, such that the $U(1)_\textsc{r}$ symmetry  is broken to a discrete subgroup. Note that, unlike 
for instance in the string dual of $\mathcal{N}=1$ \textsc{sym}~\cite{Maldacena:2000yy}, this breaking of $U(1)_R$ to a $\mathbb{Z}_{k/2}$ 
subgroup does {\it not} mean that the R-symmetry is anomalous, because the breaking occurs in the infrared ($i.e.$ for $r\to a$) rather than 
in the ultraviolet ($r\to \infty$). One has instead a spontaneous breaking of this global symmetry, in a particular point of moduli space.

\subsubsection*{Holographic duality in the blow-down limit}
From the supergravity and worldsheet data summarized above we will attempt to better characterize the four-dimensional 
$\mathcal{N}=1$ theory dual to the conifold solution under scrutiny. One actually has  to deal with two issues:  what is the theory 
dual to the singular conifold --~or, in other words, which mechanism is responsible for the singularity~-- and  what is the dual 
of the orbifoldized conifold resolved by a four-cycle. A good understanding of the former would of course help to specify the latter.

First, one expects the physics at the singularity to be different for the $Spin(32)/\zi_2$ and the $E_8\times E_8$ heterotic string theory. As recalled in the  introduction, while one does not know what happens for generic four-dimensional $\mathcal{N}=1$ 
compactifications,  the situation is well understood for small instantons in compactifications to six dimensions. 
The difference in behavior at the singularity can be understood by their different strong coupling 
limit. For $Spin(32)/\mathbb{Z}_2$ heterotic string theory, S-dualizing to type I leads to a weakly coupled description, 
corresponding to an 'ordinary' field theory. On the contrary, in
$E_8 \times E_8$ heterotic string theory, lifting the system to M-theory on $S^1/\mathbb{Z}_2\times K3$ 
leads to a theory of M5-branes with self-dual tensors, which therefore has a strongly coupled low-energy limit. Descending to four dimensions, by fibering the $K3$ on a $\mathbb{C}P^1$ base, this leads to different four-dimensional 
physics at the singularity. It corresponds to strong coupling dynamics of asymptotically-free gauge groups in 
$Spin(32)/\mathbb{Z}_2$~\cite{Kachru:1996ci} and to interacting fixed points connecting branches with different numbers of generations, 
in the $E_8\times E_8$ case~\cite{Kachru:1997rs}.

In the present context, one can also S-dualize the $Spin(32)/\mathbb{Z}_2$ solution~(\ref{sol-ansatz}) to type I. There, in the 
blow-down limit, the string coupling constant vanishes in the infrared end of the geometry ($r\to 0$), hence one expects that the low-energy 
physics of the dual four-dimensional theory admits a free-field description. In terms of these variables, the theory is also not asymptotically free, since the coupling constant blows up in the \textsc{uv}.   This theory is living on a stack of $k$ (up to order one corrections) type I D5-branes wrapping the vanishing two-cycle of the conifold. Such theories have  $Sp(k)$ gauge groups, together with a flavor symmetry coming from the D9-brane gauge symmetry. However, as seen from the supergravity solution, one has to turn on worldvolume magnetic flux on the D9-branes in order to reproduce the theory of interest. The profile of the magnetic flux in the radial direction being non-normalizable, one expects this flux to correspond to some deformation in the Lagrangian of the four-dimensional dual theory, that breaks the $SO(32)$ flavor symmetry to a subgroup set by the choice of $\vec{p}$.

Let us consider now the $E_8\times E_8$ case. There, the singularity that appears in the blow-down limit needs to be lifted to 
M-theory, where the relevant objects are wrapped M5-branes. As there is no  weakly coupled description of the \textsc{ir} physics, 
the dual theory should flow at low energies to an interacting theory, $i.e.$ to an $\mathcal{N}=1$ superconformal field 
theory. In this case one would expect naively expect an $AdS_5$-type geometry, which is not the 
case here. To understand this, first note that the little string theory decoupling limit is not a low-energy limit, 
hence the metric should not be asymptotically $AdS$.  Second, the $AdS_5$ geometry that should appear in the \textsc{ir} seems to be 
'hidden' in the strong coupling region.\footnote{In type \textsc{iia} one can construct 
non-critical strings with $\mathcal{N}=2$~\cite{Giveon:1999zm} or $\mathcal{N}=1$~\cite{Israel:2005zp} supersymmetry in four dimensions (whose worldsheet 
\textsc{cft} description is quite analogous to the present models), 
that are dual to Argyres-Douglas superconformal field theories in four dimensions. No $AdS_5$ geometry is seen in those theories, for similar reasons.}

\subsubsection*{Looking for a confining string}
The background obtained by resolution is completely smooth in the infrared, so 
one may wonder whether it is confining. 

One first notices that standard symptoms of confinement seem not to be present in our models. There is no mass gap, the  
R-symmetry is broken spontaneously to $\mathbb{Z}_{\vec{q}^{\,2}/2-2}$ only 
(rather than having an anomalous $U(1)_R$  broken further to $\mathbb{Z}_2$ by  a gaugino condensate) 
and the space-time superpotential for the blow-up mode --~that is associated to the gluino bilinear in \textsc{sym} 
duals like~\cite{Vafa:2000wi}~-- vanishes identically, see~(\ref{superpot2}). However none of these features are conclusive, 
as we are certainly dealing with theories having a complicated matter  sector.

On general grounds, a confining behavior 
can be found in holographic backgrounds by constructing  Nambu-Goto long string probes, 
attached to external quark sources  in the \textsc{uv}, and showing that 
they lead to a linear potential~\cite{Kinar:1998vq}. A confining behavior occurs whenever 
the string frame metric component $g_{tt} (r)$ has a non-vanishing 
minimum at the \textsc{ir} end of the gravitational background (forcing it to be stretched  
along the bottom of the throat). A characteristic of our solution (which is probably generic in heterotic flux backgrounds) 
is that the $\mathbb{R}^{3,1}$ part of the string frame metric is not warped, see eq.~(\ref{sol-ansatz}a). Therefore the Nambu-Goto action for a fundamental heterotic string will give simply a straight long string, as in flat space. 

In the case of  $Spin(32)/\mathbb{Z}_2$ heterotic strings, one needs to S-dualize the solution to type I, in order to study 
the low-energy physics of the dual theory after blow-up. In fact, the resolution of the conifold singularity introduces a scale $1/a$, that should correspond to some mass scale in the holographically dual 4d theory. 
The ratio of this scale over the string mass scale $1/\sqrt{\alpha'}$ is given  by 
$\sqrt{\mu/g_s}$, where $\mu$ is the double-scaling parameter that gives the effective string coupling at the bolt. Taking the doubly-scaled heterotic background in the perturbative regime, this ratio is necessarily large, meaning that one does not decouple the field theory 
and string theory modes. Therefore, in order to reach the field-theory regime,  one needs to be at strong
heterotic string coupling near the bolt. This limit is accurately  described in the type I dual, in the \textsc{ir} part of the 
geometry; however in the \textsc{uv} region $r\to \infty$ the type I solution is strongly coupled.

In type \textsc{I} the string frame metric of the solution reads:
\begin{equation}
\di s^2_\textsc{i}  = H^{-1} (r)\,  \eta_{\mu\nu}\di x^{\mu}\di x^{\nu} +\tfrac{3}{2} \di s^2 (\tilde{\mathcal{C}}_6)\, ,
\end{equation}
with $H(r) = \alpha' k/r^2$ and $r\geqslant a$. Taking a D1-brane as a confining string candidate, one would obtain exactly 
the same answer as in the heterotic frame. One can consider instead a type I fundamental string, leading  to the 
behavior expected for a confining string (as $H(r)$ has a maximum for $r=a$).\footnote{In contrast, there is no obvious candidate for a confining string in the $E_8 \times E_8$ case, suggesting again that the physics of these models is different.}  
A type I fundamental string is of course prone to breaking onto D9-branes, but this is the expected 
behavior for a gauge theory with flavor  in the confining/Higgs phase, since the 
confining string can break as quark/antiquark pairs are created. More seriously, if one tries to 'connect' this string to external sources at infinity ($i.e.$ in the 
\textsc{uv} of the dual theory), the heterotic description, which is appropriate for $r\to \infty$, does not describe at all the type I 
fundamental string.

\subsubsection*{What is the dual theory?}
Let us now summarize our findings, concentrating on the $Spin (32)/\mathbb{Z}_2$ theory. Considering first the 
blow-down limit,  the mysterious holographic dual to the supergravity background~(\ref{sol-ansatz}),  in the heterotic variables, is asymptotically free --~at least up to a scale where the little string theory description takes over~-- and flows to a strong coupling singularity. On the contrary, in the type I variables, the theory is IR-free but strongly coupled in the \textsc{uv}. 
A good field theory example of this would be $SU(N_c)$ SQCD in the free electric phase, $i.e.$ with $N_f >3N_c$ 
flavors~\cite{Seiberg:1994pq}. Then, if one identifies the electric theory with the type I description and the magnetic theory 
with the heterotic description one finds similar behaviors.

Pursuing this analogy, let us identify the resolution of the singularity in the  supergravity solution with a (full) Higgsing of 
the magnetic theory. One knows that it gives a mass term to part of the electric quark multiplets, giving an electric theory 
with $N_f=N_c$ flavors remaining massless. Then, below this mass scale (that is set by the \textsc{vev} of the blow-up modulus) 
the electric theory confines. 

In a holographic dual of such a field theory 
one would face a problem when trying to obtain a confining string solution. In fact, trying to connect
the putative string with the boundary, one would cross the threshold $1/a$ above which the electric theory has $N_f >3N_c$ flavors, hence is strongly coupled at high energies and is not described in terms of free electric quarks.

Notice that we did not claim that the field theory scenario 
described above is dual to our heterotic supergravity background, rather that it is an 
{\it example} of a supersymmetric field theory that reproduces the features implied by holographic duality. The actual construction of the 
correct field theory dual remains as an open problem.

\subsubsection*{Chiral operators in the dual theory}
A way of better characterizing the holographic duality consists in studying chiral operators in the dual 
four-dimensional theory, starting at the (singular) origin of its moduli space. 
Following~\cite{Giveon:1999zm,Israel:2005zp}, the holographic duals of these operators can be found by looking at non-normalizable operators in the 
linear dilaton background of interest. In our case, one considers the singular conifold, whose \textsc{cft} is summarized in the partition function~(\ref{conepf}). 
This provides a  definition of the dual theory at an intermediate \textsc{uv} scale, solely given in terms of the vector of magnetic charges $\vec{p}$.\footnote{The 
resolved background, obtained by adding a second gauge field corresponding to the shift vector $\vec{q}$, is interpreted in the dual theory as the result of 
giving a vacuum expectation value (\textsc{vev}) to some  space-time chiral operator, changing the \textsc{ir} of the theory, see below.} 

More specifically we look at worldsheet vertex operators of the form~:
\begin{equation}
\label{vertex-op}
\mathcal{O} = e^{-\varphi (z)} e^{ip_\mu X^\mu} e^{-QJ \rho} e^{i p_x X_\textsc{l} (z)} 
V_{j_1 \, m_1} (z)V_{\tilde{\jmath}_2\, m_2} ( z)\bar{V}_{j_1} (\bar z)\bar{V}_{j_2} (\bar z)\bar{V}_\textsc{g} (\bar z)\, .
\end{equation} 
where $e^{-\varphi}$ denotes the left superghost vacuum in the $(-1)$ picture,  $V_{j\, m}(z)$ are left-moving primaries of the 
$SU(2)/U(1)$ supercoset, $\bar{V}_j (\bar z)$ are $SU(2)_{k-2}$ right-moving primaries and $\bar{V}_\textsc{g} (\bar z)$ comes from the 
heterotic gauge sector. In order to obtain operators with the desired properties, one has to choose 
chiral or anti-chiral operators in the $SU(2)/U(1)$ super-cosets. 

Physical non-normalizable operators in a linear dilaton theory have to obey the Seiberg bound, i.e. $J<1/2$ (see~\cite{Giveon:1999zm}). Furthermore, to obtain the correct 
\textsc{gso} projection on the  left-moving side, one chooses either $(c,a)$ or $(a,c)$ 
operators of $SU(2)/U(1)\times SU(2)/U(1)$. For simplicity we make the same choice of shift vector for the non-normalizable gauge field as in the remainder of the paper, 
namely $\vec{p}=(2\ell,0^{15})$. 

Let us for instance consider $(a,c)$ operators in the twisted sector. They are characterized by $m_1 = 2j_1$ and $m_2 = 4\ell^2-2j_2$, such that $j_1+j_2 = 2\ell(M+\ell)$. 
The  left and right worldsheet conformal weights of this state read:\footnote{From the four-dimensional perspective, these operators are defined off-shell. For a given value 
of $p_\mu p^\mu$ the quantum number $J$ is chosen accordingly, in order to obtain an on-shell operator from the ten-dimensional point of view.}
\begin{subequations}
\begin{align}
\Delta_\textsc{ws} &= \frac{\alpha '}{4} p_\mu p^\mu + \frac{-2J(J-1)+j_1 +j_2}{4\ell^2} + 
\frac{(j_1- \ell M)^2}{2\ell^2} -\frac12\,,\\
\bar{\Delta}_\textsc{ws} &= \frac{\alpha '}{4} p_\mu p^\mu + \frac{-2J(J-1)+j_1(j_1+1) +j_2(j_2+1)}{4\ell^2} + \bar{\Delta}_\textsc{g} -1\,.
\end{align}
\end{subequations}
Note that the state in the gauge sector, of right-moving conformal dimension $\bar{\Delta}_\textsc{g}$, belongs to the coset $SO(32)/SO(2)=SO(30)$ 
(as one Cartan has been gauged away). This leads to the condition
\begin{equation}
j_1 = \bar{\Delta}_\textsc{g} +\frac{M^2}{2} + 2\ell M +\ell^2-\frac12\, , 
\end{equation}
and the space-time $U(1)_R$ charge reads:
\begin{equation}
R = 2\bar{\Delta}_\textsc{g} +M^2 + 2\ell M +2\ell^2-1\,.
\end{equation}
A subset of these operators transform in the singlet of the $SU(2)\times SU(2)$ 'flavor' symmetry. They are characterized by 
$j_1=j_2=0$, hence have $M=-\ell$; their space-time R charge is $R=2\ell^2$. Such an operator can always be found for any solution of the equation
\begin{equation}
\bar{\Delta}_\textsc{g} =  \frac{\ell^2+1}{2} \,,
\end{equation}
provided the state of the gauge sector $(i)$ belongs to $SO(30)_1$  and $(ii)$  is \textsc{gso}-invariant.  One 
can express its conformal dimension in terms of the modes of  the 15 Weyl  fermions as 
$\bar{\Delta}_\textsc{g} =\frac{1}{2} \sum_{i=2}^{16} (N_i)^2$. 

In order to express the solution of these constraints in a more familiar form, we introduce  the sixteen-dimensional vector $\vec{q}=(0,N_2,\ldots,N_{16})$. 
Then one finds one space-time chiral operator  for  each $\vec{q}$ such that 
$\vec{q}^{\, 2}= \ell^2+1 = \vec{p}^{\, 2}/4+1$ and $\vec{p}\cdot \vec{q}=0$ and such that it obeys the 
condition~(\ref{fwcond}), $i.e.$  $\sum_{i} q_i \equiv \ell+1 \mod 2$.

In conclusion, the four-dimensional $\mathcal{N}=1$ theory which is dual to the  warped singular conifold defined by the shift vector $\vec{p}=(2\ell,0^{15})$ contains a 
subset of chiral operators in the singlet of $SU(2)\times SU(2)$, characterized by their weight $\vec{q}$  in $\mathfrak{so}(30)$. One can give a vacuum expectation value to any of these operators without breaking supersymmetry in space-time. Following the general AdS/CFT logic, it corresponds on 
the gravity side to consider a normalizable deformation of the linear dilaton background, associated with 
the shift vector $\vec{q}$.

One describes this process on the worldsheet  by adding a Liouville potential~(\ref{Liouvintgen}) corresponding to the chosen chiral operator and satisfying $J=\ell^2$; 
this operator breaks the space-time R-symmetry to $\mathbb{Z}_{2\ell^2}$.  
For each consistent choice of $\vec{q}$, the perturbed worldsheet \textsc{cft} is given by one of the coset theories (\ref{cosetdef}) constructed in this work. 
Note that in addition to the chiral operators discussed above, many others can be found that are not singlets of $SU(2)\times SU(2)$. In principle, these operators 
can also be given a vacuum expectation value, in those cases however the worldsheet \textsc{cft} is as far as we know not solvable anymore.

As explained above we observe that, for the $E_8 \times E_8$ heterotic string theory, the singularity seems to be associated with an interacting 
superconformal fixed point. In this case the conformal dimension of these operators in space time is given by
\begin{equation}
\Delta_\textsc{st} = \frac{3}{2} | R | \, = 3\ell^2\, ,
\end{equation}
after using the $\mathcal{N}=1$ superconformal algebra.

Clearly it would be interesting to obtain a more detailed characterization of the dual theory, using for instance anomaly 
cancellation as a guideline. We leave this study for future work.

\subsection{Relation to heterotic flux compactifications}
The Klebanov--Strassler type \textsc{iib} background  serves a dual purpose. On one side, it can be used to probe 
holographically non-trivial $\mathcal{N}=1$ quantum field theories. On another side, one can engineer 
type \textsc{iib} flux compactifications which are described locally, near a conifold singularity,  by 
such a throat~\cite{Giddings:2001yu}; this allows in particular to generate large hierarchies of couplings. 
In this second context, the \textsc{ks} throat is glued 
smoothly to the compactification manifold, at some \textsc{uv} scale in the field theory dual where the 
string completion takes over. Typically the flux compactification and holographic interpretations complement each 
other. One should keep in mind however, that from the supergravity perspective, as the flux numbers are globally 
bounded from above in the orientifold compactification with flux, the curvature of the manifold is not small.

The resolved conifolds with flux constructed in this paper can also be considered from these two perspectives. 
We have highlighted above aspects of the holographic interpretation. Here we would like to discuss their embedding 
in heterotic compactifications. As outlined in the introduction, heterotic compactifications with torsion are not (in general)
conformally Calabi-Yau, and thus correspond to non-K\"ahler manifolds. This makes the global study of such compactifications, without relying on  explicit examples, problematic. 

In the absence of a known heterotic compactification for which the geometry~(\ref{sol-ansatz}) could be viewed 
as a local model, one needs to understand how to 'glue' this throat geometry to the bulk of 
a compactification. In addition the presence of a non-zero \textsc{nsns} charge at infinity makes 
it even more difficult to make sense of the integrated Bianchi identity, leading to the tadpole cancellation conditions.

Let us imagine anyway that some torsional compactification manifold contains 
a conifold singularity with \textsc{nsns} flux, leading to a non-zero five-brane charge. Heterotic 
compactifications with five-branes are  non-perturbative, as the strong coupling singularity of the five-branes sets us out of the perturbative regime. However with the particular type of resolution of the singularity used here, 
corresponding to blowing-up the point-like instantons to  finite-size, the effective string 
coupling in the throat can be chosen as small as desired. 
It corresponds, from the point of view of four-dimensional effective theory, to moving to
 another branch of moduli space which has a weakly coupled heterotic description. 

There is an important difference between the fluxed Eguchi-Hanson solution that we studied in a previous 
article~\cite{Carlevaro:2008qf} and the torsional conifold 
backgrounds constructed in this work. In the former case, there existed a subset of line bundles 
such that the geometry was globally torsion-free, $i.e.$ such that 
the Bianchi identity integrated over the four-dimensional warped Eguchi-Hanson space did not 
require a Kalb-Ramond flux. In other words, there was no net five-brane 
charge associated with the throat. Then the torsion, dilaton and warp factor of the solution 
could be viewed as 'local' corrections to this globally 
torsion-less solution near a gauge instanton, that arose because the Bianchi identity 
was not satisfied locally, $i.e.$ at the form level, as the gauge bundle departed from the standard embedding. 
In contrast, we have seen that the smooth conifold solutions considered here can 
never be made globally torsion-free, as the required shift vector $\vec{p}$ is 
not physically sensible in this case. Hence from the point of view of the full six-dimensional heterotic compactification there 
is always a net $\mathcal{H}$-flux associated with the conifold throat. This is not a problem in itself, but implies that the compactification is globally 
endowed with torsion. 

In the regime where the string coupling in the 'bulk' of the flux compactification manifold is very small, 
one expects that quantities involving only the degrees of freedom 
localized in the throat can be accurately computed in the double-scaling limit, where the conifold flux background admits a worldsheet \textsc{cft} description. 
This aspect clearly deserves further study. 

\section*{Acknowledgements}
We would like to thank David Berenstein, Pablo C\'amara, Chris Hull, Josh Lapan, Vassilis Niarchos, Francesco Nitti, Carlos Nu\~nez, \'Angel Paredes, Boris Pioline,  Cobi Sonnenschein 
and Michele Trapletti for interesting and useful discussions. L.C. is supported by the Conseil r\'egional d'Ile de France, under convention 
N$^{\circ}$F-08-1196/R and the Holcim Stiftung zur F\"orderung der wissenschaftlichen Fortbildung.

\appendix

\section{Bosonization of the heterotic gauged WZW model}
\label{AppBoson}
In this appendix we give the explicit integration of the worldsheet gauge fields for the gauged \textsc{wzw} action defined in sec.~\ref{gaugedsec}, 
equations~(\ref{S-gauge}) and~(\ref{S-Fer}).

We start by bosonizing the worldsheet fermions as follows:
\begin{equation}
\left\{ 
\begin{array}{l}
\partial \Phi_{n} =   \zeta^{2n-1} \zeta^{2n}\\ 
\bar \partial \Phi_n = \bar \xi^{2n-1}  \bar \xi^{2n} 
\end{array} \right. \ , \quad n=1,2,3 \, .
\end{equation}
Gauging the symmetry $\delta_1 A= \partial\Lambda$, we may gauge fix $\psi_1 -\psi_2=0$, renaming $\psi_1+\psi_2=\psi$. Then, by
exploiting the remaining two gauge symmetries $\delta_2 B_1=\partial M$ and $\delta_3 \bar B_2=\bar\partial N$, we can set $\phi_L=0=\phi_R$.

Taking into account anomaly cancellation~\cite{Johnson:1994jw} (requiring in particular the absence of mixed anomalies) dictates the following bosonization of the action~(\ref{S-Fer}):
\begin{multline}
\label{S-bos}
S_{\text{Fer}} (A,{\bf B}) =
\frac{1}{4\pi} \int \di^2 z \,\Big[
|\p\Phi_1-(1+2\ell)A-B_2|^2 + |\p\Phi_2+A-B_1-B_2|^2 + |\p\Phi_3-2 B_2-\ell B_1|^2 \\
+ \Phi_1\,\big(F_2+(1-2\ell)F\big) + \Phi_2 \,\big( F_2-F_1-F \big) + \Phi_3\,\big( 2 F_2-\ell F_1\big)\\
-2\ell\big(A\bar B_2 + B_2 \bar A \big) +A\bar B_1+ B_1\bar A -(1+2\ell) \big(B_1\bar B_2 -B_2 \bar B_1 \big)
\Big]\,.
\end{multline}
The complete \textsc{wzw} model~(\ref{Stot}) is recast into the form:
\begin{multline}\label{Stot2}
S_{\textsc{wzw}}(A,{\bf{B}}) = \frac{1}{8\pi}\int \di^2 z\,\Big[ \left(\tfrac{k}{2}+2\right)|\p\rho|^2
+(k-2)\Big(|\p\theta_1|^2 + |\p\theta_2|^2 + |\p\phi_1|^2+ |\p\phi_2|^2 \\
+\big(\tfrac12\p\psi +\cos\theta_1\,\p\phi_1 +\cos \theta_2\,\p\phi_2 \big)\,\bar\p\psi\Big)
+2\sum_{i=1}^3 |\p\Phi_i|^2 \Big] \\
+ \frac{1}{2\pi}\int \di^2 z\, \Big[ (k+2\ell)|A|^2 + \left(\tfrac{k-2}{2}\big(\cos\theta_1\,\p\phi_1-\cos\theta_2\,\p\phi_2\big)+\p\Phi_2-\p\Phi_1\right)\bar A
-2\ell A\,\pb\Phi_1\\
-\tfrac{k+4}{2}\,\cosh\rho\,B_1\bar B_2 +(1+2\ell)B_2\bar B_1-B_1 (\pb\Phi_2+\ell\pb\Phi_3) \\
-\left(\tfrac{k-2}{2}\big(\p\psi+\cos\theta_1\,\p\phi_1+\cos\theta_2\,\p\phi_2\big) +\p\Phi_1 +\p\Phi_2+2\p\Phi_3\right)\bar B_2 \Big]\,.
\end{multline}
Taking the large $k$ limit (or rather the large $\ell$ limit in our case) of the above, the gauge fields can
be integrated out classically, leading to the non-linear sigma model:
\begin{multline}\label{Stot3}
S_{\textsc{wzw}} = \frac{k}{8\pi}\int \di^2 z\,\Big[ \tfrac{1}{2}|\p\rho|^2
+ |\p\theta_1|^2 + |\p\theta_2|^2 + |\p\phi_1|^2+ |\p\phi_2|^2 \\
+\big(\tfrac12\p\psi +\cos\theta_1\,\p\phi_2 +\cos\theta_2\,\p\phi_2 \big)\,\bar\p\psi \Big]\\
+\frac{1}{4\pi}\int \di^2 z\, \left[
\sum_{i=1}^3 |\p\Phi_i|^2 + 2\left( \cos\theta_1\,\p\phi_1-\cos\theta_2\,\p\phi_2
+\tfrac2k\big(\p\Phi_2-\p\Phi_1\big) \right) \ell \,\pb \Phi_1 \right.\\
\left.+\frac{2}{\cosh\rho}\left(
\p\psi+\cos\theta_1\,\p\phi_1+\cos\theta_2\,\p\phi_2 +\tfrac2k\big(\p\Phi_1 +\p\Phi_2+2\p\Phi_3\big)
\right)
\big(\pb\Phi_2+\ell\,\pb\Phi_3 \big)
  \right]
\end{multline}
In order to refermionize to a standard heterotic worldsheet action, the second part of the above sigma model has to be recast, following~\cite{Johnson:1994jw}, in a sort of Kaluza-Klein form. The corresponding Lagrangian density then reads:
\begin{multline}\label{KKform}
4\pi\,{\mathcal L}(\boldsymbol{\Phi}) = \big|\p\Phi_1+\ell\big( \cos\theta_1\,\p\phi_1-\cos\theta_2\,\p\phi_2\big)\big|^2 \\+ \Big|\tfrac1\ell\,\p\Phi_2+\p\Phi_3+\frac{\ell}{\cosh\rho}\big(\p\psi+ \cos\theta_1\,\p\phi_1+\cos\theta_2\,\p\phi_2\big)\Big|^2 \\
+\ell\Big[\big(\cos\theta_1\,\p\phi_1-\cos\theta_2\,\p\phi_2\big)\,\pb\Phi_1 - \p\Phi_1
\big(\cos\theta_1\,\pb\phi_1-\cos\theta_2\,\pb\phi_2\big)\Big] \\
+ \frac{\ell}{\cosh\rho}
\Big[\big(\p\psi+\cos\theta_1\,\p\phi_1+\cos\theta_2\,\p\phi_2\big)\big(\pb\Phi_2+\ell\pb\Phi_3)\\ -
\big(\p\Phi_2+\ell\p\Phi_3)\big(\pb\psi+\cos\theta_1\,\p\phi_1+\cos\theta_2\,\pb\phi_2\big)
\Big] \\
-\ell^2 \big| \cos\theta_1\,\p\phi_1-\cos\theta_2\,\p\phi_2 \big|^2
- \frac{\ell^2}{\cosh^2\rho} \big|\p\psi+ \cos\theta_1\,\p\phi_1+\cos\theta_2\,\p\phi_2 \big|^2\\
+ \frac{4\ell}{k}\big(\p\Phi_2-\p\Phi_1\big)\,\pb\Phi_1
+\frac{4}{k\,\cosh\rho}\big(\p\Phi_1 +\p\Phi_2+2\p\Phi_3 \big)\big( \pb\Phi_2+\ell\pb\Phi_3\big)\,.
\end{multline}
Then, upon refermionization, one arrives at the non-linear sigma-model given in eq.~(\ref{Sbos-final}) and~(\ref{Sfer-final}).

\section{${\mathcal N}=2$ characters and useful identities}
\label{appchar}
\boldmath
\subsection*{$\mathcal{N} =2$ minimal models}
\unboldmath
The characters of the $\mathcal{N} =2$ minimal models, i.e.  the supersymmetric $SU(2)_k / U(1)$ gauged \textsc{wzw} models, are conveniently defined through the characters $C^{j\ (s)}_{m}$ of the $[SU(2)_{k-2} \times U(1)_2] / U(1)_k$ bosonic 
coset, obtained by splitting the Ramond and Neveu--Schwartz 
sectors according to the fermion number mod 2~\cite{Gepner:1987qi}. These characters are determined implicitly through the
identity:
\begin{equation}
\chi^{j} (\tau,\nu)
\Theta_{s,2}(\tau,\nu-\nu') = \sum_{m \in \zi_{2k}} C^{j\ (s)}_{m} (\tau,\nu')  \Theta_{m,k} (\tau,\nu-\tfrac{2\nu'}{k}) \, ,
\end{equation}in terms of the theta functions of $\widehat{\mathfrak{su} (2)}$
at level $k$, defined as 
\begin{equation}
\Theta_{m,k} (\tau,\nu) = \sum_{n \in \mathbb{Z}}
q^{k\left(n+\tfrac{m}{2k}\right)^2}
e^{2i\pi \nu k \left(n+\tfrac{m}{2k}\right)}  \qquad m \in \mathbb{Z}_{2k}\, ,
\end{equation}
and $\chi^j (\tau,\nu)$ the  characters of the $\widehat{\mathfrak{su}(2)}$ affine 
algebra  at level $k-2$. Highest-weight representations are labeled by  $(j,m,s)$, corresponding to primaries of 
$SU(2)_{k-2}\times U(1)_k \times U(1)_2$. The following identifications apply:
\begin{equation}
(j,m,s) \sim (j,m+2k,s)\sim
 (j,m,s+4)\sim
 (k/2-j-1,m+k,s+2)
\end{equation}
as  the selection rule $2j+m+s =  0  \mod 2$. The spin $j$ is restricted to $0\leqslant j \leqslant \tfrac{k}{2}-1$.  
The conformal weights of the superconformal primary states are:
\begin{equation}
\begin{array}{cclccc}
\Delta &=& \frac{j(j+1)}{k} - \frac{n^2}{4k} + \frac{s^2}{8} \ & \text{for} & \ -2j \leqslant n-s \leqslant 2j \\
\Delta &=& \frac{j(j+1)}{k} - \frac{n^2}{4k} + \frac{s^2}{8} + \frac{n-s-2j}{2}
\ & \text{for} & \ 2j \leqslant n-s \leqslant 2k-2j-4 \\
\end{array}
\end{equation}
and their $R$-charge reads:
\begin{equation}
Q_R = \frac{s}{2}-\frac{m}{k} \mod 2 \,. 
\end{equation}
A {\it chiral} primary state is obtained for $m=2(j+1)$ and 
$s=2$ (thus odd fermion number). It has conformal dimension
\begin{equation}
\Delta= \frac{Q_R}{2} = \frac{1}{2} - \frac{j+1}{k}\, .
\end{equation}
An {\it anti-chiral} primary state is obtained for $m=2j$ and $s=0$ 
(thus even fermion number). Its conformal dimension reads:
\begin{equation}
\Delta= -\frac{Q_R}{2} = \frac{j}{k}\, .
\end{equation}
Finally we have the following modular S-matrix for the $\mathcal{N}=2$ minimal-model characters:
\begin{equation}
S^{jm s}_{j' m' s'} = \frac{1}{2k} \sin \pi
\frac{(1+2j)(1+2j')}{k} \ e^{i\pi \frac{mm'}{k}}\ e^{-i\pi ss'/2}.
\end{equation}
The usual Ramond and Neveu--Schwarz characters, that we use in the bulk of the paper, are  obtained as:
\begin{equation}
\mathcal{C}^{j}_{m} \oao{a}{b} =  e^{\frac{i\pi ab}{2}} \left[ \mathcal{C}^{j\, (a)}_{m} 
+(-)^b \mathcal{C}^{j\, (a+2)}_{m} \right]
\end{equation}
where $a=0$ (resp. $a=1$) denote the \textsc{ns} (resp. \textsc{r}) sector, and characters 
with $b=1$ are twisted by $(-)^F$. They are related to $\widehat{\mathfrak{su}(2)}_k$ characters 
through:
\begin{equation}
\chi^j  \vartheta \oao{a}{b} = \sum_{m \in \zi_{2k}} C^j_m \oao{a}{b} \Theta_{m,k}\,.
\end{equation}
In terms of those one has the reflexion symmetry:
\begin{equation}
C^j_m \oao{a}{b} = (-)^b C^{\tfrac{k}{2}-j-1}_{m+k} \oao{a}{b}\, . 
\label{reflsym}
\end{equation} 

\subsection*{Supersymmetric $\slc$}
The characters of the $\slc$ super-coset
at level $k'$ come in different categories corresponding to
irreducible unitary representations of  $SL(2,\mathbb{R})$.

\noindent The \emph{continuous representations} correspond to $J = 1/2 + iP$,
$P \in \mathbb{R}^+$. Their characters are denoted by
 ${\rm ch}_c (\tfrac{1}{2}+ip,M) \oao{a}{b}$, where the $U(1)_R$ charge of the primary is $Q=2M/k'$. They read:
\begin{equation}
{\rm ch}_c (\tfrac{1}{2}+ip,M;\tau,\nu) \oao{a}{b} = \frac{1}{\eta^3 (\tau)} q^{\frac{p^2+M^2}{k'}} \vartheta \oao{a}{b} (\tau,\nu)
e^{2i\pi\nu \frac{2M}{k'}}\, .
\end{equation}
The \emph{discrete representations}, of characters $\mathrm{ch}_d (J,r) \oao{a}{b}$,
have a real $\slr$ spin in the range $1/2 < J < (k'+1)/2$. Their  $U(1)_R$  charge reads  
\begin{equation}
Q_R=\frac{2(J+r+a/2)}{k'}\quad , \qquad r\in \zi\, .
\label{Rchargecoset}
\end{equation}  
Their  characters are given by 
\begin{equation}
{\rm ch}_d (J,r;\tau,\nu) \oao{a}{b} =  \frac{
  q^{\frac{-(J-1/2)^2+(J+r+a/2)^2}{k'}}
e^{2i\pi\nu \frac{2J+2r+a}{k'}}}{1+(-)^b \,
e^{2i\pi \nu} q^{1/2+r+a/2} } \frac{\vartheta \oao{a}{b} (\tau, \nu)}{\eta^3 (\tau)}.
\label{idchar}
\end{equation}
One gets a {\it chiral} primary for $r=0$, i.e. $M=J$, in the \textsc{ns} sector  (with even fermion number). Its conformal dimension reads
\begin{equation}
\Delta= \frac{Q_R}{2} = \frac{J}{k'}\, . 
\end{equation}An {\it anti-chiral} primary is obtained for $r=-1$ (with odd fermion number). Its conformal dimension reads
\begin{equation}
\Delta= -\frac{Q_R}{2} =\frac{1}{2}-\frac{J-1}{k'}\, . 
\end{equation} 
\emph{Extended characters} are defined for $k'$ integer by summing
over $k'$ units of spectral flow~\cite{Eguchi:2003ik}.\footnote{One can extend their definition to the case of rational $k'$, which 
is not usefull here.} For instance, the extended continuous characters are:
\begin{multline}
{\rm Ch}_c (\tfrac{1}{2}+ip,M;\tau,\nu) \oao{a}{b}=   \sum_{w \in \zi} 
{\rm ch}_c (\tfrac{1}{2}+ip,M+k'w;\tau,\nu) \oao{a}{b} \\
\hspace{-1cm}= \frac{q^{\frac{p^2}{k'}}}{\eta^3 (\tau)} \vartheta \oao{a}{b} (\tau,\nu)
\Theta_{2M,k'} (\tau,\tfrac{2\nu}{k'})
\label{extcontchar}
\end{multline}
where discrete $\mathcal{N}=2$  R-charges are chosen: $2M \in \zi_{2k'}$. 
These characters close among themselves under the action of the modular group. 
For instance, the S transformation gives:
\begin{equation}
{\rm Ch}_{c} (\tfrac{1}{2}+ip,M;-\tfrac{1}{\tau}) \oao{a}{b}
= \frac{1}{2k'}\int_0^\infty \!\!\! \di p' \, \cos \frac{4\pi p p'}{k'}\!\! \sum_{2M' \in \mathbb{Z}_{2k'}} \!\!
e^{-\frac{4i\pi M M'}{k'}}
{\rm Ch}_{c} (\tfrac{1}{2}+ip',M';\tau) \oao{b}{-a}\, .
\end{equation}
The same holds for discrete representations, whose modular transformations are more involved 
(see~\cite{Eguchi:2003ik,Israel:2004xj}).

\bibliography{bibbundle}

\end{document}